\documentclass{emulateapj}
\usepackage{graphics,graphicx,rotating}
\usepackage[colorlinks,allcolors=blue]{hyperref}
\usepackage[all]{hypcap} 
\usepackage{xspace}
\usepackage{amssymb}
\usepackage{amsmath}      
\usepackage{epstopdf}
\usepackage[titletoc]{appendix}
\usepackage{subfigure}
\usepackage{xcolor}
\usepackage{natbib}
\newcommand{\hst}{{\it HST}~}

\begin{document}
\title{The most massive galaxies with large depleted cores: structural parameter relations and black
  hole masses}
  \shorttitle{Depleted cores, galaxy scaling relations and SMBHs}
\shortauthors{Dullo} \author{\color{blue} Bililign T. Dullo}
\affil{Departamento de F\'isica de la Tierra y Astrof\'isica,
  Instituto de F\'isica de Part\'iculas y del Cosmos IPARCOS,
  Universidad Complutense de Madrid, E-28040 Madrid, Spain;
   {\color{blue} bdullo@ucm.es}}

\begin{abstract}
                    
  Luminous spheroids ($M_{V} \la - 21.50 \pm 0.75$ mag) contain
  partially depleted cores with sizes ($R_{\rm b}$) typically
  \mbox{0.02 -- 0.5 kpc}. However, galaxies with $R_{\rm b} >$ 0.5 kpc
  are rare and poorly understood. Here we perform detailed
  decompositions of the composite surface brightness profiles,
  extracted from archival {\it Hubble Space Telescope} and
  ground-based images, of 12 extremely luminous ``large-core''
  galaxies that have $R_{\rm b} >$ 0.5 kpc and
  \mbox{$M_{V} \la -23.50 \pm 0.10$ mag}, fitting a core-S\'ersic
  model to the galaxy spheroids.  Using 28 ``normal-core'' (i.e.,
  $R_{\rm b} < $ 0.5 kpc) galaxies and 1 ``large-core'' (i.e.,
  $R_{\rm b} >$ 0.5 kpc) galaxy from the literature, we constructed a
  final sample of 41 core-S\'ersic galaxies. We find that large-core
  spheroids (with stellar masses $M_{*} \ga 10^{12}M_{\sun}$) are not
  simple high-mass extensions of the less luminous normal-core
  spheroids having $ M_{*} \sim 8\times 10^{10} -
  10^{12}M_{\sun}$. While the two types follow the same strong
  relations between the spheroid luminosity $L_{V}$ and $R_{\rm b}$
  ($R_{\rm b} \propto L_{V}^{1.38 \pm 0.13}$), and the spheroid
  half-light radius $R_{\rm e}$
  ($R_{\rm e} \propto L^{1.08 \pm 0.09}_{V}$, for ellipticals plus
  BCGs), we discover a break in the core-S\'ersic $\sigma-L_{V}$
  relation occurring at \mbox{$M_{V} \sim -23.50 \pm 0.10$ mag}.
  Furthermore, we find a strong log-linear $R_{\rm b}-M_{\rm BH}$
  relation for the 11 galaxies in the sample with directly determined
  SMBH masses $M_{\rm BH}$---3/11 galaxies are large-core
  galaxies---such that $R_{\rm b} \propto M_{\rm BH}^{0.83 \pm
    0.10}$. However, for the large-core galaxies the SMBH masses
  estimated from the $M_{\rm BH}-\sigma$ and core-S\'ersic
  $M_{\rm BH}-L$ relations are undermassive, by up to a factor of 40,
  relative to expectations from their large $R_{\rm b}$ values,
  confirming earlier results. Our findings suggest that large-core
  galaxies harbour overmassive SMBHs
  ($M_{\rm BH} \ga 10^{10} M_{\sun}$), considerably
  ($\sim 3.7-15.6 \sigma$ and $\sim 0.6-1.7 \sigma$) larger than
  expectations from the spheroid $\sigma$ and $L$, respectively.  We
  suggest that the $R_{\rm b}-M_{\rm BH}$ relation can be used to
  estimate SMBH masses in the most massive galaxies.

\end{abstract}


\section{Introduction}

It is now believed that all massive galaxies contain a supermassive
black hole (SMBH) at their centre
(\citealt{1998AJ....115.2285M,1998Natur.395A..14R,2005SSRv..116..523F}).
The connection between SMBHs and the properties of their host galaxies
has been a subject of ongoing interest (see
\citealt{2013ARA&A..51..511K,2016ASSL..418..263G} for recent reviews).
SMBH masses ($M_{\rm BH} $) scale with a wide range of host galaxy
properties such as the stellar velocity dispersion ($\sigma$;
\citealt{2000ApJ...539L...9F,2000ApJ...539L..13G}) and bulge
luminosity ($L$;
\citealt{1995ARA&A..33..581K,1998AJ....115.2285M,2003ApJ...589L..21M}). In
addition, observations reveal partially depleted cores---a flattening
in the inner stellar light distributions of galaxies---that are
explained as an imprint left by binary SMBHs on the central structures
of their host galaxies.

Luminous spheroids ($M_{V} \la - 21.50 \pm 0.75$ mag), for the most
part, possess depleted cores. In the hierarchical structure formation
model, the brightest and most massive galaxies are built through
generations of galaxy merger events (e.g.,
\citealt{1972ApJ...178..623T,1978MNRAS.183..341W,1982ApJ...252..455S,1988ApJ...331..699B,1993MNRAS.264..201K,2007MNRAS.375....2D,2013MNRAS.435..901L}). SMBH
binaries invariably form in such galaxy mergers (e.g.,
\citealt{2003ApJ...582L..15K,2006ApJ...646...49R,2008MNRAS.386..105B,2011MNRAS.410.2113B,2012ApJ...744....7B,2015ApJ...806..219C,2019arXiv190703757G}). \mbox{N-body}
simulations suggest that depleted cores are generated via three-body
interactions between inner stars from the galaxy core regions and
orbitally decaying SMBH binaries which form in major, $\lq\lq$dry"
(gas-poor) mergers of galaxies (e.g.,
\citealt{1980Natur.287..307B,1991Natur.354..212E,2001ApJ...563...34M,2006ApJ...648..976M,2009ApJS..181..486H}).
Due to only radial orbits being capable of bringing inner stars in
close proximity to the SMBH binary, the binary scouring process leaves
a relative excess of tangential orbits in the galaxy cores
(\citealt{1997NewA....2..533Q,2001ApJ...563...34M,
  2003ApJ...583...92G,2014ApJ...782...39T,2016Natur.532..340T,2018ApJ...864..113R}). Also,
in \citet{2015ApJ...798...55D} we found a tendency for the depleted
core regions of luminous galaxies to be round.  In this paper we focus
on the brightest galaxies which are expected to host the most massive
SMBHs, making them excellent structural probes of extreme cases of
core depletion caused by the cumulative actions of massive binary
SMBHs.

Earlier studies of depleted cores using ground-based observations
lacked the spatial resolution to reveal cores with small angular sizes
in enough detail (e.g.,
\citealt{1966ApJ...143.1002K,1978ApJ...222....1K,1978ApJ...221..721Y,1982MNRAS.200..361B,1985ApJS...57..473L}). The
availability of high-resolution {\it Hubble Space Telescope (\it HST)}
imaging has subsequently allowed us to resolve such small cores and
properly characterise depleted cores of ``core-S\'ersic'' galaxies
(e.g.,
\citealt{1993AJ....106.1371C,1994ESOC...49..147K,1994AJ....108.1567J,
  1994AJ....108.1598F,1994AJ....108.1579V,1995AJ....110.2622L,1996AJ....111.1889B,1996AJ....112..105G,1997AJ....114.1771F,2001AJ....122..653R,2001AJ....121.2431R,2003AJ....125..478L,2003AJ....125.2951G,2006ApJS..164..334F,2007ApJ...664..226L,2007ApJ...662..808L,2011MNRAS.415.2158R,2012ApJ...755..163D,2013ApJ...768...36D,2014MNRAS.444.2700D,2017MNRAS.471.2321D,2013AJ....146..160R}). However,
as done in a few of these studies (e.g.,
\citealt{2003AJ....125.2951G,2006ApJS..164..334F,2012ApJ...755..163D,2013ApJ...768...36D,2014MNRAS.444.2700D,2013AJ....146..160R})
reliably determining if a flat inner core in a galaxy actually
reflects a deficit of stars relative to the inward extrapolation of
the spheroid's outer profile relies on careful modelling of the galaxy
light profile using the core-S\'ersic model.  Applying the
core-S\'ersic model \citet[see also
\citealt{2013ApJ...768...36D}]{2012ApJ...755..163D} showed that 18 per
cent of ``cores'' according to the Nuker model
(\citealt{1995AJ....110.2622L,2007ApJ...664..226L}) were actually
misidentified S\'ersic spheroids with low S\'ersic indices and no
depleted cores (see also \citealt{2003AJ....125.2951G};
\citealt{2004AJ....127.1917T}; \citealt{2014MNRAS.444.2700D}). Cores
measured using the Nuker model break
radii\footnote{\citet{2003AJ....125.2951G} revealed that the Nuker
  model parameters are unstable and deviate from the true values as
  larger radial extents are probed by the light profile fitting.}  are
also typically 2$-$3 times larger than the core-S\'ersic model break
radii, $R_{\rm b}$, (e.g.,
\citealt{2004AJ....127.1917T,2006ApJS..164..334F,2011MNRAS.415.2158R,2012ApJ...755..163D,2013ApJ...768...36D,2014MNRAS.444.2700D}). \citet[Appendix
C]{2007ApJ...662..808L} advocated the use of the \lq\lq cusp radius'',
i.e., the radius where the negative logarithmic slope of the fitted
Nuker model equals 1/2 ($r_{{\gamma}^{'}=1/2}$,
\citealt{1997ApJ...481..710C}), as a measure of the core size. While
the ``cusp radius'' somewhat agrees with the core-S\'ersic break
radius (\citealt{2012ApJ...755..163D}), its application fails to
discriminate whether galaxies contain a partially depleted core or
not, since all galaxy light profiles have a cusp radius.

In the past few decades, a few papers (e.g.,
\citealt{1997AJ....114.1771F,2007ApJ...662..808L,2014MNRAS.444.2700D,2013AJ....146..160R,2016Natur.532..340T})
have shown strong scaling relations involving the structural
parameters of core-S\'ersic spheroids.  Unfortunately, these scaling
relations were established using core-S\'ersic spheroids having
``normal'' size cores (i.e., $R_{\rm b} \sim 20 - 500$ pc) and SMBH
masses $M_{\rm BH} \la 3 \times 10^{9} M_{\sun}$. The behavior of
such scaling relations remains unknown for the most luminous
core-S\'ersic galaxies with ``large'' size cores (i.e., $R_{\rm b} >$
0.5 kpc).  In addition, SMBHs with masses of order $10^{10} M_{\sun}$
are hosted by high-luminosity quasars at high redshift
\citep[e.g.,][]{2015Natur.518..512W} and recent observations have
found them in a few extremely massive, present-day galaxies
\citep{2011Natur.480..215M,2016Natur.532..340T,2019arXiv190710608M}. These are important
to constrain the SMBH scaling relations at high masses
($M_{\rm BH} \ga 3 \times 10^{9} M_{\sun}$). Given the luminosity
function of galaxies
(\citealt[e.g.,][]{1988MNRAS.232..431E,2001ApJ...560..566K,2003ApJ...599...38B,2012MNRAS.421..621B}),
it implies that ``large-core'' galaxies are rare, although they are
becoming increasingly common as more galaxies with high luminosity are
modelled (e.g., NGC 6166, \citealt{2007ApJ...662..808L}; 4C +74.13,
\citealt{2009ApJ...698..594M}; A2261-BCG,
\citealt{2012ApJ...756..159P,2016ApJ...829...81B}; NGC 4486 and NGC
4889, \citealt{2013AJ....146..160R}; NGC 1600,
\citealt{2016Natur.532..340T}; A2029-BCG,
\citealt{2017MNRAS.471.2321D}; A1689-BCG,
\citealt{2017ApJ...849....6A}).  However and as noted above, caution
should be exercised when interpreting large depleted cores,
particularly those identified by the Nuker model (e.g.,
\citealt{2007ApJ...662..808L,2014ApJ...795L..31L}). For example,
\citet{2014ApJ...795L..31L} claimed that the BCG Holm 15A has the
largest core size measured in any galaxy to date
($r_{{\gamma}^{'}=1/2} \sim 4.6$ kpc) based on their Nuker model
analysis.  In contrast,
\citet{2015ApJ...807..136B,2016ApJ...819...50M} did not identify a
central stellar deficit relative to the spheroid's outer S\'ersic
profile in their modeling of the  galaxy's CFHT and Gemini data,
respectively, while \citet{2019arXiv190710608M} fit the 2D
\mbox{core-S\'ersic}+S\'ersic+GaussianRing3D model to their Wendelstein
image of the galaxy and reported a core size \mbox{$R_{\rm b } \sim 2.8$
  kpc}.

Accurate extension of the galaxy structural scaling relations to the
most massive galaxies carries valuable clues about the supposed joint
evolution of SMBHs and their host spheroids.  Of particular relevance
is the observed correlation between the mass of the SMBH
($M_{\rm BH}$) and the size of the depleted core ($R_{\rm
  b}$). Absence of a bend/offset in the $R_{\rm b}-M_{\rm BH}$
relation for the most massive spheroids would imply the core size is a
good predictor of SMBH masses at the high mass end, more reliable than
$\sigma$ (e.g., \citealt{2016Natur.532..340T}).
\citet{2007ApJ...662..808L} noted that SMBH masses for the most
luminous galaxies are overmassive relative to the inference from the
high mass end of the \mbox{$M_{\rm BH}-\sigma$} relation, but they are
in better agreement with those from \mbox{$M_{\rm BH}-L$} relation
(see also \citealt{2013ApJ...768...29V}). The $M_{\rm BH}-\sigma$
relation predicts $M_{\rm BH}$ for the most massive galaxies (i.e.,
\mbox{$\sigma \sim 300 -390$ km s$^{-1}$},
\citealt{2007ApJ...662..808L,2007AJ....133.1741B}) cannot exceed
$M_{\rm BH} \sim 5 \times 10^{9} M_{\sun}$, whereas predicted
$M_{\rm BH}$ from the $M_{\rm BH}-L$ relation can surpass
$M_{\rm BH} \sim 10^{10} M_{\sun}$
(\citealt{2007ApJ...662..808L}). The rationale for this discrepancy is
that the most massive spheroids are expected to undergo a larger
number of dry major mergers that increase their stellar mass, black hole
mass and size, while keeping their velocity dispersion relatively
unaffected (e.g.,
\citealt{2003MNRAS.342..501N,2007ApJ...658...65C,2012ApJ...744...63O,2013MNRAS.429.2924H}). This
is also evident from the broken $\sigma-L$ relation of elliptical
galaxies \citep{1976ApJ...204..668F} which displays a shallower slope
at bright magnitudes $M_{V} \la -21.5$ mag
(\mbox{$\sigma \propto L^{1/(5-8)}$}, e.g.,
\citealt{1981ApJ...251..508M,2007ApJ...662..808L,2013ApJ...769L...5K}),
whereas at fainter magnitudes \mbox{$\sigma \propto L^{1/2}$} (e.g.,
\citealt{1983ApJ...266...41D,1992AJ....103..851H,2005MNRAS.362..289M}). This
change in the slope of the $\sigma-L$ relation matches the
core-S\'ersic versus S\'ersic structural divide (e.g., \citealt{2019arXiv190806838S}).

Here, we perform careful, multi-component (halo/intermediate-scale
component/spheroid/nucleus) decompositions of the new light profiles
of 12 ``large-core'' galaxies (i.e., 9 BCGs, 2 second brightest cluster galaxies
and 1 brightest group galaxy) using a
core-S\'ersic model and a S\'ersic model.  Together with the
large-core BCG IC~1101 \citep{2017MNRAS.471.2321D}, these 13 galaxies
constitute the largest sample of ``large-core'' galaxies studied to
date. We aim to revisit the structural scaling relations of our full
sample of 41 core-S\'ersic galaxies (13 ``large-core'' galaxies plus
28 ``normal-core'' galaxies, \citealt{2014MNRAS.444.2700D}) over a
comprehensive dynamic range in core size ($R_{\rm b} \sim 0.02 - 4.2$ kpc),
spheroid luminosity ($ - 20.70$ mag $\ga M_{V} \ga - $25.40 mag),
spheroid stellar mass
($ M_{*} \sim 8\times 10^{10} - 7\times 10^{12}M_{\sun}$) and SMBH
mass ($M_{\rm BH} \sim 2\times 10^{8} - 2\times 10^{10}M_{\sun}$).

The paper is organised as follows. Data and photometry for our new
sample of 12 large-core galaxies are presented in
Section~\ref{Sec2}. We then discuss the analytical models employed and
multi-component decompositions of these 12 large-core galaxies in
Section~\ref{Sec3}. Accurate structural relations for our full sample
of 41 core-S\'ersic galaxies are presented in Section~\ref{Sec4}. We
go on to discuss the stellar mass deficits, galaxy environment and
formation of normal- and large-core galaxies in
Section~\ref{Sec5}. Section~\ref{ConV} summarizes our main
conclusions. There are two appendices at the end of this paper
(Appendices \ref{AppA} and \ref{AppB}). Appendix~\ref{AppA} shows the
multi-component decompositions of the new major-axis surface
brightness profiles of the 12 large-core galaxies. In
Appendix~\ref{AppB} we show the spatial distribution of large-core
galaxies and their nearest neighbors.

Throughout this paper, we assume a cosmology with $H_{0}$ = 70 km
$\rm{s}^{-1}$ Mpc$^{-1}$, $\Omega_{\Lambda}$ = 0.7, and $\Omega_{m}$ =
0.3 and quote magnitudes in the Vega system, unless noted otherwise.

 \begin{figure*}
\hspace*{-1.706846230145072599cm}    
 \includegraphics[angle=0,scale=.708]{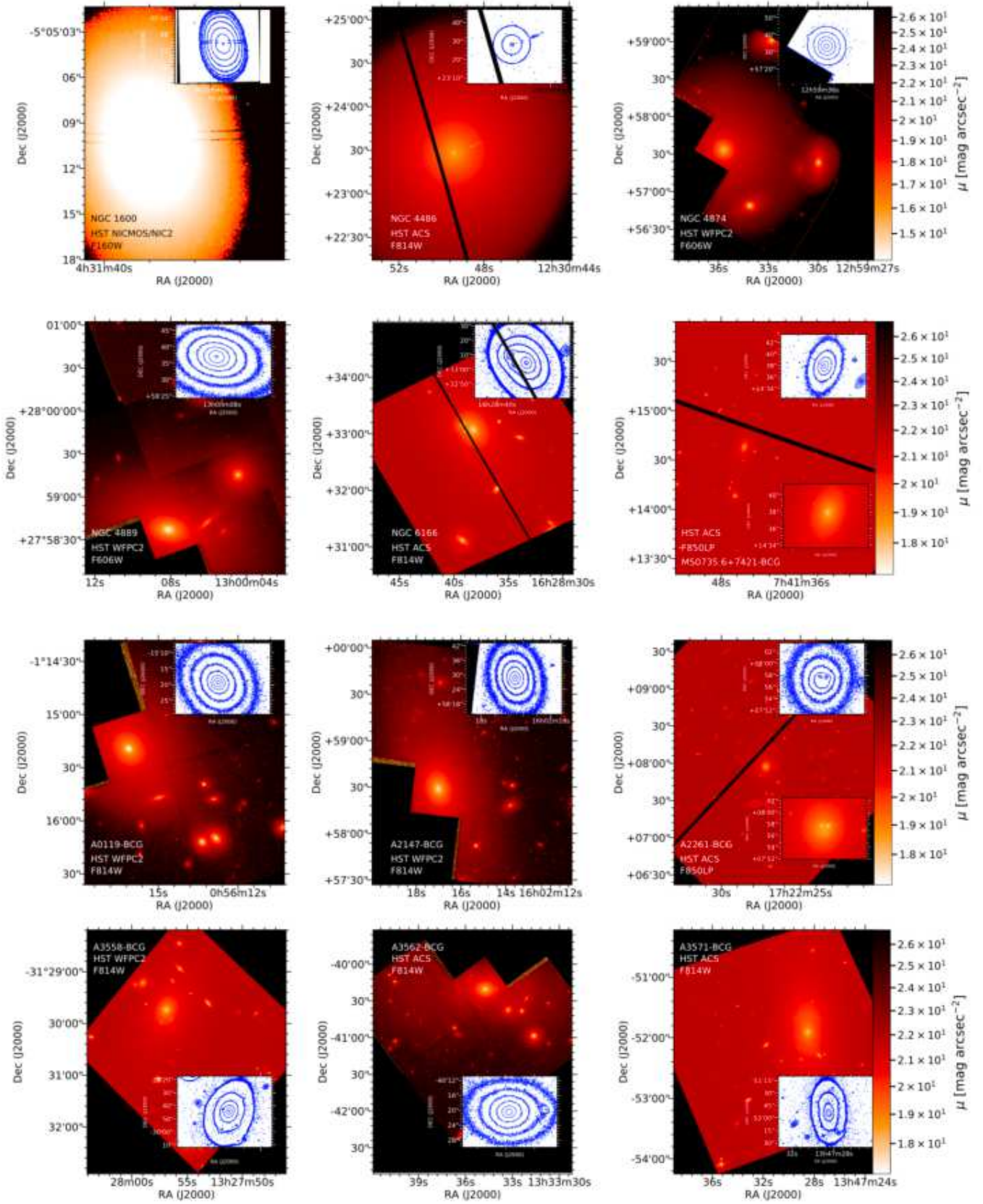}
\caption{ {\it HST} images of our sample of 12 massive galaxies with
   large break radii $R_{\rm b} > 0.5$ kpc (Tables~\ref{Table2} and
  \ref{Table3}).  The insets show the contour levels in steps of 0.6
   mag arcsec$^{-2}$. For 4C +74.13 and A2261-BCG, we show zoomed-in
   regions centered on the BCGs. North is up and east is to the left.}
 \label{Fig1} 
  \end{figure*}

\begin{center}
\begin{table} 
\setlength{\tabcolsep}{0.038191209948in}
\caption{Global properties for our full sample of 13 core-S\'ersic galaxies
with large break radii (i.e., $R_{\rm b} > 0.5$ kpc).}
\label{Table1}
\begin{tabular}{@{}llcccccc@{}}
\hline
\hline
Cluster or Group&BCG or BGG& Type & D &$\sigma $\\
&&&(Mpc)&(km s$^{-1}$)&\\
(1)&(2)&(3)&(4)&(5)&\\
\multicolumn{1}{c}{} \\              
\hline                           
NGC 1600                      & NGC 1600                 & E3                      &66.0         &331\\
Virgo                              &   NGC 4486$^{\dagger}$      & cD pec         &    23.0 &323\\
Coma/Abell 1656          &NGC 4874$^{\dagger}$         & cD0          &  106.4   &271\\
Coma/Abell 1656          &NGC 4889                  &cD4      & 96.6 &347[2a]\\
Abell 2199                     &  NGC 6166               & cD2 pec & 130.4   &300[4a]&\\
MS0735.6+7421            & 4C +74.13          &cD         & 925.3   &239[1a]&\\
Abell 0119                     & UGC 579                  & E     &
                                                                    185.7&287\\
Abell 2029                     & IC 1101         &E             &363.0 & 378\\
Abell 2147                     & UGC 10143              &cD           & 153.4&276\\
Abell 2261                     & A2261-BCG            &cD                               &   958.8 &387[3a]\\
Abell 3558                     & ESO 444-G46          & cD4            & 204.9  &248\\
Abell 3562                     & ESO 444-G72          &  $\rm SAB(rs)0^{0}$      &213.3 &236\\
Abell 3571                     & ESO 383-G76          &cD5             &169.0& 322\\

\hline
\end{tabular} 

Notes.---Col.\ (1): cluster or group name.  All the galaxies in
our sample except for the dominant group galaxy NGC~1600 reside in
clusters.  Col.\ (2): the superscript  `$\dagger$'  denotes the
two second brightest cluster elliptical galaxies in the sample which inhabit the
centers of their clusters (Fig.~\ref{Fig1}).  An alternative designation for A2261-BCG  is 
2MASX J17222717+3207571.
Col.\ (3): classification came from
NED\footnote{http://ned.ipac.caltech.edu}.  Col.\ (4): distances (D) came from
NED (3K CMB), assuming $H_{0} = 70$ km s$^{-1}$ Mpc$^{-1}$. Col.\ (5): central
velocity dispersions ($\sigma$) are taken from
HyperLeda\footnote{(http://leda.univ-lyon1.fr)} \citep{2003A&A...412...45P}
unless the source is indicated. Sources:  [1a]
\citet{2009ApJ...698..594M}; [2a]
\citet{2011Natur.480..215M}; [3a] \citet{2012ApJ...756..159P}; and [4a] \citet{2015ApJ...807...56B}.
\end{table}
\end{center}

\section{Data and photometry}\label{Sec2}

\subsection{Sample Selection}\label{Sec2.1}


We searched the
literature for galaxies which were suspected to have a core-S\'ersic
break radius $R_{\rm b} > 0.5$ kpc and with archival high-resolution
{\it Hubble Space Telescope (HST)} imaging. This resulted in a
selected sample of 13 galaxies (see Table~\ref{Table1}). The case of
the BCG IC~1101 is published in \citet{2017MNRAS.471.2321D} and  the
remaining 12 galaxies consist of 9 galaxies with $r_{\gamma} > 0.5$ kpc
drawn from the \citet{2007ApJ...664..226L}, 4C +74.13 \citep[$r_{\gamma} = 1.54$
kpc]{2009ApJ...698..594M}, the BCG of Abell A2261
\citep[$r_{\gamma} = 3.95$ kpc]{2012ApJ...756..159P} and the giant cD
galaxy NGC 4486 (\citealt{2011MNRAS.415.2158R}, $ R_{\rm b} \sim 0.5$
kpc; \citealt{2013AJ....146..160R}, $R_{\rm b} \sim 0.7$ kpc). Basic
data for the full sample of 13 galaxies are listed in Table~\ref{Table1}.

Fitting the `sharp-transition' ($\alpha \rightarrow \infty$)
core-S\'ersic model to the inner 10$\arcsec$  {\it HST} ACS/HRC light profiles of 23
massive galaxies, \citet[their
table 2]{2008MNRAS.391.1559H} measured $R_{\rm b} > 0.5$ kpc for 5/23
galaxies with high velocity dispersion ($\sigma >$ 350 km s$^{-1}$).
These five galaxies are not included here, suspecting that the
sharp-transition core-S\'ersic model used by
\citet{2008MNRAS.391.1559H} may have caused the break radii of the
galaxies to be overestimated, as was the case for the BCG SDSS
J091944.2+562201.1. \citet{2008MNRAS.391.1559H} reported a large break
radius of $R_{\rm b} \sim 1.54$ kpc for this galaxy which was later
found to have a much smaller break radius ($R_{\rm b} \sim 0.55$ kpc)
after fitting a smoother transition (i.e., $\alpha \sim 1.2$)
core-S\'ersic model to the galaxy's two-dimensional light distribution
\citep{2016ApJ...829...81B}.  After we complete the analysis in this paper,
\citet{2017ApJ...849....6A} reported a large break radius of 3.8 kpc
for the brightest cluster galaxy of Abell 1689. We did not include
this galaxy in the paper.

\subsection{Classification}\label{Sec2.2}

All the galaxies in our sample except for three (NGC 1600, NGC 4486
and NGC 4874) are classified as BCGs (\citealt{2007ApJ...662..808L,
  2017MNRAS.471.2321D}), see Table~\ref{Table1}. The elliptical galaxy
\mbox{NGC 1600} is the brightest member of the poor \mbox{NGC 1600}
group. The giant elliptical NGC 4486, which resides at the heart of
the Virgo cluster, is the second brightest galaxy in the cluster, only
$\sim$ 0.2 mag fainter than the cluster's brightest elliptical galaxy
NGC 4472 (\citealt{2011Natur.480..215M}).  Akin to NGC 4486, the giant
elliptical NGC 4874 is the second brightest galaxy sitting at the
center of the Coma cluster.

\subsection{ Archival galaxy images and surface brightness profiles}

We used high-resolution {\it HST} imaging of the galaxies obtained
mainly in the Wide Field Planetary Camera 2 (WFPC2) F814W, Advanced
Camera for Surveys (ACS) WFC F814W/F850LP and Near Infrared Camera and
Multi-Object Spectrometer (NICMOS) NIC2 F160W filters in order to
minimise the obscuring effects of dust.  {\it HST} WFPC2 F606W images
were used for 2/12 galaxies with no obvious dust absorption (NGC 4874
and NGC 4889).  These {\it HST} images were all retrieved from the
Hubble Legacy Archive (HLA\footnote{https://hla.stsci.edu}) and
processed through the standard HLA data reduction
pipeline. Table~\ref{Table2} lists the {\it HST} programs, instruments
and filters for the sample of 12 galaxies studied in this paper.  The
{\it HST} F814W, F850LP and F160W filters are analog to the
Johnson-Cousins $I$-band, $SDSS~z$-band and Johnson-Cousins
\mbox{$H$-band} filters, respectively. The full WFPC2 detector is a
mosaic of three wide field cameras plus a smaller high-resolution
planetary camera, yielding a 160$\arcsec$ $\times$ 160$\arcsec$
L-shaped field-of-view (FOV).  The mosaic of the two ACS WFC CCD
cameras covers a $\sim$202$\arcsec$ $\times$ 202$\arcsec$ rhomboidal
area.  The NIC2 images have a relatively smaller field of view of
19$\arcsec$.2 $\times$ 19$\arcsec$.2.  The spatial scales of the final
processed images are 0$\arcsec$.05, 0$\arcsec$.05 and 0$\arcsec$.1 for
the ACS WFC, NICMOS NIC2 and combined WFPC2 images, respectively
(Table~\ref{Table2}). Fig.~\ref{Fig1} shows the {\it HST} images for
the 12 sample galaxies.

BCGs\footnote{For BCGs, the faint stellar envelopes are due to the
  intracluster light (ICL).} and the most luminous elliptical galaxies
tend to have extended, low surface brightness stellar envelopes,
halos, \citep{1965ApJ...142.1364M,
  1974ApJ...194....1O,1986ApJS...60..603S}.  Therefore, the extraction
of accurate surface brightness profiles for such galaxies requires
spatially extended images. The high-resolution \hst ACS WFC images
were sufficiently extended in radius to robustly determine the shapes
of the outer parts of the light profiles for the two farthest (D $>$
925 Mpc) galaxies in our sample (4C +74.13 and A2261-BCG),
Fig.~\ref{Fig1}. For one sample galaxy (A3558-BCG), we extracted a
composite surface brightness profile from the ACS/F814W images
obtained as a part of the \hst observing program 10429 under visits 17
and 21 (PI: J. Blakeslee), probing a large range in radius
($R \sim 150\arcsec$).

\begin{table*} 
\begin{center}
\caption{Data source}
\label{Table2}
\begin{tabular}{@{}llcccccc@{}}
\hline
\hline
Galaxy& {\it HST}  Program & {\it HST} Filter &{\it HST} image
                                                scale&Data at large
                                                       radii&Field of View\\
&&&(arcsec pixel$^{-1}$)&&(arcmin)\\
(1)&(2)&(3)&(4)&(5)&(6)\\
\multicolumn{1}{c}{} \\              
\hline                           
NGC 1600                       &   7886         & NICMOS/F160W
                                              &  0$\farcs$05 &CGS
                                                               $I$-band
                                                            &$8.90
                                                              \times 8.90$\\
NGC 4486                       &  10543         &  ACS/F814W
                                              &  0$\farcs$05 & SDSS
                                                               $i$-band
                                                               &$13.52
                                                              \times 9.83$\\
NGC 4874                       &   6104          &  WFPC2/F606W  &
                                                                   0$\farcs$10 & SDSS $r$-band &$13.52
                                                              \times 9.83$\\
NGC 4889                        &  5997           & WFPC2/F606W &0$\farcs$10  & SDSS $r$-band &$13.52
                                                              \times 9.83$\\
NGC 6166                        &   9293         &ACS/F814W         & 0$\farcs$05 & SDSS $i$-band &$13.52
                                                              \times 9.83$  \\
4C +74.13                        &   10495        & ACS/F850LP  & 0$\farcs$05 &---  &--- \\
A0119-BCG                     &  8683           & WFPC2/F814W  & 0$\farcs$10 &SDSS $i$-band &$13.52
                                                              \times 9.83$ \\
A2147-BCG                     &  8683           & WFPC2/F814W& 0$\farcs$10  &SDSS $i$-band  &$13.52
                                                              \times 9.83$ \\
A2261-BCG                      &  12066        & ACS/F850LP  & 0$\farcs$05 &--- &--- \\
A3558-BCG                     &  10429            & ACS/F814W      & 0$\farcs$05 &   {\it HST} ACS/F814W&$3.37
                                                              \times 3.37$\\
A3562-BCG                     &   8683             &WFPC2/F814W   & 0$\farcs$10 &2MASS  J-band&$8.53
                                                              \times 17.07$ \\
A3571-BCG                     &    10429          &ACS/F814W       &0$\farcs$05   &2MASS  J-band&$8.53
                                                              \times 17.07$ \\

\hline
\end{tabular} \\
Notes.--- Space- and ground-based imaging used for our sample
of 12 extremely massive galaxies with $R_{\rm b} > 0.5$ kpc (i.e.,
`large-core galaxies'). Col.~(1): galaxy name. Col.~(2):
{\it HST} programs.  Program ID: GO-5997  (PI:  J.\ Lucey);
GO-6104 (PI: W.\ Harris); 
GO-7886 (PI: A.\ Quillen); 
GO-8683  (PI: R.\ van der Marel);
GO-9293 (PI: H.\ Ford); 
GO-10429 (visits 17 and 21, PI: J.\ Blakeslee);
GO-10495  (PI: B.\ McNamara);
GO-10543  (PI: E.\ Baltz); and
GO-12066 (PI: M.\ Postman). Cols.~(3) and (4): {\it HST} filters and image
scales. Cols.~(5) and (6): data at
large radii and the associated field of view. For NGC~1600,
we combine the  inner {\it HST}
F160W/NICMOS NIC2 light profile ($ R \la 4\arcsec$) with the 
ground-based, Carnegie-Irvine 
Galaxy Survey (CGS) $I$-band data at $ R > 4\arcsec$ obtained using
the du Pont 2.5 m telescope \citep{2011ApJS..197...22L}. For
A3558-BCG,  we used ACS/F814W images  obtained as a part of  the {\it
  HST} observing
program 10429 under visits 17 and 21 (PI: J.\ Blakeslee).
\end{center}

\end{table*}

For the remaining 9/12 sample galaxies ({\mbox NGC ~1600}, {\mbox
  NGC~4486}, NGC~4874, NGC~4889, NGC~6166, {\mbox A0119-BCG}, {\mbox
  A2147-BCG}, {\mbox A3562-BCG} and {\mbox A3571-BCG}), the
high-resolution \hst NICMOS, WFPC2 and ACS images were limited in
radius to accurately describe the outer part of the galaxies' light
profiles and to perform reliable sky background subtraction (see
Fig.~\ref{Fig1}). These galaxies' \hst light profiles at small radii
(i.e., typically $R\la 40\arcsec$) were matched with low-resolution,
ground-based data at larger radii ($R > 40\arcsec$) determined from
extended images from the SDSS\footnote{https://www.sdss.org} and
2MASS
archives\footnote{https://irsa.ipac.caltech.edu/Missions/2mass.html} (see
Table~\ref{Table2}). An exception is {\mbox NGC 1600}, where we
combine the high-resolution \hst profile at small radii
($R\la 4\arcsec$) with ground-based surface brightness profile at
large radii ($R > 4\arcsec$) that was extracted by
\citet{2011ApJS..197...22L} from extended $I$-band images obtained
with the du Pont 2.5 m telescope. When possible, we used \hst and
ground-based images taken with similar filters. This
was not the case for {\mbox NGC 1600}, {\mbox A3562-BCG} and {\mbox
  A3571-BCG}, where we matched light profiles extracted from images
obtained using different filters (Table~\ref{Table2}, see also
\citealt{2007ApJ...662..808L,2016Natur.532..340T}). We find an
excellent overlap between the \hst light profiles of the galaxies and
the corresponding ground-based data over the
$R \sim 2\arcsec-60\arcsec$ radial range, except for {\mbox NGC 1600}
where the overlap between the \hst and ground-based data is over
$R \sim 2\arcsec-6\arcsec$.

The full details of our data reduction steps and the surface
 brightness profile extraction techniques are given in
 \citet{2017MNRAS.471.2321D, 2018MNRAS.475.4670D}. The extraction of
 accurate surface brightness profiles depends on the careful masking
 of the bright foreground stars, low-luminosity neighbouring and
 background galaxies, and chip defects in the image.  Initial masks
 were generated for the galaxies by running SEXTRACTOR
 \citep{1996A&AS..117..393B}, which were then complemented with manual
 masks. We follow the steps in \citet{2019ApJ...871....9D} to subtract
 the model images of the target galaxies from the science images and
 to improve our initial masks.  The composite surface brightness
 ($\mu$) and ellipticity ($\epsilon$) profiles of the full sample
 galaxies, extracted using the {\sc iraf} task {\sc ELLIPSE}
 \citep{1987MNRAS.226..747J}, are available in
 Appendix~\ref{AppA}. Given that we shift the ground-based profiles to the
 match \hst data, the composite surface brightness profiles are given
 in their respective \hst filters (Fig.~\ref{Fig1} and Appendix~\ref{AppA}).

We quote all magnitudes in the VEGA magnitude system.

\section{Light profile modeling of BCGs and central dominant galaxies}\label{Sec3} 

As noted above, BCGs and the most luminous elliptical galaxies tend to
exhibit faint stellar envelopes at large radii.  This is evident by a
light excess in their outer light distribution with respect to the
\citet{1948AnAp...11..247D} $R^{1/4}$ model, which is fit to the
spheroidal\footnote{The term ``spheroid'' is used here to refer to the
  underlying host galaxy in case of elliptical galaxies and the bulge
  for disc galaxies.}  components of the galaxies (e.g.,
\citealt{1974ApJ...194....1O,1977MNRAS.178..137C,1981ApJ...243...26D,1984ApJ...286..106L,1986ApJS...60..603S,
  2003Ap&SS.285...67G, 2005ApJ...618..195G, 2005MNRAS.358..949Z}).

The \citet{1963BAAA....6...41S, 1968adga.book.....S} $R^{1/n}$ model describes
the stellar light distributions of faint and intermediate-luminosity
($M_{B} \ga-$ 20.5 mag) spheroids (see
\citealt{2012ApJ...755..163D,2016MNRAS.462.3800D,2019ApJ...871....9D}). This model is
written as
\begin{equation}
I(R) = I_{\rm e} \exp \left\{ - b_{n}\left[ 
\left(\frac{R}{R_{\rm e}}\right)^{1/n} -1 \right]\right\},
\label{Eq1}
\end{equation}
where $ I_{\rm e}$ is the intensity at the half-light (effective)
radius ($R_{\rm e}$). The quantity $b_{n}\approx 2n- 1/3$ for $1\la
n\la 10$ \citep{1993MNRAS.265.1013C}, is expressed as a
function of the S\'ersic index $n$, and ensures that $R_{\rm e}$
encloses half of the total luminosity. For $n= 0.5$ and 1, the
S\'ersic model is a Gaussian function and an exponential function,
respectively.

The surface brightness profiles of luminous ($M_{B} \la 20.5$ mag)
spheroids are well described by the core-S\'ersic model, which is a
combination of an inner power-law core and an outer S\'ersic profile
with a transition region
\citep{2003AJ....125.2951G, 2004AJ....127.1917T}. This model,
discussed in detail in \citet{2012ApJ...755..163D}, is given by
 
\begin{equation}
I(R) =I' \left[1+\left(\frac{R_{\rm b}}{R}\right)^{\alpha}\right]^{\gamma /\alpha}
\exp \left[-b\left(\frac{R^{\alpha}+R^{\alpha}_{\rm b}}{R_{\rm e}^{\alpha}}
\right)^{1/(\alpha n)}\right], 
\label{Eq2}
 \end{equation}
with 
\begin{equation}
I^{\prime} = I_{\rm b}2^{-\gamma /\alpha} \exp 
\left[b (2^{1/\alpha } R_{\rm b}/R_{\rm e})^{1/n}\right], 
 \end{equation}
where $I_{\rm b}$ is intensity at the core break radius $R_{\rm b}$, $\gamma$ is
the slope of the inner power-law region, and $\alpha$ regulates the
sharpness of the transition between the inner power-law core and the outer
S\'ersic profile. The parameters $R_{\rm e}$ and $b$  are defined as in the S\'ersic
model.

\begin{figure*}
\hspace*{1.72562932599cm}
  \includegraphics[angle=270,scale=0.7]{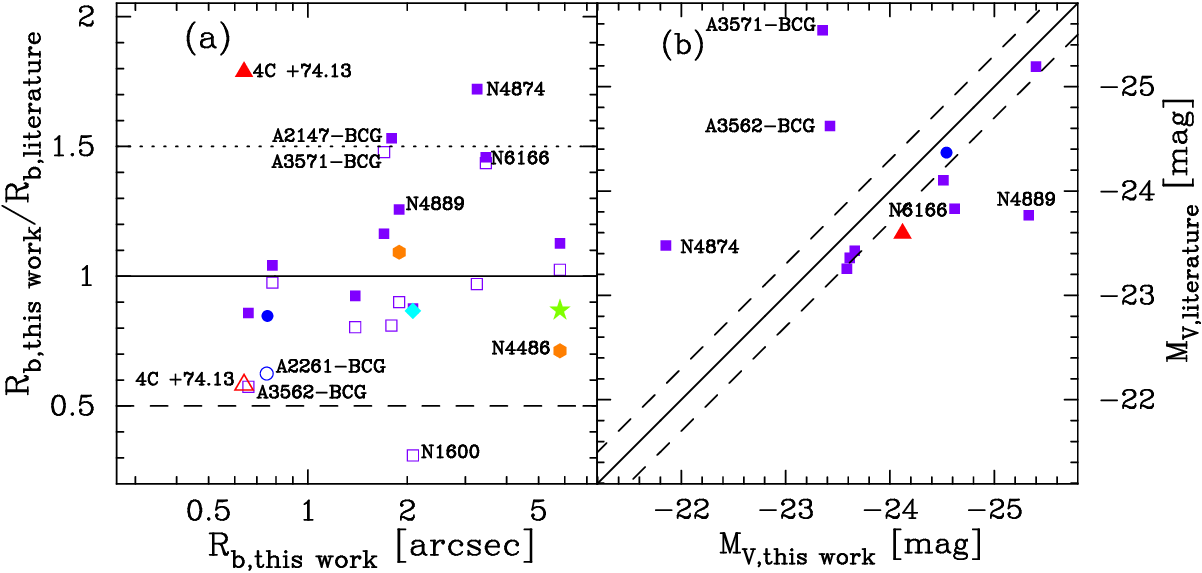}
  \caption{Left: comparison of our core-S\'ersic break radii (Table
    \ref {Table3}) and the cusp radii ($r_{\rm \gamma}$), Nuker break
    radii and core-S\'ersic break radii from the literature
    (\citealt{2007ApJ...664..226L,
      2009ApJ...698..594M,2011MNRAS.415.2158R,
      2012ApJ...756..159P,2013AJ....146..160R,
      2016Natur.532..340T}). For 4C +74.13, the
    \citet{2009ApJ...698..594M} Nuker break radius is converted to
    $r_{\rm \gamma}$ using their Nuker model fit parameters. The
    filled (and open) boxes are the cusp (and Nuker break) radii for
    galaxies in common with \citet{2007ApJ...664..226L}. The filled
    (and open) triangle and circle denote the cusp (and Nuker break)
    radii for 4C +74.13 \citep{2009ApJ...698..594M} and A2261-BCG
    \citep{2012ApJ...756..159P}, respectively (See
    Section~\ref{Litcomp}). The filled hexagons are the core-S\'ersic
    break radii for NGC~4486 and NGC~4889 from
    \citet{2013AJ....146..160R}, whereas the filled diamond and star
    denote the core-S\'ersic break radii for NGC~1600
    \citep{2016Natur.532..340T} and NGC~4486
    \citep{2011MNRAS.415.2158R}, respectively.  Right: comparison of
    our absolute $V$-band spheroid magnitudes $M_{V}$ (Table
    \ref{Table3}) with previous spheroid magnitudes in the
    literature. We adjusted the absolute magnitudes from the
    literature using our distances given in Table~\ref{Table1} (see
    the text for further detail). The solid line is a one-to-one
    relation, while the dashed lines are
    $M_{V,\rm literature} = M_{V,\rm this work} \pm 0.30$ mag.
    Symbolic representations are as in panel (a). }
\label{Fig2} 
 \end{figure*}

\subsection{Multi-component decomposition of large-core  galaxies}\label{Sec3.1} 

In \citet{2017MNRAS.471.2321D}, the 1D and 2D decompositions of the
high-resolution \hst F702W and F450W data of the \mbox{BCG IC 1101}
revealed a S\'ersic intermediate-scale component, an outer exponential
halo and an inner Gaussian component that are additional to the
core-S\'ersic spheroid light distribution. This galaxy has the largest
core size measured in any galaxy to date ($R_{\rm b} \sim 4.2$ kpc).
Following \citet{2017MNRAS.471.2321D}, we fit, here, the major-axis
surface brightness profiles of the remaining 12 ``large-core'' (i.e.,
$R_{\rm b} > 0.5$ kpc) core-S\'ersic galaxies (Table~\ref{Table1})
using a point-spread function (PSF)-convolved core-S\'ersic model.
Fig.~\ref{FigA1} shows the fit residual profiles and the corresponding
root-mean-square (rms) residual values ($\Delta$).  A single
core-S\'ersic model was adequate only for two of the 12 sample
galaxies (NGC 4889 and A3558-BCG). Additional nuclear light components
(i.e., AGN or nuclear star clusters) were identified in three sample
galaxies (NGC 4486, NGC 6166 and A2147-BCG) and were modelled using a
Gaussian or a S\'ersic model.  Of the 12 sample galaxies, nine have an
outer stellar halo light which was well described by an exponential
function. We find that the light profiles for two sample galaxies
(NGC~4874 and A3571-BCG) are well described by a core-S\'ersic
spheroid, a S\'ersic intermediate-scale component and an outer
exponential halo model. In Section~\ref{Sec5.3}, we show that the
colors for these two objects gradually turn bluer towards larger
radii, consistent with our multi-component light profile
decompositions.  In addition, \citet{2017MNRAS.464..356V} showed that
the velocity dispersion of NGC~4874 increases outward from $240$ km
s$^{-1}$ near the center to 350 km s$^{-1}$ at $R \sim 60 \arcsec$,
akin to that of the large-core galaxy \mbox{IC 1101} having an
intermediate-scale component.  We also note that
\citet{2007MNRAS.378.1575S} modelled the $R$-band ground-based data of
NGC~4874 with poor seeing conditions ($\sim 1\farcs5$), excluding the
inner $\sim3\arcsec$ data points, using a S\'ersic spheroid plus a de
Vaucouleurs stellar halo model. Due to the intermediate-scale
component of the galaxy which was missed in the
\citet{2007MNRAS.378.1575S} modelling, their fit resulted in an
incorrect, high S\'ersic index ($n=4$) for the outer exponential halo.

The best-fitting parameters are determined by iteratively minimizing
the rms residuals using the Levenberg-Marquardt optimisation algorithm
\citep{2017MNRAS.471.2321D}. For each iteration, the profiles of
individual model components were convolved with the Gaussian
point-spread function (PSF) and then summed to create the final model
profile. Our PSF implementation is as in \citet{2017MNRAS.471.2321D}.
The full widths at half-maximum (FWHMs) of the PSFs have been measured
from bright stars in the \hst images of the
galaxies. \citet{2001MNRAS.328..977T} noted that Moffat functions are
numerically better suited for modelling the PSFs in {\it HST} images
than Gaussian functions when the inner galaxy light profiles are
steep. Fitting both Gaussian and Moffat PSF-convolved models to the
{\it HST} light profiles of the BCG IC~1101 with a flat (i.e., an
inner slope $\gamma \le 0.05-0.08$) depleted core,
\citet{2017MNRAS.471.2321D} however showed that the best-fitting
structural parameters remain unchanged irrespective of the choice of
PSF. This should also be the case for the bulk (10/12) of our sample
galaxies with flat cores (i.e., $\gamma \la 0.15$). While the
remaining two galaxies NGC~4486 and \mbox{4C +74.13} have
$\gamma > 0.15$, our Gaussian PSF-convolved models give excellent fits
to the galaxies' {\it HST} profiles (Appendix~\ref{AppA}). However,
we admit the possibility that the inner S\'ersic model component of
NGC~4486 may be a poor fit to the nuclear source in the galaxy.

\subsection{Spheroid  magnitudes and stellar masses}\label{Bul_mag}

The total integrated spheroid fluxes for the sample galaxies were
computed using the best-fitting major-axis, core-S\'ersic structural
parameters and the ellipticities of the spheroids (Table~\ref{Table3}
and Fig~\ref{FigAII}).  We corrected these magnitudes for foreground
Galactic extinction using reddening values taken from NED
\citep{2011ApJ...737..103S}. The magnitudes were also corrected for
(1+$z$)$^{4}$ surface brightness dimming. To compare directly with
previously published magnitudes (\citealt{2007ApJ...662..808L};
\citealt{2009ApJ...698..594M};
\citealt{2012ApJ...756..159P,2014MNRAS.444.2700D,2017MNRAS.471.2321D}),
our magnitudes are converted into the $V$-band vega magnitudes (see
Table~\ref{Table3}). To achieve this, first we extracted light
profiles of five sample galaxies from archival {\it HST} images obtained in the
$V$-band filter or filters closer to the $V$-band---\mbox{NGC 4486},
ACS/F606W (GO-10543, PI: E. Baltz); \mbox{NGC 6166}, WFPC2/F555W
(GO-7265, PI: D. Geisler); \mbox{A2261-BCG}, ACS/F606W (GO-12066, PI:
M. Postman); \mbox{A3571-BCG}, ACS/F475W (GO-10429, PI: J. Blakeslee);
\mbox{A3558-BCG}, ACS/F475W (GO-12238, PI: W. Harris). Next, the total
integrated magnitudes in the corresponding filters, calculated in the
same manner as noted above, were converted into the $V$-band
magnitudes, when necessary using \citet[their
Table~3]{1995PASP..107..945F}, see Table~\ref{Table3}.  For the two
most distant galaxies in our sample (MS0735-BCG and A2261-BCG),
evolution- and K-corrections were performed utilizing the values
reported in the literature (\citealt{2009ApJ...698..594M};
\citealt{2012ApJ...756..159P}). Furthermore, the galaxy distances in
\citet{2007ApJ...664..226L, 2009ApJ...698..594M, 2012ApJ...756..159P}
are somewhat different from those adopted here (see
Table~\ref{Table1}). Therefore, we adjusted the magnitudes from the
literature to our distances.

We convert the spheroid luminosities ($L$) into stellar masses
($M_{*}$) using stellar mass-to-light ($M/L$) ratios which are
obtained from \citet{1994ApJS...95..107W} assuming an old ($\sim$12
Gyr) stellar population (see Table~\ref{Table5}).

\begin{figure}
  
\includegraphics[angle=270,scale=0.5369]{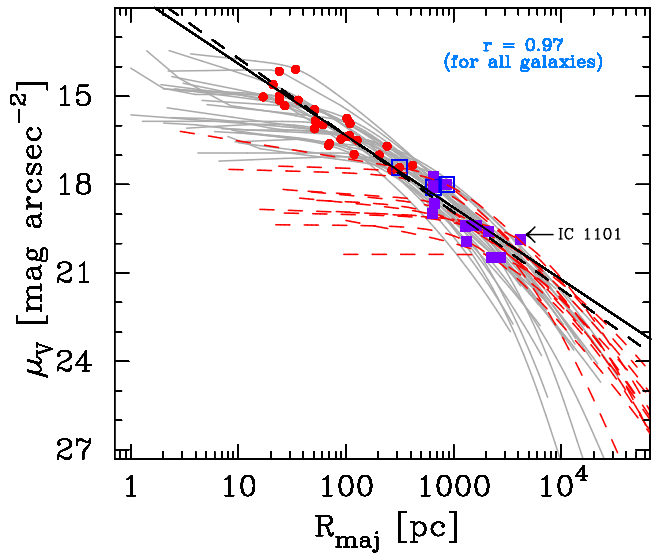}
\caption{Core-S\'ersic fits to the major-axis surface brightness
  profiles of 28+13(=41) core-S\'ersic galaxies (solid  and
  dashed curves). Filled red circles mark the core-S\'ersic break
  radii ($R_{\rm b}$) of the 28 core-S\'ersic early-type galaxies with
  $R_{\rm b} < 0.5$ kpc \citep[their
  Table~2]{2014MNRAS.444.2700D}. Filled purple boxes indicate the
  break radii for the 13 core-S\'ersic galaxies with $R_{\rm b} > 0.5$
  kpc (i.e., the 12 core-S\'ersic galaxies from this work,
  Table~\ref{Table3}, plus IC~1101,
  \citealt{2017MNRAS.471.2321D}). Blue squares enclose the three
  galaxies (NGC 3842, NGC 1600 and NGC 4889) with directly measured
  SMBH masses $M_{\rm BH} \ga 10^{10} M_{\rm \sun}$
  (\citealt{2011Natur.480..215M}; \citealt{2016Natur.532..340T}). The
  dashed line is a symmetric least-squares fit to the ($R_{\rm b}$,
  $\mu_{\rm b}$) data set for the full sample of 41 core-S\'ersic galaxies, while the
  solid line is a least-squares fit the 28 core-S\'ersic early-type
  galaxies with $R_{\rm b} < 0.5$ kpc \citep{2014MNRAS.444.2700D}.
  The correlation between $R_{\rm b}$ and $\mu_{\rm b}$ for the 41
  core-S\'ersic galaxies is extremely strong, with Pearson correlation
  coefficient r $\sim 0.97$.}
\label{Fig3} 
 \end{figure}

 \subsection{Comparison to previous fits  }\label{Litcomp}

 Here, we compare the values of our core-S\'ersic $R_{\rm b}$ and
 $M_{V}$ with those from the literature for the 12 galaxies
 (Table~\ref{Table3}) and discuss notable discrepancies (see
 Fig.~\ref{Fig2}). It should be noted that this work has, for the
 first time, modeled the light profiles for full sample of 12 galaxies
 using a core-S\'ersic model.
 
 In general, the core-S\'ersic break radii are in fair agreement with
 the cusp radii, but there are five discrepant data points
 (Fig.~\ref{Fig2}a). Our core-S\'ersic break radii are ($\ga25$\%)
 larger than the cusp radii reported for four galaxies in common with
 \citet{2007ApJ...664..226L}, \mbox{NGC 4874}, \mbox{NGC 4889},
 \mbox{NGC 6166} and \mbox{A2147-BCG}, and for \mbox{4C +74.13}
 \citep{2009ApJ...698..594M}. For comparison, the 1$\sigma$
 uncertainty range of $R_{\rm b}$ adopted in this paper is $2.5$\%. We
 believe the discrepancies in the break radii arise from differences
 in the (1) fitting models employed, (2) fitted radial extent of the
 galaxy light profiles and (3) treatment of distinct galaxy structural
 components.  Our break radii are obtained from careful,
 multi-component (halo/intermediate-scale component/spheroid/nucleus)
 decompositions of spatially extended (i.e., typically
 $R\ga100\arcsec$) light profiles. In contrast, \citet[][and
 references therein]{2007ApJ...664..226L}; \citet{2009ApJ...698..594M}
 determined the cusp radii fitting the Nuker model to their
 10$\arcsec$ galaxy light profiles, without accounting for any
 additional light components.

 As noted in the Introduction, the Nuker break radii poorly match the
 core-S\'ersic break radii (see Fig.~\ref{Fig2}a). Of the ten galaxies
 in common with \citet{2007ApJ...664..226L}, six (\mbox{NGC 1600},
 \mbox{NGC 6166}, \mbox{A2147-BCG}, \mbox{A3558-BCG}, \mbox{A3562-BCG}
 and \mbox{A3571-BCG}) have Nuker break radii that differ from our
 core-S\'ersic $R_{\rm b}$ by more than $25$\%. Also, the Nuker break
 radii for 4C+74.13 \citep{2009ApJ...698..594M} and A2261-BCG
 \citep{2012ApJ...756..159P} are roughly $ 65$\% larger than our
 $R_{\rm b}$ values. We note that the Nuker\footnote{The S\'ersic
   $R^{1/n}$ model can approximate a power law profile for large
   values of $n$. This means that the surface brightness distributions
   of some brightest cluster galaxies (BCGs) with large $n$ may be
   well described by a core-S\'ersic model and also by a Nuker model
   (e.g., see \citealt{2003AJ....125.2951G}). }  break radii tend to
 be bigger (smaller) than our core-S\'ersic break radii for spheroids
 with a Sersic index $n\la 6$ ($n\ga 9$).

 We have also included a comparison of our break radii with those from
 similar studies in the literature which performed core-S\'ersic model
 fits (Fig.~\ref{Fig2}a, \mbox{NGC 4486},
 \citealt{2011MNRAS.415.2158R}; \mbox{NGC 4486} and \mbox{NGC 4889},
 \citealt{2013AJ....146..160R}; NGC~1600,
 \citealt{2016Natur.532..340T}). The agreement between our
 core-S\'ersic break radii and those from the literature is remarkably good,
 except for the break radius of \mbox{NGC 4486}
 \citep{2013AJ....146..160R} which is $\sim$40\% larger than ours.  It
 appears that the \citet{2013AJ....146..160R}
 $R_{\rm b} \sim 8\farcs14$ and $n \sim 8.9$ for \mbox{NGC 4486} are
 biased high due to the outer halo light of the galaxy that was not
 separately modeled. While \citet{2009ApJS..182..216K} identify excess
 halo light at large radii ($R \sim 400-1000\arcsec$) with respect to
 their S\'ersic model fit to the main body of the \mbox{NGC 4486},
 \citet{2013AJ....146..160R} fit the core-S\'ersic model to the entire
 radial extent ($R \sim 1000\arcsec$) of the galaxy light profile
 from \citet{2009ApJS..182..216K}. We find that our
 $R \sim 400\arcsec$ light profile for the galaxy is better described
 using a core-S\'ersic model for spheroid plus a Gaussian function for
 the AGN (Fig.~\ref{FigA1}); this fit yields
 $R_{\rm b} \sim 5\farcs80$ and $n \sim 6.2$.
 
Fig.~\ref{Fig2}b shows our magnitudes (Table~\ref{Table3}) disagree
with those from past works (\citealt{2007ApJ...662..808L};
\citealt{2009ApJ...698..594M}; \citealt{2012ApJ...756..159P}), 
by typically more than $\sim$ 0.30 mag. Our spheroid  luminosities are
brighter than those from \citet{2007ApJ...664..226L} for seven of the
10 galaxies that we have in common (\mbox{NGC 1600}, \mbox{NGC 4486},
\mbox{NGC 4889}, \mbox{NGC 6166}, \mbox{A0119-BCG}, \mbox{A2147-BCG}
and \mbox{A3558-BCG}).
This discrepancy arises primarily because all these seven galaxies
have spheroids with $n > 4$ (Table~\ref{Table3}). For `regular'
elliptical galaxies, \citet[their Section 2.1]{2007ApJ...662..808L}
used total $V_{T}$ or $B_{T}$ galaxy magnitudes from the Third
Reference Catalogue of Bright Galaxies RC3
(\citealt{1991rc3..book.....D}), which are determined by
fitting the galaxy light profiles using  the \citet{1948AnAp...11..247D}
$R^{1/4}$ function (i.e., $n$ = 4 S\'ersic model). For BCGs, \citet{2007ApJ...662..808L} estimated the spheroid 
luminosities fitting the $R^{1/4}$ model to the inner $R \la 50$ kpc
light profiles and they argued that the halo light contribution to the
BCGs' light profiles over the fitted radial extents was
insignificant. However, we find three BCGs in common with
\citet{2007ApJ...662..808L}, \mbox{A2147-BCG}, \mbox{A3562-BCG} and
\mbox{A3571-BCG}, where the outer halo light contributes inside
$R \sim 50$ kpc (Appendix~\ref{AppA}). This in part has caused the
\citet{2007ApJ...662..808L} spheroid  magnitudes for \mbox{A3562-BCG}
and \mbox{A3571-BCG} to be brighter than ours. Not surprisingly, the
\citet{2007ApJ...662..808L} spheroid  magnitudes for the two galaxies
in the sample with an intermediate light component (NGC 4874 and
\mbox{A3571-BCG}) are brighter than ours (see Tables~\ref{Table3}  and
\ref{Table4}).

\setlength{\tabcolsep}{0.03020860in}
\begin{table*}
\begin {minipage}{180mm}
~~~~~~~~~~~\caption{ Best fitting core-S\'ersic parameters for our
  large-core galaxies.}
\label{Table3}
\begin{tabular}{@{}lllcccccccccccccccccccccccccccccccccccccccccccccc@{}}
\hline
\hline
Galaxy&Type&{\it HST} Filter &$ \mu_{\rm b} $ &
                                                                   $R_{\rm b}$
  &$R_{\rm b}$ &$ \gamma$&$\alpha$&$n$&$R_{\rm e}$&$R_{\rm e}$&$\epsilon_{\rm b}$&$M_{\rm
                                                                uncorr}$&$M_{\rm
                                                                corr}$&$\Delta
                                                                        M$
                                                                        &$M_{V,
                                                                           \rm
             corr}$\\
&& &(mag arcsec$^{-2}$)&(arcsec)&(kpc)&&&&(arcsec)&(kpc) &&(mag) &(mag) &&(mag)\\
(1)&(2)&(3)&(4)&(5)&(6)&(7)&(8)&(9)&(10)&(11)&(12) &(13)&(14)&(15) &(16)\\
\multicolumn{6}{c}{} \\ 
\hline                                           

NGC	1600	&	E	&	F160W/NICMOS&	15.14	&	2.08	&	0.65	&	0.04	&	2	&	6.3	&	72.6	&	22.8	&	0.270	&	-26.53	&	-26.61	&2.95	&-23.66	\\
NGC	4486	&	E$^{\dagger}$  	&	F814W/ACS	&	16.41	&	5.80	&	0.64	&	0.24	&	5	&	6.2	&	185.9	&	20.6	&	0.021	&	-24.86	&	-24.92&1.33	&	-23.59	\\
NGC	4874	&	E$^{\dagger}$  	&	F606W/WFPC2	&	19.05	&	3.25	&	1.63	&	0.13	&	2	&	4.0	&	4.9	&	2.5	&	0.099	&	-22.06	&	-22.18	&0.33 &	-21.85	\\
NGC	4889	&	BCG	&	F606W/WFPC2	&	17.68	&	1.89	&	0.86	&	0.04	&	2	&	13.3	&	563.9	&	256.7	&	0.065	&	-25.54	&	-25.66	&0.33&	-25.33	\\
NGC	6166	&	BCG	&	F814W/ACS	&	18.26	&	3.46	&	2.11	&	0.05	&	2	&	9.0	&	136.5	&	83.1	&	0.135	&	-25.80	&	-25.95	&1.33&	-24.62	\\
4C +74.13	&	BCG	&	F850LP/ACS	&	18.83	&	0.64	&	2.24	&	0.28	&	2	&	3.7	&	6.0	&	20.9	&	0.100	&	-25.00	&	-25.76	&1.64&	-24.12	\\
A0119-BCG	&	BCG	&	F814W/WFPC2	&	17.34	&	0.78	&	0.67	&	0.10	&	5	&	6.8	&	112.5	&	96.1	&	0.074	&	-25.58	&	-25.82	&1.31&	-24.51	\\
A2147-BCG	&	BCG	&	F814W/WFPC2	&	18.09	&	1.79	&	1.28	&	0.14	&	2	&	6.4	&	44.6	&	31.8	&	0.180	&	-24.73	&	-24.92	&1.31&	-23.61	\\
A2261-BCG	&	BCG	&	F850LP/ACS	&	18.69	&	0.75	&	2.71	&	0.00	&	5	&	2.1	&	4.9	&	17.6	&	0.036	&	-25.33	&	-26.34	&1.80&	-24.54	\\
A3558-BCG	&	BCG	&	F814W/ACS	&	18.08	&	1.39	&	1.30	&	0.03	&	2	&	5.4	&	131.9	&	123.7	&	0.029	&	-26.45	&	-26.73	&1.33&	-25.40	\\
A3562-BCG	&	BCG	&	F814W/WFPC2	&	17.66	&	0.66	&	0.64	&	0.06	&	2	&	3.6	&	18.9	&	18.4	&	0.099	&	-24.44	&	-24.73	&1.31&	-23.42	\\
A3571-BCG	&	BCG	&	F814W/ACS	&	18.56	&	1.70	&	1.33	&	0.01	&	2	&	10.2	&	68.9	&	53.8	&	0.062	&	-24.47	&	-24.72	&1.37&	-23.35	\\

\hline   
\end{tabular} 

Notes.--- Structural parameters from the core-S\'ersic model fits to
the major-axis surface brightness profiles of the spheroids of our
large-core galaxies.  Col. (1) galaxy name. The superscript
`$\dagger$' indicates the two second brightest cluster elliptical
galaxies in the sample. Col. (2) morphological type.  Col. (3) {\it
  HST} filters and instruments.  Cols. (4$-$11) best-fitting
core-S\'ersic model parameters. Col. (12) galaxy ellipticity at the
break radius $R_{\rm b}$. Col. (13) spheroid absolute magnitudes
derived using our fit parameters (cols.\ 4$-$11). Spheroid magnitudes
corrected for Galactic dust extinction using NED \citep{2011ApJ...737..103S}, and $(1+z)^{4}$ surface
brightness dimming are given in col. (14). For 4C~+74.13 and
A2261-BCG, we also carried out evolution- and $K$-corrections using
the values taken from \citet{2009ApJ...698..594M} and
\citet{2012ApJ...756..159P}, respectively.  Col. (15) $\Delta M$=({\it
  HST} filter)-$V$. Col. (16) corrected, $V$-band spheroid absolute
magnitude. We estimate that the uncertainties on the core-S\'ersic
parameters $R_{\rm b}$, $\gamma$, $n$ and $R_{\rm e}$ are $\sim$
2.5\%, 10\%, 20\% and 25\%, respectively. The uncertainty on
$\mu_{\rm b}$ is $\sim$ 0.02 mag arcsec$^{-2}$. These errors on the
fit parameters were estimated following the techniques in \citet[their
Section~3.1]{2019ApJ...871....9D}.
\end{minipage}
\end{table*}

\setlength{\tabcolsep}{0.0360in}
\begin{table*}
\begin {minipage}{180mm}
~~~~~~~~~~~\caption{Parameters associated with additional light components}
\label{Table4}
\begin{tabular}{@{}lllcccccccccccccccccccccccccccccccccccccccccccccc@{}}
\hline
\hline
Galaxy&{\it HST} Filter&$\mu_{\rm
                                                              0,h}$/$\mu_{\rm
                                                              e,S}$&$\rm
                                                                     h$/$R_{\rm
                                                                     e,S}$&$n_{\rm
                                                                            S}$&$\epsilon_{\rm h}/\epsilon_{\rm S}$& $m_{\rm pt}$&$m_{\rm
                                                                                    h,uncorr}$/$m_{\rm
                                                                                     S,uncorr}$&$M_{\rm
                                                                                    h,corr}$/$M_{\rm
                                                                                     S,corr}$&$M_{V,\rm
                                                                                               h}$/$M_{V,\rm
                                                                                               S}$\\
&&(mag arcsec$^{-2}$)&(arcsec)&&&(mag) &(mag) &(mag) &(mag)\\
(1)&(2)&(3)&(4)&(5)&(6)&(7)&(8)&(9)&(10)\\
\multicolumn{6}{c}{} \\ 
\hline                                           
NGC	1600	&	F160W/NICMOS	&	20.14/---	&	29.1/---	&	---	&	0.30/---	&	---	&	11.21/	&	-22.97/	&	-20.02/---	\\
NGC	4486	&	F814W/ACS	&	---/---	&	---/---	&	---	&	---/---	&	15.9	&	---/---	&	---/---	&	---/---	\\
NGC	4874	&	F606W/WFPC2	&	23.56/23.2	&	175.3/46.5	&	0.80	&	0.2/0.00	&	---	&	10.59/12.48	&	-24.67/-22.78	&	-24.34/-22.45	\\
NGC	4889	&	F606W/WFPC2	&	---/---	&	---/---	&	---	&	---/---	&	---	&	---/---	&	---/---	&	---/---	\\
NGC	6166	&	F814W/ACS	&	23.22/13.44	&	110.2/0.02	&	0.66	&	0.40/0.00	&	---	&	11.57/19.60	&	-24.16/-16.12	&	-22.83/-14.79	\\
4C	74.13	&	F850LP/ACS	&	23.53/---	&	43.3/---	&	---	&	0.50/---	&	---	&	14.11/---	&	-26.49/---	&	-24.85/---	\\
A0119-BCG	&	F814W/WFPC2	&	23.24/---	&	82.4/---	&	---	&	0.32/---	&	---	&	12.08/---	&	-24.50/---	&	-23.19/---	\\
A2147-BCG	&	F814W/WFPC2	&	22.75/---	&	85.1/---	&	---	&	0.50/---	&	20.2	&	11.86/	&	-24.27/---	&	-22.96/---	\\
A2261-BCG	&	F850LP/ACS	&	22.30/---	&	20.7/---	&	---	&	0.20/---	&	---	&	13.97/---	&	-26.95/---	&	-25.15/---	\\
A3558-BCG	&	F814W/ACS	&	---/---	&	---/---	&	---	&	---/---	&	---	&	---/---	&	---/---	&	---/---	\\
A3562-BCG	&	F814W/WFPC2	&	22.22/---	&	68.9/---	&	---	&	0.50/---	&	---	&	11.79/---	&	-25.16/---	&	-23.85/---	\\
A3571-BCG	&	F814W/ACS	&	21.17/21.05	&
                                                                  373.0/25.3
                                                                          &
                                                                            0.54	&	0.60/---	&	---	&	7.31/12.23	&	-29.08/-24.16	&	-27.71/-22.79	\\
\hline  
\end{tabular} 

Notes.---Structural parameters for additional light
components. Col.~(1) galaxy name.  Col.~(2) {\it HST} filters and
instruments. Cols.~(3-5) best-fitting parameters of the exponential
halo/S\'ersic model component. Cols.~(6) ellipticity of the
halo/S\'ersic model component. Col.~(7) apparent magnitude of the nucleus.  Col.~(8) apparent halo/S\'ersic magnitude derived using
our best-fitting exponential/S\'ersic model parameters (cols.~$3-6$).
Col.~(9) absolute halo/S\'ersic model component magnitude corrected
for Galactic dust extinction and $(1+z)^{4}$ surface brightness
dimming. For 4C~+74.13 and A2261-BCG, we also
carried out evolution- and $K$-corrections using the values taken from
\citet{2009ApJ...698..594M} and \citet{2012ApJ...756..159P},
respectively.  Col.~(10) corrected, $V$-band halo/S\'ersic
model component absolute  magnitude.

\end{minipage}
\end{table*}

\begin{center}
\begin{table*}
\setlength{\tabcolsep}{0.00058808in}
\begin {minipage}{180mm}
\caption{Large-core galaxy data}
\label{Table5}
\begin{sideways}
\begin{tabular}{@{}llccccccccc@{}}
\hline
\hline
Galaxy&{\it HST} Filter&$M/L$&log~($M_{*}/M_{\sun}$) &log~($M_{*}/M_{\sun}$)&log~($L_{\rm def}/L_{\sun}$)&log~($M_{\rm def}/M_{\sun}$)&log~($M_{\rm BH}/M_{\sun}$)&log~($M_{\rm BH}/M_{\sun}$)&log~($M_{\rm BH}/M_{\sun}$)&$M_{\rm def}/M_{\rm BH}$\\
&&&(spheroid )&(halo/S\'ersic
               comp)&&&($\sigma$-based)&(L-based)&($R_{\rm b}$-based)&($\sigma$-based/L-based/$R_{\rm b}$-based)\\
(1)&(2)&(3)&(4)&(5)&(6)&(7)&(8)&(9)&(10)&(11)\\
\multicolumn{1}{c}{} \\              
\hline
NGC 1600 &F160W [N]      &1.60&12.18&   10.73 &10.49   & 10.67       &10.23$^{+0.04}_{-0.04}$[d1]&10.23$^{+0.04}_{-0.04}$[d1]&10.23$^{+0.04}_{-0.04}$[d1]&2.8/2.8/2.8 \\
NGC  4486 &F814W [A]      &3.20& 12.12&   ---&     9.96   & 10.46       &9.76$^{+0.03}_{-0.03}$[d2]&$9.76^{+0.03}_{-0.03}$[d2]&$9.76^{+0.03}_{-0.03}$[d2] &5.1/5.1/5.1\\
NGC  4874 &F606W [W]    &3.80& 11.32& 12.31/11.56  &10.23  & 10.81   & 8.91$^{+0.42}_{-0.42}$&8.98$^{+0.34}_{-0.34}$& 10.54$^{+0.45}_{-0.45}$&79.8/67.6/1.9\\ 
NGC  4889 &F606W [W]     &3.80& 12.71&---& 10.39   &10.97     & 10.30 $^{+0.25}_{-0.62}$[d3]&10.3$^{+0.25}_{-0.62}$[d3]&10.3$^{+0.25}_{-0.62}$[d3]&4.7/4.7/4.7\\
NGC  6166 &F814W [A]      &3.20& 12.53&   11.81/8.60& 10.71   &11.21   &9.14$^{+0.43}_{-0.43}$&10.47$^{+0.47}_{-0.47}$& 10.68$^{+0.46}_{-0.46}$&117.3/5.5/3.4 \\
MS0735-BCG &F850LP [A]  &3.00&12.38&  12.67&10.33    &10.81        &8.61$^{+0.41}_{-0.41}$&10.20$^{+0.43}_{-0.43}$&  10.71$^{+0.46}_{-0.46}$&157.0/4.1/1.3\\
A0119-BCG &F814W [W]     &3.20&12.49&11.97& 9.66      &10.17       &9.04$^{+0.43}_{-0.43}$&10.41$^{+0.46}_{-0.46}$& 10.08$^{+0.42}_{-0.42}$ &13.5/0.6/1.2\\  
A2147-BCG &F814W [W]   &3.20&12.13&11.87  &10.23    &10.74        &8.95$^{+0.42}_{-0.42}$& 9.93$^{+0.40}_{-0.40}$ &10.42$^{+0.44}_{-0.44}$&61.2/6.4/2.1\\
A2261-BCG &F850LP [A]     &3.00&12.61&  12.86& 10.17     &10.65       &9.73$^{+0.48}_{-0.48}$&10.43$^{+0.48}_{-0.48}$&10.81$^{+0.45}_{-0.45}$&  8.2/1.7/0.7\\ 
A3558-BCG &F814W [A]     &3.20& 12.84&--- &10.13     &10.63       &8.70$^{+0.41}_{-0.41}$&10.89$^{+0.54}_{-0.54}$&10.43$^{+0.44}_{-0.44}$&85.7/0.6/1.7\\
A3562-BCG &F814W [W]     &3.20&12.06&  12.23 &9.62      &10.12        &8.59$^{+0.41}_{-0.41}$&9.83$^{+0.39}_{-0.39}$&10.05$^{+0.42}_{-0.42}$&34.6/2.0/1.2\\ 
A3571-BCG &F814W [A]      &3.20& 12.04&  13.78/11.81&10.25     &10.76      &9.31$^{+0.44}_{-0.44}$&9.79$^{+0.39}_{-0.39}$&10.44$^{+0.43}_{-0.43}$&28.0/9.3/2.1\\ 
\hline
\end{tabular} 
\end{sideways}
Notes.---Col.~(1) galaxy name. Col.~(2) {\it HST} filters and
instruments (NICMOCS, N; ACS, A and WFPC2, W). Col.~(3) stellar
mass-to-light ($M/L$) ratios are obtained from
\citet{1994ApJS...95..107W} assuming a $\sim 12$ Gyr old stellar
population. Cols.~(4) and (5) stellar mass of the spheroid, halo and
any additional S\'ersic components. Col.~(6) central stellar luminosity
deficit in units of solar luminosity.  Col.~(7) central stellar mass
deficit determined using cols.~(3) and (6). Cols.~(8), (9) and (10) SMBH masses
predicted using the \citet[their Fig.~2 and Table
3]{2013ApJ...764..151G} ``non-barred $M-\sigma$'' relation, their
$B$-band core-S\'ersic $M_{\rm BH}-L$ relation and the
``(direct SMBH)-based $R_{\rm b}-M_{\rm BH}$''  relation, see the text for
details. For three galaxies, we
use direct SMBH mass measurements obtained from three
different sources and adjusted them to our distances:
d1=\citet{2016Natur.532..340T}; d2=\citet{2011ApJ...729..119G}; d3= \citet{2011Natur.480..215M}. 
We assume a 10\% uncertainty on $\sigma$
to estimate the error on the predicted SMBH mass. Col.~(11) ratios
between mass deficit (col.~7) and black hole masses (cols.~8, 9 and 10), $M_{\rm def}/M_{\rm BH}$. 
\end {minipage}
\end{table*}

\end{center}

\setlength{\tabcolsep}{0.040in}
\begin{table*}
\begin {minipage}{168mm}

\caption{Scaling relations for core-S\'ersic galaxies} 
\label{Table6}
\begin{tabular}{@{}lllccccccccccccccccccccccccccccccc@{}}
\hline
\hline
Relation &OLS bisector fit& $\Delta$ &\\
\hline
 \multicolumn{4}{c}{This work (12 galaxies) + \citet[1 core-S\'ersic galaxy,
                                             IC~1101]{2017MNRAS.471.2321D}
  + \citet[28 core-S\'ersic
                                     galaxies]{2014MNRAS.444.2700D}}\\\\
$R_{\rm b}-\mu_{\rm b}$&$\mbox{log}\left(\frac{R_{\rm
                          b}}{\mbox{pc}}\right)= (0.38\pm
                          0.02)\left(\mu_{\rm b} - 17.50\right)
                          +~(2.45~ \pm 0.03)$&0.18 dex\\
$R_{\rm b}-M_{V}$&$\mbox{log}\left(\frac{R_{\rm b}}{\mbox{pc}}\right)= (-0.55\pm 0.05)\left(M_{V}+23.4\right) +~(2.65~ \pm 0.09)$&0.47 dex\\
$R_{\rm b}-M_{\rm BH}$  (11 direct $M_{\rm BH}$ masses)
        &$\mbox{log}\left(\frac{R_{\rm b}}{\mbox{pc}}\right)= (0.83\pm
           0.10)~\mbox{log}\left(\frac{M_{\rm BH}}{10^{9.30}
          M_{\sun}}\right) +~(  2.18~ \pm 0.08)$&0.24
 dex\\
$R_{\rm b}-M_{\rm BH}$  ($M-\sigma$ derived $M_{\rm BH}$ for 30 galaxies
        &$\mbox{log}\left(\frac{R_{\rm b}}{\mbox{pc}}\right)= (1.19\pm
          0.14)~\mbox{log}\left(\frac{M_{\rm BH}}{10^{9.30}
          M_{\sun}}\right) +~(2.64~ \pm 0.13)$&0.61
 dex\\
~~~~plus 11 direct $M_{\rm BH}$ masses)&&\\

$R_{\rm b}-M_{\rm BH}$ ($M-L$ derived $M_{\rm BH}$ for 30 galaxies
        &$\mbox{log}\left(\frac{R_{\rm b}}{\mbox{pc}}\right)= (1.05\pm 0.09)~\mbox{log}\left(\frac{M_{\rm BH}}{10^{9.85} M_{\sun}}\right) +~(2.66~ \pm 0.09)$&0.44 dex\\
~~~~plus 11 direct $M_{\rm BH}$ masses)\\
$\sigma - M_{V}$&$\mbox{log}\left(\sigma\right)= (-0.08\pm
                  0.01)\left(M_{V}+23.40\right) +~(2.49~ \pm
                  0.02)$&0.09 dex\\ 
$M_{V}-R_{\rm e}$& $M_{V}
                   = (-2.32 \pm 0.19)~\mbox{log}\left(\frac{R_{\rm
                   e}}{2\times10^{4.00} \rm pc}\right) +~( -23.19 ~ \pm 0.12)$&0.70   \\
$M_{V}-n$& $M_{V}
                   = (-4.13 \pm 1.87)~\mbox{log}\left(\frac{n}{5.0}\right) +~(-22.70 ~ \pm  0.21)$& 1.25   \\\\

\hline

\hline
\multicolumn{4}{c}{\citet[28 core-S\'ersic
                                     galaxies]{2014MNRAS.444.2700D}}\\\\
$R_{\rm b}-\mu_{\rm b}$&$\mbox{log}\left(\frac{R_{\rm
                          b}}{\mbox{pc}}\right)= (0.41\pm
                          0.04)\left(\mu_{\rm b} - 16.00\right) +~(1.86~ \pm 0.04)$&0.18 dex\\
$R_{\rm b}-M_{V}$&$\mbox{log}\left(\frac{R_{\rm b}}{\mbox{pc}}\right)=
                   (-0.45\pm 0.05)\left(M_{V}+22\right) +~(1.79~ \pm
                   0.06)$&0.30 dex\\
$R_{\rm b}-M_{\rm BH}$  ($M-\sigma$ derived $M_{\rm BH}$ for 23
  galaxies &$\mbox{log}\left(\frac{R_{\rm b}}{\mbox{pc}}\right)= (0.80\pm 0.10)~\mbox{log}\left(\frac{M_{\rm BH}}{10^{9} M_{\sun}}\right) +~(2.01~ \pm 0.05)$&0.27 dex\\
plus 8 direct $M_{\rm BH}$ masses)&&\\
$R_{\rm b}-M_{\rm BH}$ ($M-L$ derived $M_{\rm BH}$ for 23 galaxies
        &$\mbox{log}\left(\frac{R_{\rm b}}{\mbox{pc}}\right)= (0.79\pm 0.08)~\mbox{log}\left(\frac{M_{\rm BH}}{10^{9} M_{\sun}}\right) +~(1.75~ \pm 0.06)$&0.27 dex\\
~~~~plus 8 direct $M_{\rm BH}$ masses)\\

\hline
\hline
Relation &OLS bisector fit& r&\\
\hline

  $\Sigma_{5}-R_{\rm b}$  (for 9 large-core
  galaxies excluding  &$\mbox{log}\left(
                        \Sigma_{5}\right)= (1.51 \pm
                        0.47)~\mbox{log}\left(\frac{R_{\rm b}}{\mbox{pc}}\right) +~(0.78~ \pm  0.11)$&0.54
  \\
  the large-core, group galaxy\footnote{If we include the only sample
large-core, group galaxy NGC~1600  (with a low
  $\Sigma_{5}$ value $\sim 0.31$ Mpc$^{-2}$, see Figs.~\ref{Fig10A}
  and ~\ref{FigAII}), then for the 10 large-core
  galaxies the OLS bisector  finds a relation between $\Sigma_{5}$ and
  $R_{\rm b}$ with a slope of 2.40 $\pm$  1.03, an intercept of 0.68
  $\pm$ 0.17 and r $\sim$ 0.57.} and the three  \\
 distant, D$\ga 360$ Mpc, sample large-core galaxies) \\
  $\Sigma_{10}-R_{\rm b}$  (for 10 large-core
  galaxies excluding  &$\mbox{log}\left(\Sigma_{10}\right)= (   1.61\pm
                        0.73)~\mbox{log}\left(\frac{R_{\rm b}}{\mbox{pc}}\right) +~(1.05~ \pm  0.15)$&0.32
  \\
  the three distant, D$\ga 360$ Mpc,  sample\\
  large-core galaxies) \\
  \hline
\end{tabular} 
Notes.---Scatter in the vertical direction ($\Delta$). Pearson correlation coefficient (r).

\end{minipage}
\end{table*}

\begin{figure}
\hspace*{-.17909142562932599cm}   
\vspace*{.109142562932599cm}   
\includegraphics[angle=270,scale=0.685369]{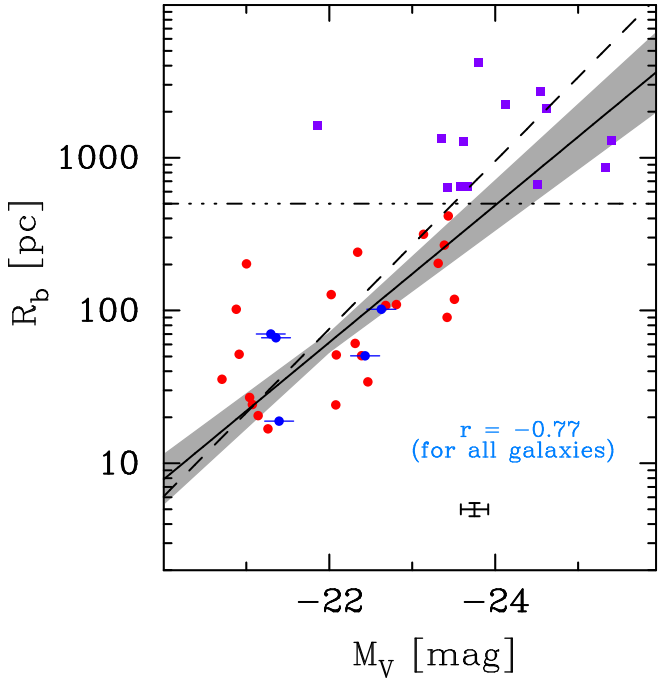}
\caption{The correlation between the core-S\'ersic break radius
  ($R_{\rm b}$) and $V$-band spheroid  absolute  magnitude 
  ($M_{V}$) for our sample of 41 core-S\'ersic galaxies.  Filled red
  circles and disk symbols show the 23 core-S\'ersic elliptical
  galaxies and 5 core-S\'ersic S0 galaxies, respectively, with
  $R_{\rm b} < 0.5$ kpc \citep[their Table~2]{2014MNRAS.444.2700D},
  while filled purple boxes indicate the 13 core-S\'ersic galaxies
  with $R_{\rm b} > 0.5$ kpc (i.e., the 12 core-S\'ersic galaxies from
  this work, Table~\ref{Table3}, plus IC~1101,
  \citealt{2017MNRAS.471.2321D}).  The horizontal, dashed-dotted line
  indicates the $R_{\rm b} = 0.5$ kpc demarcation. The dashed line is
  a symmetric least-squares fit to the 41 core-S\'ersic galaxies,
  while the solid line is a symmetric OLS fit to the 28 `normal-core'
  galaxies \citep{2014MNRAS.444.2700D}. The shaded region shows the
  $1 \sigma$ uncertainty on the regression fit. Pearson correlation
  coefficient (r) and representative error bars for the 41
  core-S\'ersic galaxies are shown at the bottom. }
\label{Fig4}
 \end{figure}

\begin{figure}
\hspace*{-.17909142562932599cm}   
\vspace*{.109142562932599cm}   
\includegraphics[angle=270,scale=0.7585369]{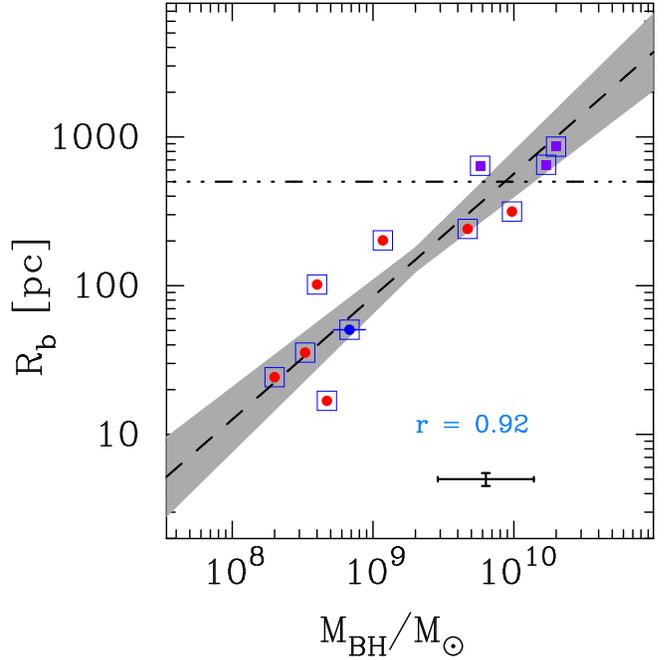}
\caption{Similar to Fig.~\ref{Fig4}, but showing here the correlation between
  core-S\'ersic break radius ($R_{\rm b}$) and  SMBH mass ($M_{\rm BH}$)
  for 11 galaxies in our sample with directly measured
  $M_{\rm BH}$. The dashed line shows our symmetric OLS fit.    }
\label{Fig45}
 \end{figure}

\section{Scaling relations for core-S\'ersic galaxies}\label{Sec4}

Luminous early-type galaxies are known to exhibit tight scaling
relations involving their central and global structural parameters
(e.g.,
\citealt{1997AJ....114.1771F,2012ApJ...755..163D,2013AJ....146..160R,2014MNRAS.444.2700D,2016Natur.532..340T}). \citet{2013AJ....146..160R}
investigated the correlation between the core-S\'ersic break radius
($R_{\rm b}$) and the velocity dispersion, spheroid absolute magnitude
and SMBH mass for a sample of 23 core-S\'ersic elliptical galaxies
with $R_{\rm b}\la 0.8$ kpc. In \cite{2014MNRAS.444.2700D}, we
explored a number of scaling relations involving $R_{\rm b}$ and the
break surface brightness ($\mu_{\rm b}$) for a sample of 28
core-S\'ersic early-type spheroids with $R_{\rm b} \la 0.5$ kpc and
$-20.70$ mag $\ga M_{\rm V} \ga$ $-23.60$ mag.  In this paper we
combine these 28 core-S\'ersic spheroids with $R_{\rm b} \la 0.5$ kpc
and $8\times10^{10} \la M_{*} \la 10^{12} M_{\sun}$ (henceforth
`normal-core spheroids', \citealt{2014MNRAS.444.2700D}) and the 13
extremely massive spheroids ($M_{*} \ga 10^{12} M_{\sun}$) with
$R_{\rm b} \ga 0.5$ kpc and \mbox{$M_{\rm V} \la -23.50 \pm 0.10$} mag
(henceforth `massive large-core spheroids', Tables~\ref{Table1} and
\ref{Table3}) to explore whether or not extremely massive, large-core
spheroids adhere to the structural scaling relations established by
the relatively less massive, normal-core spheroids. Our
41(=28+13) galaxies represent the hitherto largest sample of
core-S\'ersic galaxies with detailed multi-component decompositions of
the high-resolution, extended light profiles.

 \begin{figure*}
 \hspace{3.964 cm}   
\includegraphics[angle=270,scale=0.7]{Rb_vs_Black_Hole_Mass_2.ps}
 \caption{Similar to Figs.~\ref{Fig4} and \ref{Fig45}, but shown here are correlations
  between the core-S\'ersic model break radius, $R_{\rm b}$,
  (Table~\ref{Table3} and \citealt[their
  Table~2]{2014MNRAS.444.2700D}) and SMBH mass, $M_{\rm BH}$,   (Table~\ref{Table5} and \citealt[their
  Table~4]{2014MNRAS.444.2700D}). The dotted line is our best  OLS
  fit to the 11 
  galaxies with dynamically determined SMBH masses (enclosed in
  boxes, see also Fig.~\ref{Fig45}). For the
  remaining 30 galaxies, the SMBH masses are estimated using the
  \citet{2013ApJ...764..151G} non-barred $M_{\rm BH}-\sigma$ relation
  (a) and their $B$-band core-S\'ersic $M_{\rm BH}-L$ relation
  (b).  }
\label{Fig5} 
 \end{figure*}

\begin{figure}
\hspace*{-.164 cm}   
\includegraphics[angle=270,scale=0.625]{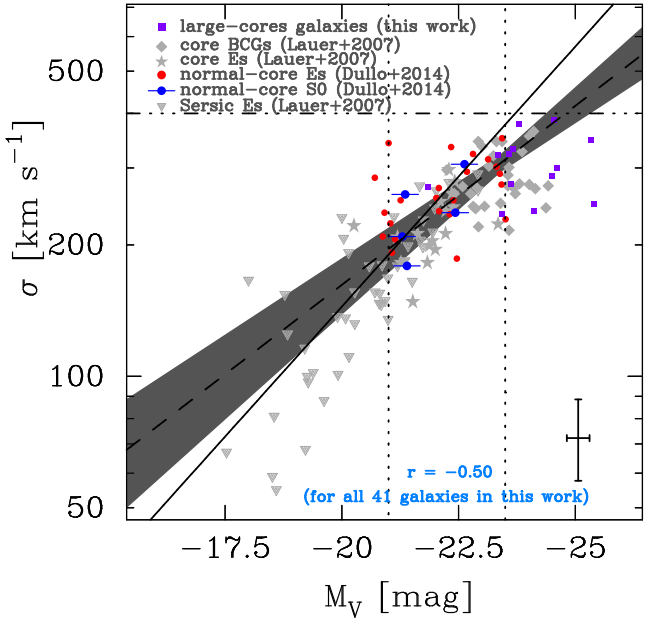}
\caption{Similar to Fig.~\ref{Fig4}, but showing here the correlation
  between $\sigma$ and $M_{V}$\citep{1976ApJ...204..668F}. The
  vertical dotted lines indicate the $M_{V} = -21.0$ mag and
  $M_{V} = -23.50$ mag demarcations, corresponding to the S\'ersic
  versus core-S\'ersic and normal-core versus large-core divides,
  respectively.  The horizontal dashed-dotted line indicates the
  velocity dispersion function cutoff for local galaxies of
  $\sigma = 400$ km s$^{-1}$ (e.g.,
  \citealt{2003ApJ...594..225S,2007ApJ...662..808L,2007AJ....133.1741B}). The
  S\'ersic and core-S\'ersic galaxies from
  \citet{2007ApJ...662..808L}, shown here to better reveal the breaks
  in the $\sigma-L$ relation, are not included in the least-squares
  fits. }
\label{Fig6} 
 \end{figure}

\begin{figure}
\hspace*{-.34 cm}   
\includegraphics[angle=270,scale=0.45]{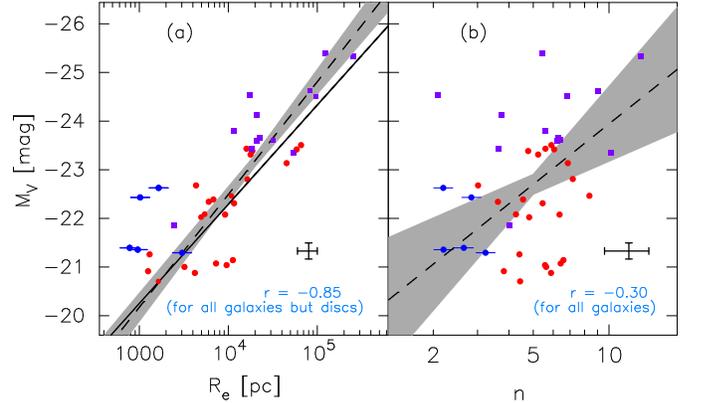}
\caption{Similar to Fig.~\ref{Fig4}, but shown here are correlations
  between $V$-band spheroid absolute magnitude  and effective
  radius $R_{\rm e}$ (a) and S\'ersic index $n$ (b). For the
$L_{V}-R_{\rm e}$  relation (a), the dashed line
  is a symmetric least-squares fit to the 36(=41-5) core-S\'ersic galaxies
  after excluding the 5 S0 galaxies in the sample, while the solid line is a
  symmetric least-squares fit to the 23 normal-core
  elliptical galaxies \citep{2014MNRAS.444.2700D}. For the
$L_{V}-n$  relation (b), we show a
  symmetric least-squares fit to full sample of 41 core-S\'ersic
  galaxies (dashed line).}
\label{Fig7} 
 \end{figure}

\subsection{Structural correlations with the core's size: 
  large-core versus normal-core spheroids}\label{Sec4.1}

Fig.~\ref{Fig3} shows a compilation of major-axis core-S\'ersic model
profiles for the 41 core-S\'ersic spheroids (see also Appendix
\ref{AppA}). Also shown in this figure is a strong correlation between
$R_{\rm b}$ and $\mu_{\rm b}$ with Pearson
correlation coefficient r $\sim 0.97$ (see also
\citealt{1997AJ....114.1771F,2007ApJ...662..808L}). We find that our
massive large-core spheroids follow the tight $R_{\rm b}-\mu_{\rm b}$
sequence defined by the relatively less massive normal-core spheroids (Table
\ref{Table6}).

In Fig.~\ref{Fig4} we expand on Fig.~6a from
\citet{2014MNRAS.444.2700D} and plot the relation between the
core-S\'ersic break radius ($R_{\rm b}$) and $V$-band spheroid
absolute magnitude ($M_{\rm V}$). The dashed and solid lines are
ordinary least squares (OLS) bisector regression fits
\citep{1992ApJ...397...55F} to the ($R_{\rm b}, M_{\rm V}$) data set
for the full sample of 41 core-S\'ersic spheroids and the 28
normal-core spheroids, respectively. Because our full galaxy sample
spans a wider range in $M_{\rm V}$ than that of
\citet{2014MNRAS.444.2700D}, the new, well constrained
$R_{\rm b}-L_{\rm V}$ relation for the full sample
($R_{\rm b} \propto L_{\rm V}^{1.38 \pm 0.13}$) has a slope
$\sim 22\%$ steeper than the near-linear relation for the normal-core
galaxies alone ($R_{\rm b} \propto L_{\rm V}^{1.13 \pm 0.13}$,
\citealt{2014MNRAS.444.2700D}). It is worth noting that the larger
cores of extremely massive galaxies are consistent with these
galaxies' extremely bright spheroid magnitudes
(\mbox{$M_{\rm V} \la -23.50 \pm 0.10$} mag), although the inclusion
of the 13 large-core galaxies has increased the vertical rms scatter
around the $R_{\rm
  b}$$-$$L_{\rm V}$ relation in the log $R_{\rm b}$ direction
($\Delta$) by 57\%, $\Delta_{\rm normal-core} \sim 0.30$ dex and
$\Delta_{\rm full\_sample} \sim 0.47$ dex (see Table~\ref{Table6}).
For reference, the slopes of the $R_{\rm b}-L_{\rm V}$ relations
published by
\citet{1997AJ....114.1771F,2003AJ....125..478L,2005A&A...439..487D,2007ApJ...662..808L,
  2013AJ....146..160R} are 1.15, 0.72, 1.05 $\pm$ 0.10, 1.32 $\pm$
0.11, and 1.28 $\pm$ 0.18.

Finally, in Figs.~\ref{Fig45}, \ref{Fig5}(a) and \ref{Fig5}(b) we
investigate whether there is an offset or break in the
$R_{\rm b}-M_{\rm BH}$ relation due to normal-core versus large-core
spheroids.  Of the full sample, 11/44 (3 large-core and 8 normal-core)
galaxies have SMBH mass determined dynamically from stellar or gas
kinematic measurements (e.g. \citealt{2005SSRv..116..523F}; see
Table~\ref{Table5} and \citealt[their Table
4]{2014MNRAS.444.2700D}). We refer to such SMBH masses as `direct'
SMBH masses. The OLS bisector finds a remarkably tight
$R_{\rm b}-M_{\rm BH,direct}$ relation for these 11 galaxies
(\mbox{$R_{\rm b} \propto M_{\rm BH, direct}^{0.83 \pm 0.10}$}, r
$\sim 0.92$ and $\Delta \sim$ 0.24 dex, see
Fig.~\ref{Fig45}). Although we only have 11 galaxies with direct SMBH
masses, the $R_{\rm b}-M_{\rm BH,direct}$ relation found here is in
excellent agreement with the core-S\'ersic
$R_{\rm b}-M_{\rm BH,direct}$ relation reported by
\citet[Fig.~4]{2016Natur.532..340T} for 20 core-S\'ersic galaxies
with direct SMBH mass measurements: \mbox{log ($R_{\rm b}/$pc) = (0.85
  $\pm$ 0.10) log ($M_{\rm BH}/10^{9.30}$)} + (2.17 $\pm$ 0.45). This
strongly suggests that the \mbox{$R_{\rm b}-M_{\rm BH,direct}$}
relation (Table~\ref{Table6}) is well constrained.

However, combining these 11 direct SMBH masses with predicted SMBHs
for the remaining 30 galaxies without direct SMBH mass measurements,
we find that large-core spheroids are offset systematically from the
\mbox{$R_{\rm b}-M_{\rm BH, direct}$} relation defined by the galaxy
sample with measured $M_{\rm BH}$ (dotted lines) and from the
\mbox{$R_{\rm b}-M_{\rm BH}$} sequence traced by the normal-core
spheroids alone (solid lines, Figs.~\ref{Fig5}(a), \ref{Fig5}(b) and
Table~\ref{Table6}). The slopes of the $R_{\rm b}-M_{\rm BH}$
relations for the normal-core spheroids are different from those for
the full sample (see Table~\ref{Table6}). In Fig.~\ref{Fig5}(a), we
used the \citet[their Table 3]{2013ApJ...764..151G} non-barred
$M_{\rm BH}-\sigma$ relation to predict the black hole masses for the
remaining 30 galaxies without direct SMBH masses, while in
Fig.~\ref{Fig5}(b) the predicted SMBH masses were based on the
near-linear \citet[their Table 3]{2013ApJ...764..151G} $B$-band
core-S\'ersic $M_{\rm BH}-L$ relation transformed here into the
$V$-band using $B-V = 1.0$ (\citealt{1995PASP..107..945F}). The offset
nature of large-core spheroids at the high-mass end of the
$R_{\rm b}-M_{\rm BH}$ relations (Figs.~\ref{Fig5}a and \ref{Fig5}b)
suggest that either extremely massive galaxies have unusually large
break radii or the $M_{\rm BH}-\sigma$ and core-S\'ersic
$M_{\rm BH}-L$ relations underestimate SMBH masses in extremely
massive galaxies, or both cases are true. However, given the tight
$R_{\rm b}-\mu_{\rm b}$, $R_{\rm
  b}$$-$$L_{\rm V}$ and $R_{\rm b}-M_{\rm BH,direct}$ relations
(Figs.~\ref{Fig3}, \ref{Fig4} and \ref{Fig45}), the observed offsets
are likely due to underestimated SMBH masses (e.g.,
\citealt{2011Natur.480..215M,2012ApJ...756..179M,2013ApJ...768...29V,
  2016Natur.532..340T,2018MNRAS.474.1342M}). Indeed, the direct SMBH
measurements in three large-core galaxies
(NGC~1600, \citealt{2016Natur.532..340T}, NGC~4889,
\citealt{2013ApJ...764..184M} and Holm~15A, 
\citealt{2019arXiv190710608M}) reveal that, relative to the best-fitting
$M_{\rm BH}-\sigma$ relation, these galaxies are offset by $0.7-1$ dex
towards large $M_{\rm BH}$.

Also, combining the $R_{\rm b}$$-$$L_{\rm V}$ relation
(Fig.~\ref{Fig4}) with the $R_{\rm b}-M_{\rm BH, direct}$
(Fig.~\ref{Fig45}) results in a steeper \mbox{$M_{\rm BH}-L$} relation
(\mbox{$M_{\rm BH} \propto L_{V}^{1. 66\pm 0.25}$}, see
Table~\ref{Table6}) than the \citet{2013ApJ...764..151G}
$B$-band core-S\'ersic $M_{\rm BH}-L$ relation
(\mbox{$M_{\rm BH} \propto L_{B}^{1. 35\pm 0.30}$}), although the
slopes of these two relations are consistent within the $1\sigma$
uncertainty.  Extrapolating the $R_{\rm b}-M_{\rm BH,direct}$ relation
(Fig.~\ref{Fig45}) to high masses, we find that the $M_{\rm BH}-L$
relation is a better predictor of $M_{\rm BH}$ than the
$M_{\rm BH}-\sigma$ relation for BCGs and central dominant galaxies,
confirming the finding by other authors
(\citealt{2007AJ....133.1741B,2007ApJ...662..808L,2011Natur.480..215M,2012ApJ...756..179M,2013ApJ...768...29V,2018MNRAS.474.1342M,2019arXiv190309965P}).

\subsection{ Global structural relations for core-S\'ersic galaxies}

 In a continued endeavor to determine if the scaling relations for
 normal-core spheroids continue to the large-core spheroids, here we
 investigate global structural relations involving $M_{ V}$,  the effective (half-light) radius
 ($ R_{\rm e}$) and the S\'ersic index ($n$) for
 our full sample of 41 core-S\'ersic spheroids. 

\subsubsection{ $\sigma-L_{V}$}
 
In Fig.~\ref{Fig6}, we show the $\sigma-L_{V}$ relation
\citep{1976ApJ...204..668F} relation for core-S\'ersic spheroids. The
OLS bisector yields \mbox{$\sigma \propto L_{V}^{1/(5.00 \pm 0.60)}$}
for the 41 core-S\'ersic spheroids, compared to the well-known
\citep{1976ApJ...204..668F} relation ($\sigma \propto L^{1/4}$).  It
has been shown that bright core-S\'ersic galaxies follow a shallow
$\sigma-L$ relation \mbox{$\sigma \propto L^{1/(4-8)}$} (e.g.,
\citealt{1981ApJ...251..508M,2007ApJ...662..808L,2013ApJ...769L...5K,2019arXiv190806838S}),
whereas S\'ersic galaxies with low luminosities ($M_{V} \ga -21.5$
mag) define a steeper relation \mbox{$\sigma \propto L^{1/2}$} (e.g.,
\citealt{1992AJ....103..851H,2005MNRAS.362..289M}), see
Fig.~\ref{Fig6}. We have identified here a previously unreported
substructure in the $\sigma-L_{V}$ relation for core-S\'ersic
spheroids. This relation seems to have a break at
\mbox{$M_{V} \sim -23.50 \pm 0.10$} mag\footnote{Assuming an old
  stellar population of $M/L_{V}=5.6$, $M_{V} \sim -23.50 \pm 0.10$
  mag yields a stellar mass of
  $M_{*} \sim 1.2 \times 10^{12} M_{\sun}$ for $M_{V,\rm sun} =4.83$
  mag.}(Fig.~\ref{Fig6}). At magnitudes brighter than this transition
$M_{V}$ value, the relation flattens and exhibits larger scatter which
may suggest a breakdown.  The slope of our $\sigma-L$ relation for the
full sample ($1/(5.00 \pm 0.63)$) is shallower than that found for the
normal-core spheroids alone ($1/(3.50\pm 0.61)$). In Fig.~\ref{Fig6},
we also show S\'ersic and core-S\'ersic galaxies from
\citet{2007ApJ...662..808L} to better demonstrate the breaks in the
$\sigma-L_{V}$ relation but these data points
(\citealt{2007ApJ...662..808L}) are not included in the least-squares
fits. The $\sigma-L_{V}$ relation for S\'ersic, normal-core and
large-core core-S\'ersic galaxies will be investigated in a
forthcoming paper.  

Our finding is consistent with the notion that major, dry mergers add
the stellar mass, black hole mass and sizes in equal proportion while
increasing the velocity dispersion only slightly (see also
Section~\ref{Sec4.2.2}).  In general, the velocity dispersion for
local galaxies does not exceed $\sigma \sim 400$ km s$^{-1}$
(\citealt{2003ApJ...594..225S,2007ApJ...662..808L,2007AJ....133.1741B}). As
such, the luminosity of large-core galaxies, which are thought to have
undergone multiple successive dry mergers, would be unusually bright
for the galaxies' $\sigma$, compared to the normal-core galaxies.  We
agree with \citet{2007AJ....133.1741B} who found that BCGs follow a
shallower $\sigma-L$ relation than most (other) early-type galaxies
(see also \citealt{1991ApJ...375...15O,2006MNRAS.369.1081B}).  In
contrast, \citet[their eq.~7]{2007ApJ...662..808L} and
\citet{2013ApJ...769L...5K} advocated for a single power-law relation
for BCGs and other bright ellipticals with $M_{V} \la -20.5$
mag. However, the break in the $\sigma-L$ relation near
\mbox{$M_{V} \sim -23.50 \pm 0.10$} can easily be seen in
Fig.~\ref{Fig6} and \citet[their Figs.~1 and
2]{2013ApJ...769L...5K}. In fact \citet{2013ApJ...769L...5K} wrote
that their core and coreless galaxies overlap over the luminosity
range \mbox{$-20.50$ mag $>$ $M_{V}$ $>$ $-22.85$ mag}. As such their
shallow core-S\'ersic $\sigma-L_{V}$ relation
(\mbox{$\sigma \propto L_{V}^{1/(8.33 \pm 1.24)}$)} is mainly driven by
galaxies with $M_{V} \la -22.85$ mag.

\subsubsection{ $L_{V}-R_{\rm e}$ and $L_{V}-n$ }\label{Sec4.2.2}

In Figs.~\ref{Fig7}(a) and \ref{Fig7}(b), we investigate the behaviour
of $M_{ V}$ as a function of the effective radius ($R_{\rm e}$) and
S\'ersic index ($n$), Table~\ref{Table6}. As noted by
\citet{2014MNRAS.444.2700D}, for a given $M_{ V}$ the spheroids of
core-S\'ersic lenticular galaxies tend to be compact
($R_{\rm e} \la 2$ kpc), see also
\citet{2013ApJ...768...36D,2013pss6.book...91G,2015ApJ...804...32G,2016MNRAS.457.1916D}. Excluding
the 5 lenticular galaxies in our sample, we find that large-core and
normal-core spheroids follow the same tight correlation between
$L_{V}$ and $R_{\rm e}$.  A symmetrical OLS fit to the 36(=41$-$5)
core spheroids gives a near-linear $L_{V}-R_{\rm e}$ relation
\mbox{$R_{\rm e} \propto L_{V}^{1.08\pm 0.09}$} with r $\sim -0.85$;
our slope is slightly steeper than the $L-R_{\rm e}$ relation for BCGs
in \citet{2007AJ....133.1741B} \mbox{$R_{\rm e} \propto L^{0.88}$}. 

We find a much weaker correlation between $M_{ V}$ and
$n$ for our 41 core-S\'ersic spheroids. A symmetrical OLS fit finds
\mbox{$L_{V} \propto n^{1.65\pm 0.75}$} with r $\sim -0.30$
(Table~\ref{Table6}). Interestingly, however, the $M_{\rm BH}-n$
relation that we derive combining our $L_{V}-n$, $R_{\rm b}-L_{\rm V}$
and \mbox{$R_{\rm b}-M_{\rm BH, direct}$} relations for the full
sample (\mbox{$M_{\rm BH} \propto n^{2.75\pm 1.31}$}) has a slope
which is consistent with those of the relations in \citet[slope
$\sim 2.68 \pm 0.40$]{2007ApJ...655...77G} and \citet[slope
$\sim 2.69 \pm 0.33$]{2019ApJ...873...85D}.

 \section{Discussion}\label{Sec5}

\begin{figure*}
\hspace*{3.64 cm}   
\includegraphics[angle=270,scale=0.647]{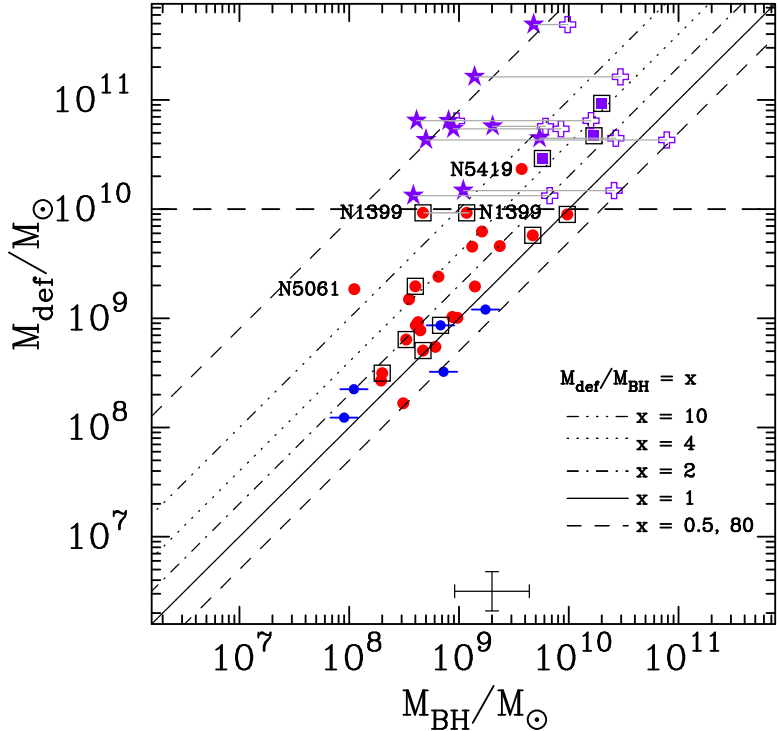}
  \caption{Central mass deficit ($M_{\rm def}$) plotted as a function of
  black hole mass ($M_{\rm BH}$). For 11 galaxies enclosed in boxes,
  we used their dynamically determined SMBH masses (Table~\ref{Table5}
  and \citealt[their Table~4]{2014MNRAS.444.2700D}). For NGC 1399, we
  plot two direct SMBH mass measurements.  For the remaining 30
  galaxies the SMBH masses were estimated using the $M-\sigma$ (filled
  disks, filled circles and filled stars) or $M-L$ relations (open
  crosses). All 13 large-core galaxies shown in purple (filled boxes,
  filled stars and open crosses) have
  $M_{\rm def} \ga 10^{10} M_{\sun}$, whereas the normal core galaxies
  (filled red circles and blue disks) typically have
  $M_{\rm def} \la 10^{10} M_{\sun}$. A representative error bar is
  shown at the bottom of the panel. }
\label{Fig9} 
 \end{figure*}
 
\subsection{Central stellar mass deficit}\label{Sec5.1}

The central stellar mass deficits in core-S\'ersic spheroids
($M_{\rm def}$) are useful to explain if the large cores of BCGs and
central dominant galaxies reflect the actions of overmassive SMBHs or
intense core scouring via large amount of galactic merging or a
combination of both. As mentioned in the introduction, the favoured
mechanism for the creation of central stellar mass deficits is the three-body
encounters of the core stars with the inspiraling binary SMBHs that
form from dry major mergers
(\citealt{1980Natur.287..307B,1991Natur.354..212E,1997AJ....114.1771F,2001ApJ...563...34M,2002MNRAS.331L..51M}).
Using of \mbox{$N$-body} simulations, \citet{2006ApJ...648..976M}
first revealed that the accumulated stellar mass deficit that is
generated by binary SMBHs after $\mathcal{N}$ numbers of successive
dry major mergers scales as
$M_{\rm def} \approx 0.5 \mathcal{N} M_{\rm BH}$, where $M_{\rm BH}$
is the total sum of the masses of the binary SMBHs.

In order to derive the  central stellar mass deficits for the large-core
galaxies, we follow the same prescription in
\citet{2014MNRAS.444.2700D,2017MNRAS.471.2321D,2018MNRAS.475.4670D}.
The central stellar luminosity deficit ($L_{\rm def}$) is computed as
the difference in luminosity between inwardly-extrapolated outer
S\'ersic profile of the complete core-S\'ersic model fit to the
spheroid (Eq.~\ref{Eq1}) and the core-S\'ersic model itself
(Eq.~\ref{Eq2}). For each galaxy this luminosity deficit is converted
into $M_{\rm def} $ using the stellar mass-to-light ratios ($M/L$)
given in Table~\ref{Table5}.  We measure stellar mass deficits for the
large-core galaxies that are $M_{\rm def} \ga 10^{10} M_{\sun}$,
larger than those for normal-core galaxies
$M_{\rm def} \la 10^{10} M_{\sun}$ except for NGC~5419. The
brightest cluster galaxy, NGC~5419, of the poor cluster Abell S753 is
a normal-core galaxy with $R_{\rm b} \sim 416$ pc and
$M_{\rm def} \sim 2.3 \times 10^{10} M_{\sun}$ (\citealt{2014MNRAS.444.2700D}).

Fig.~\ref{Fig9} shows the mass deficits as a function of directly
measured or predicted SMBH masses.  For the 11 core-S\'ersic galaxies
in the sample with directly measured $M_{\rm BH}$ we find
$M_{\rm def} \sim 0.5-5~M_{\rm BH}$, consistent with spheroid
formation via  a reasonably
large number of dry major merger events (1$-$10, \citealt{2006ApJ...648..976M}). This figure is consistent with
$M_{\rm def} /M_{\rm BH}$ ratio we find for the normal-core galaxies
without direct $M_{\rm BH}$; using SMBH masses predicted based on the
$M_{\rm BH}-\sigma$ relation (\citealt[their
Table~4]{2014MNRAS.444.2700D}) gives
$M_{\rm def} \sim 0.5-4~M_{\rm BH,\sigma-based}$.  Our findings also
agree with previous work which derived
$M_{\rm def} \sim 0.5-4~M_{\rm BH}$ for normal-core galaxies using
similar methods (e.g., \citealt{2004ApJ...613L..33G,
  2006ApJS..164..334F,2013AJ....146..160R,2013ApJ...768...36D,2014MNRAS.444.2700D,2018MNRAS.475.4670D})
and with \citet{2010MNRAS.407..447H} who calculated $M_{\rm def}$ via
a model-independent analysis of the light profiles, rather than
determining the difference in luminosity between a S\'ersic fit and a
core-S\'ersic fit as done here, finding
$M_{\rm def} \sim 2~M_{\rm BH}$. Our results can also be compared with
studies which estimated the merger rates for massive galaxies from
observations of close galaxy pairs, finding that massive galaxies have
undergone 0.5 to 4 major mergers since $z\sim 3$
\citep[e.g.,][]{2006ApJ...652..270B,2007IAUS..235..381C,2009MNRAS.396.2003L,2012MNRAS.425..287E,2012ApJ...747...34B,2013MNRAS.433..825L,2016ApJ...830...89M,2017MNRAS.470.3507M,2019arXiv190312188D}.

Of particular importance here is that the large-core galaxies without
direct $M_{\rm BH}$ have predicted SMBH masses
($M_{\rm BH,\sigma-based}$ and $M_{{\rm BH},L{\rm -based}}$) that are
undermassive for their stellar mass deficits (Fig.~\ref{Fig9} and
Table~\ref{Table5}). This echos the offset nature of the predicted
SMBH masses of large-core spheroids shown in Fig.~\ref{Fig5}. We find
$M_{\rm def} \sim (10-160) ~M_{\rm BH,\sigma-based}$ and
$M_{\rm def} \sim (2-70)~M_{{\rm BH},L{\rm -based}}$. These figures
correspond to unrealistically high number of major dry mergers
($\mathcal{N} \sim 5-320$) for the bulk ($\sim 70$\%) of the
large-core spheroids. We argue that the excessive merger rates have
arisen because the $M_{\rm BH}-\sigma$ and $M_{\rm BH}-L$ relations
significantly underestimate the SMBH masses for large-core galaxies
(see Section~\ref{Sec4.1}). On the other hand, the
$R_{\rm b}-M_{\rm BH,direct}$ relation for the core-S\'ersic spheroids
(Fig.~\ref{Fig45}) is such that the predicted SMBH masses for the
large-core spheroids based on $R_{\rm b}$ are typically a factor of
1.7$-$4.5 (and 10$-$43), i.e., $\sim 0.6-1.7 \sigma$ (and
$\sim 3.7-15.6 \sigma$), larger than expectations from the spheroid
$L$ (and $\sigma$), Table~\ref{Table5}. If we use these $R_{\rm b}$
based SMBH masses then the derived number of major, dry mergers for
large-core galaxies would be $\mathcal{N} \sim 2-7$
(Table~\ref{Table5}), in good agreement with observations and
theoretical expectations in a hierarchical Universe.

It is worthwhile noting that enhanced core scouring can occur due to a
gravitational radiation-recoiled SMBH\footnote{Interested readers are
  referred to Section 5.4 of \citet{2014MNRAS.444.2700D} and Section
  6.1 of \citet{2013ApJ...768...36D} for further discussions on
  alternative mechanisms for the formation of enhanced depleted
  cores.} (e.g.,
\citealt{1989ComAp..14..165R,2004ApJ...613L..37B,2004ApJ...607L...9M,2008ApJ...678..780G}). While
substantial SMBH recoiling events would lower the inferred merger
rates by a few, this process does not account for the unrealistically
high values of $\mathcal{N}$ quoted above. We also note the binary
SMBH core scouring scenario assumes that the SMBH binaries coalesce in
most merged galaxies via the emission of gravitational wave. The
alternative scenario is multiple SMBH systems form \citep[e.g.,][]{2019arXiv190710639L}, generating large
stellar mass deficits due to the gravitational sling-shot ejection of
the SMBHs \citep{2012MNRAS.422.1306K}. This process would lead to a
smaller SMBH mass, at odds with the tight $R_{\rm b}-M_{\rm BH}$
correlation for core-S\'ersic spheroids
(Fig.~\ref{Fig4}). Furthermore, in Section~\ref{Sec5.2}, we also show
the excessive merger rates cannot be explained by the galaxy
environment.

\subsection{Impact of environment on the break radius and galaxy
  merger rate}\label{Sec5.2}

 Given the rarity of large-core galaxies, it is of interest to explore
 the impact of the environment on their break radii and major merger
 histories. All the  large-core galaxies in our sample except for three (NGC 1600,
 NGC 4486 and NGC 4874) are classified as BCGs (Section~\ref{Sec2.2}).
 \mbox{NGC 1600} is the brightest member of the poor \mbox{NGC 1600}
 group, whereas NGC 4486 and NGC 4874 are the second brightest cluster
 galaxies sitting at the centre of their host clusters.

\begin{figure}
\includegraphics[angle=270,scale=0.58440299354]{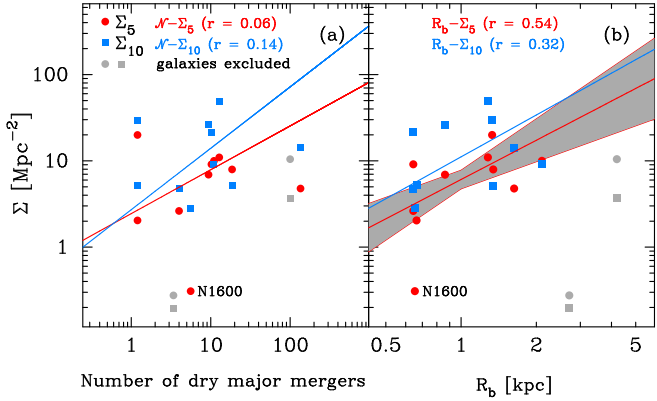}
\caption{Comparison of environmental measures ($\Sigma_{5}$ and
  $\Sigma_{10}$) with the number of major dry mergers
  $\mathcal{N}$($\approx 2 M_{\rm def}/M_{\rm BH}$,
  \citealt{2006ApJ...648..976M}) (a) and break radius $R_{\rm b}$ (b)
  for our 9 large-core galaxies, excluding two of the
  three most distant sample galaxies in our sample with D $\ga 360$ Mpc
  (A2261-BCG and IC~1101, grey points) and the BCG 4C +74.13 with no
  robust data for its nearest neighbours in NED (Table~\ref{Table1}).
  We use $\mathcal{N}$ derived using $M_{{\rm BH},L{\rm-based}}$ or
  direct $M_{\rm BH}$ when available. $\Sigma_{5}$ which is based on 5
  nearest neighbour galaxies with $M_{B} \la -19.5$ mag is less
  affected by distance than $\Sigma_{10}$ calculated using 10 nearest
  neighbours with $M_{B} \la -18$ mag. Overall, the galaxy merger rate
  exhibits no significant dependence the environment estimates (see
  the text). On the
  other hand, large-core galaxies in high-density regions tend to
  exhibit larger depleted cores than those in relatively low-density
  environments. }
\label{Fig10A}
 \end{figure}

\begin{figure}
\hspace*{0.043099004259cm}   
 \vspace*{.2342341330 cm}   
  \includegraphics[angle=0,scale=0.273689]{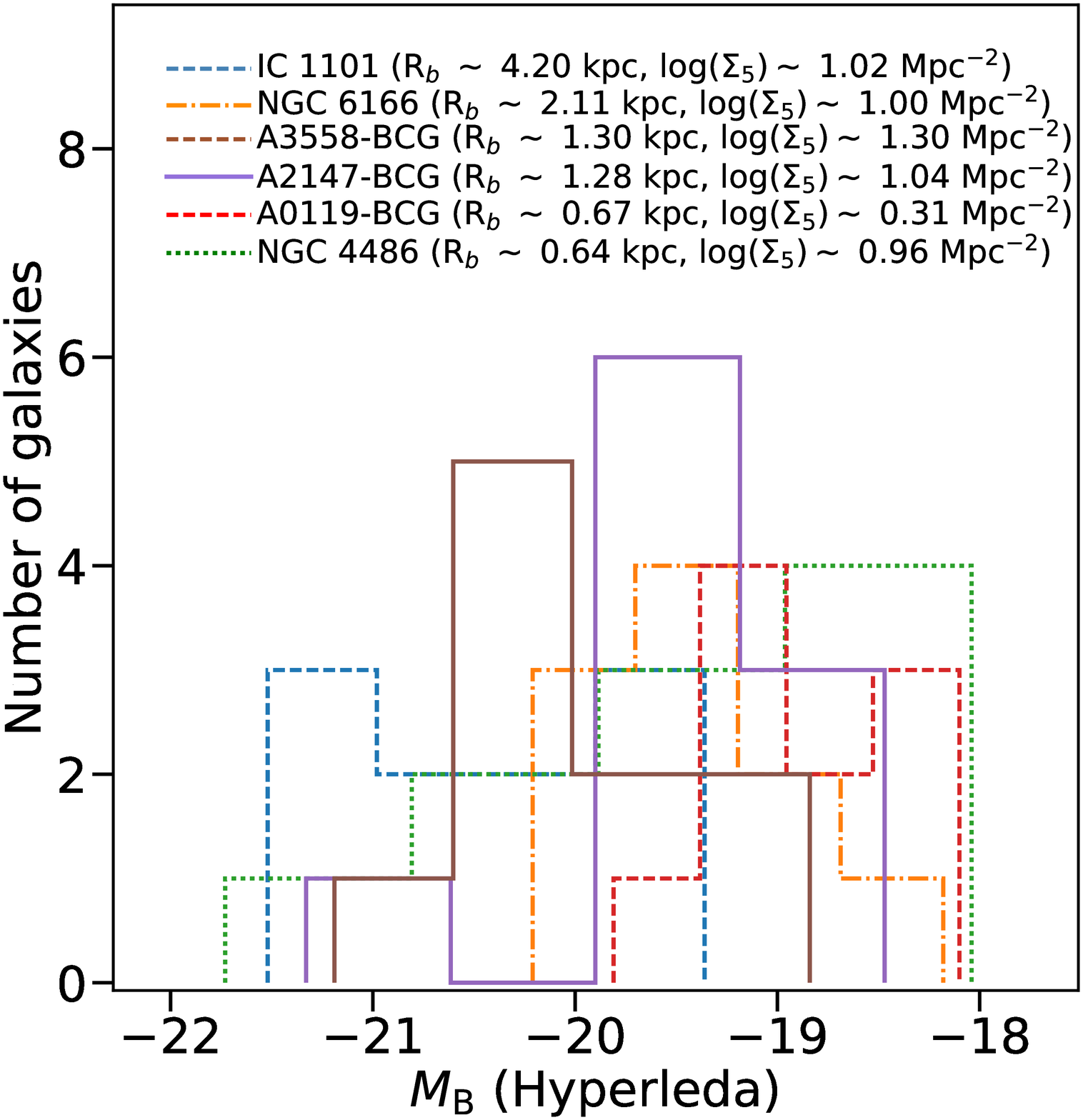}
\caption{Histograms of absolute $B$-band galaxy magnitudes ($M_{B}$,
  Hyperleda) of the 10 nearest neighbours (with $M_{B} \la -18$ mag)
  for six large-core galaxies (NGC~4486, NGC~6166, A0119-BCG,
  A2147-BCG, A3558-BCG and IC~1101), Table~\ref{Table1}. These six
  galaxies have representative break radii for the large-core galaxy
  population, allowing us to explore the trend between the break radii
  of large-core galaxies ($R_{\rm b}$) and the luminosities of their
  10 nearest neighbours. Large-core galaxies with brighter neighbours
appear to have larger depleted cores.}
\label{Fig10} 
 \end{figure}

Following \citet[their Section~3.1]{2011MNRAS.416.1680C}, we make use
of two parameters ($\Sigma_{5}$ and $\Sigma_{10}$) to obtain an
estimate of the galaxy environment. The surface density $\Sigma_{10}$
is defined as ${\rm N_{gal}}$/($\pi R_{10}^{2}$), where $R_{10}$ is
the radius, centred on a large-core galaxy, that encloses the 10 nearest
neighbours with $M_{B} \la -18.0$ mag and the relative recession velocity of the galaxies 
\mbox{$| V_{\rm hel,large-core} -V_{\rm hel,neighbour}|<$ 300 km
  s$^{-1}$}. Similarly, $\Sigma_{5}$=${\rm N_{gal}}$/($\pi R_{5}^{2}$),
where $R_{5}$ is the radius centred on a large-core galaxy enclosing
the 5 nearest neighbours with $M_{B} \la -19.5$ mag and
\mbox{$| V_{\rm hel,large-core} -V_{\rm hel,neighbour}|<$ 300 km
  s$^{-1}$}. The nearest neighbour identification, recession
velocities ($V_{\rm hel}$) are based on NED, while the $B$-band
absolute galaxy magnitudes are from Hyperleda. 
We excluded the
large-core galaxy \mbox{4C +74.13} with no robust data for its nearest
neighbours in NED. The caveat here is that galaxies with
\mbox{$M_{B} \ga -19.5$ mag} may be too faint for detection at the
distances of \mbox{4C +74.13} (D $\sim$ 925 Mpc), A2261-BCG (D $\sim$
959 Mpc) and IC 1101 (D $\sim$ 363 Mpc), see Figs.~\ref{Fig10A},
\ref{Fig10} and Appendix~\ref{AppB}. Therefore, we caution that
A2261-BCG and IC 1101 can be biased toward having brighter nearest
neighbours and low $\Sigma_{10}$, compared to the other large-core
galaxies with D $\la$ 213 Mpc (see Appendix~\ref{AppB}).
 
 Fig.~\ref{Fig10A} shows the trends between environmental measures
 ($\Sigma_{5}$ and $\Sigma_{10}$) and (a) number of major dry mergers
 $\mathcal{N}$ and (b) break radius $R_{\rm b}$ for the 12 large-core
 galaxies in our sample (see also Appendix~\ref{AppB}). The values of
 $\mathcal{N}$ were derived using $L$-based SMBH masses
 ($M_{{\rm BH},L{\rm-based}}$) or direct $M_{\rm BH}$ when available
 (see Table~\ref{Table5}).  Excluding A2261-BCG and IC 1101, we find
 significant correlations between $R_{\rm b}$ and $\Sigma_{5}$ and
 $\Sigma_{10}$, in the sense that large-core galaxies in high-density
 regions tend to exhibit larger $R_{\rm b}$ than those in relatively
 low-density environments (Table~\ref{Table6}).  These correlations
 also reveal that more massive SMBHs are hosted by large-core galaxies
 that reside in denser environments. The Pearson correlation
 coefficients for the $\Sigma_{5}-R_{\rm b}$ and
 $\Sigma_{10}-R_{\rm b}$ relations are r~$\sim 0.54-0.57$ and
 r~$\sim 0.32$, respectively.

 We do not witness a correlation between the number of dry major
 mergers $\mathcal{N}$ and the environment estimates $\Sigma_{5}$ and
 $\Sigma_{10}$ (r~$\sim 0.06-0.14$). Accordingly, the excessive amount
 of major mergers that we obtained for a couple of large-core galaxies
 ($\mathcal{N} \ga 5-320$, Section~\ref{Sec5.1} and Fig.~\ref{Fig9})
 cannot be readily explained by their local projected environmental
 densities. Instead, the $L$-based SMBH masses---which we used to
 calculate the merger rates ($\mathcal{N}$
 $\approx 2 M_{\rm def}/M_{\rm BH}$, \citealt{2006ApJ...648..976M})
 for the bulk (7/10) of the galaxies in Fig.~\ref{Fig10A}---may
 explain the high $\mathcal{N}$ values as well as the poor
 $\Sigma-\mathcal{N}$ correlation, since $M_{{\rm BH},L{\rm-based}}$
 appear to be undermassive for the break radii ($R_{\rm b}$) and mass
 deficits ($M_{\rm def}$) of large-core galaxies (Figs.~\ref{Fig5} and
 \ref{Fig9}).

In Fig.~\ref{Fig10}, we show histograms of absolute $B$-band galaxy
magnitudes ($M_{B}$) of the 10 nearest neighbours with $M_{B} \la -18$
mag for six large-core galaxies (NGC 4486, NGC 6166, A0119-BCG,
A2147-BCG, A3558-BCG and IC 1101). These six large-core galaxies are
selected to have representative break radii for the large-core galaxy
population. We cannot presently reach a firm conclusion, but there is
a hint that large-core galaxies with larger break radii and high
$\Sigma_{5}$ have brighter companions (see also
Appendix~\ref{AppB}). This reinforces the idea that the most massive
galaxies experience a higher proportion of mergers between massive
spheroidal systems.

\begin{figure*}
\hspace*{1.2693072599cm}   
\includegraphics[angle=0,scale=0.55299354]{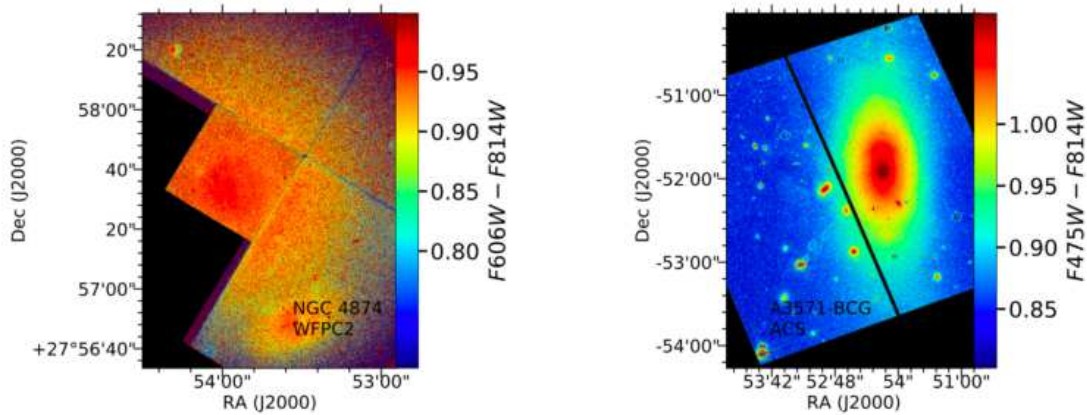}
\caption{{\it HST} WFPC2 F606W $-$ F814W and {\it HST} ACS F475W $-$
  F814W colour maps for the two large-core galaxies in our sample with
  intermediate-scale components (NGC~4874 and A3571-BCG). The maps
  reveal that the galaxies turn bluer with increasing radius. North is
  up and east is to the left. }
\label{Fig14} 
 \end{figure*}

\subsection{Formation of  core-S\'ersic galaxies: 
  ``large-core''  versus ``normal-core'' galaxies
  }\label{Sec5.3}

  We postulate that both ``large-core'' and ``normal-core''
  core-S\'ersic spheroids are built through a reasonably large number
  of successive dry major mergers ($\mathcal{N} \sim 1-10$) involving
  SMBHs (e.g.,
  \citealt{1997AJ....114.1771F,1999ASPC..182..124K,2003AJ....125..478L,2004AJ....127.1917T,2006ApJ...652..270B,2006ApJS..164..334F,2007ApJ...662..808L,2009ApJS..182..216K,
    2009MNRAS.396.2003L,2011MNRAS.412L...6B,2013AJ....146..160R,2013ApJ...768...36D,
    2014MNRAS.444.2700D,2015ApJ...798...55D,2017MNRAS.471.2321D,2018MNRAS.475.4670D}). In
  the previous sections we have shown that ``large-core'' spheroids of
  BCGs and central dominant galaxies with $R_{\rm b} \ga 0.5$ kpc,
  $M_{V} \la -23.50 \pm 0.1$ mag, $M_{\rm def} \ga 10^{10} M_{\sun}$
  are extremely massive (i.e., $M_{*} \ga 10^{12}M_{\sun}$). The same
  $\mu_{\rm b}-R_{\rm b}$, $R_{\rm b}-L_{V}$, $L_{V}-R_{\rm e}$ and
  $L_{V}-n$ relations defined by the ``normal-core'' core-S\'ersic
  spheroids with $R_{\rm b} \la 0.5$ kpc, $-20.70$ mag $\ga M_{V} \ga$
  $-23.60$ mag, $M_{\rm def} \la 10^{10} M_{\sun}$ and stellar masses
  $ M_{*} \sim 8\times 10^{10} - 10^{12}M_{\sun}$ hold up at higher
  masses for the large-core spheroids. This is also the case for the
  $R_{\rm b}-M_{\rm BH}$ relation when using directly measured SMBH
  masses.

  The bulk ($\sim$77\%) of our large-core galaxies are BCGs, which is
  unsurprising as BCGs are predicted to experience a more intense
  merging and accretion events than galaxies with relatively low
  luminosities. Our findings hint that large-core spheroids are more
  likely to undergo a higher proportion of dry major mergers than the
  normal-core spheroids.  Although we only have three large-core
  spheroids with measured $M_{\rm BH}$, we find
  $\mathcal{N} \sim 6-10$ for these spheroids, whereas the bulk (6/8)
  of our normal-core spheroids with measured $M_{\rm BH}$ have
  $\mathcal{N} \sim 2-4$, see Fig.~\ref{Fig9}. This is in line with
  the analytic and semi-analytic study by \citet{2013ApJ...768...29V},
  however see \citet{2015MNRAS.446.2330S}. \citet{2018MNRAS.477.5327K}
  also found that about half of the most massive galaxies
  ($M_{*} \ga 10^{12}M_{\sun}$) exhibit prolate-like rotation,
  consistent with these galaxies being products of dry major mergers.
  The depleted cores and stellar mass deficits in large-core spheroids
  likely reflect the cumulative effect of multiple dry mergers and the
  ensuing excavation of inner stars from the core by coalescing,
  overmassive black hole binaries with a final mass
  $M_{\rm BH} \ga 10^{10} M_{\sun}$ (Sections~\ref{Sec4.1} and
  \ref{Sec5.1}), typically a factor of 1.7$-$4.5 (and 10$-$43) larger
  than the SMBH masses estimated using the spheroids' $L$ (and
  $\sigma$).  In passing we note that \citet[see also
  \citealt{2012MNRAS.424..224H}]{2018MNRAS.474.1342M} wrote that 40\%
  of their sample of 72 BCGs at redshift of $z \sim 0.006- 0.300$ must
  have $M_{\rm BH} \ga 10^{10} M_{\sun}$ to lie on the fundamental
  plane of black hole accretion. The analytic arguments by
  \citet{2019MNRAS.487.4827K} predict present-day galaxies with
  overmassive black holes that lie above the $M_{\rm BH}-\sigma$
  relation may be descendants of the compact blue nuggets formed at
  $z \ga 6$, suggesting an evolutionary link between the high-redshift
  blue nuggets and large-core galaxies.

  The higher prevalence of dry mergers associated with large-core
  spheroids can explain the break that we identified for the first
  time in the core-S\'ersic $\sigma-L_{V}$ relation occurring at
  \mbox{$M_{V} \sim -23.50 \pm 0.10$ mag}, not to be confused with the
  change in the slope of the $\sigma-L_{V}$ relation due to
  \mbox{core-S\'ersic} versus S\'ersic galaxies. The prediction that
  major, dry mergers add the stellar mass, black hole mass and sizes
  in equal proportion while increasing the velocity dispersion only
  slightly (e.g., \citealt{2003MNRAS.342..501N,2007ApJ...658...65C,
    2012ApJ...744...63O,2013MNRAS.429.2924H}) would imply that the
  $\sigma-L_{V}$ relation may not hold across the full mass range of
  core-S\'ersic spheroids. We find a flattening of the slope of the
  $\sigma-L_{V}$ relation and larger scatter at the extremely massive
  end, where the velocity dispersions of the large-core spheroids
  increase only slightly with $L_{V}$, compared to the relatively
  steep $\sigma-L_{V}$ relation for normal-core galaxies
  (Fig.~\ref{Fig6}).

  Turning to the intermediate- and large-scale components, as noted in
  Section~\ref{Sec3.1}, of the 13 large-core galaxies in our sample 10
  have low surface brightness outer stellar halos with
  exponential-like distribution of stars and with physical scales of
  $R_{\rm e} \sim 10-300$ kpc (Appendix~\ref{FigA1} and
  \citealt{2017MNRAS.471.2321D}, see also
  \citealt{2007MNRAS.378.1575S,2008A&A...483..727P,2017ApJ...849....6A}).
  The spheroid-to-halo effective radius ratios for the large-core
  galaxies are $R_{\rm e,spheroid}/R_{\rm e,halo} \sim 0.05-1$
  (Tables~\ref{Table3} and \ref{Table4}). Three out of these 10
  large-core galaxies (NGC~4874, A3571 and IC~1101) also exhibit
  intermediate-scale components. The color maps for these three
  galaxies become gradually bluer towards larger radii (see
  Fig.\ref{Fig14} and \citealt[their
  Fig.~6]{2017MNRAS.471.2321D}). The color maps of NGC~4874 and A3571
  (Fig.~\ref{Fig14}) were created adopting the same prescription
  described in \citet[their Section 3.3]{2017MNRAS.471.2321D}.
  Coupled with our decompositions, these color maps suggest different
  origins for the inner, red spheroid with a high concentration of old
  stars and the intermediate- and large-scale components with
  relatively bluer colors. We find that, when present, the fractional
  contributions of the outer halos and intermediate-scale components
  are typically $\sim$15$-$60\% and $\sim$20\% of the total (i.e.,
  spheroid + intermediate-component + halo) flux, respectively
  (similar results have been previously reported by e.g.,
  \citealt{2005ApJ...618..195G,2007MNRAS.378.1575S,2010ApJ...725.2312O,2016ApJ...820...49J}).
 
  It appears plausible that the natural assembly pathways for the
  intermediate-scale and outer halo components of large-core galaxies
  are significant accretion of less massive neighbours and the
  accumulation of stars stripped during galaxy-galaxy encounters
  occurring at $z < 1$ (e.g.,
  \citealt{2007MNRAS.378.1575S,2009ApJ...702.1058Z,2009ApJ...699L.178N,2010ApJ...725.2312O,2011ApJS..195...15D,2012MNRAS.425.3119H,2013MNRAS.429.2924H,2012ApJ...754..115J,2015MNRAS.449.2353B,2015MNRAS.451.2703C,2016MNRAS.458.2371R,2017MNRAS.471.2321D,2018MNRAS.475.4670D,2018MNRAS.474..917M,2018MNRAS.475..648P}). A
  small contribution to the intermediate-components and halo lights
  could come from core stars which are gravitationally ejected by the
  SMBH binaries and accumulate at large radii outside the core or
  escape from the spheroid at high velocities
  \citep[e.g.,][]{1988Natur.331..687H,2008MNRAS.383...86O,2012ApJ...754L...2B}. The
  masses of these ejected stars (i.e., the stellar mass deficits) in
  large-core galaxies are typically $\sim$10$-$20\% (and $\sim$1$-$3\%
  ) of the stellar masses of their intermediate- (and outer halo)
  components (Table~\ref{Table5}). Moreover, the observed trend of
  outwardly rising ellipticity for the large-core galaxies, except for
  the only group galaxy in our sample, NGC~1600 (Appendix~\ref{AppA})
  suggests that the halo (and perhaps also the intermediate-scale
  component) stars at large radii trace the global cluster potential
  rather than the host spheroid potential
  \citep[e.g.,][]{1979A&AS...37..591D,1991AJ....101.1561P,1995ApJ...440...28P,1998ApJ...502..141D,2005ApJ...618..195G,2006MNRAS.372L..68K,2017NatAs...1E.157W}.

\section{Conclusions}\label{ConV}

Motivated by the need to re-investigate the formation and structural
scaling relations of massive galaxies over a wide dynamic range in
spheroid luminosity and stellar mass, we have extracted composite
({\it HST} WFPC2, ACS and NICMOS plus ground-based) major-axis surface
brightness and ellipticity profiles for 12 extremely massive
core-S\'ersic galaxies (9 BCGs, 2 second brightest cluster galaxies
and 1 brightest group galaxy) with core sizes $R_{\rm b} > 0.5$ kpc.
We perform careful, multi-component (halo/\mbox{intermediate-scale
  component}/spheroid/nucleus) decompositions of these composite light
profiles which typically cover a large radial range $R \ga 100$
arcsec. In so doing, we modelled the spheroid with a core-S\'ersic
profile and, when present, we fit an exponential function to the outer
halo, a S\'ersic model to the intermediate-scale component and a
Gaussian or a S\'ersic profile to the nucleus. This is the first time
this has been done for the full sample of 12 galaxies. The
decompositions yield an excellent fit to the light profiles of the
galaxies with a median rms scatter of 0.031 mag arcsec$^{-2}$.

We additionally included the galaxy with largest depleted core
detected to date IC~1101 ($R_{\rm b} \sim 4.2$ kpc) from
\citet{2017MNRAS.471.2321D} and 28 core-S\'ersic early-type galaxies
with $R_{\rm b} < 0.5$ kpc from \citet{2014MNRAS.444.2700D}. This
resulted in the largest sample of 41 core-S\'ersic galaxies studied to
date consists of 13 ``large-core'' galaxies  having $R_{\rm b} > 0.5$ kpc and the
remaining 28 ``normal-core'' galaxies  with $R_{\rm b} < 0.5$ kpc. Our
principal conclusions are as follows:\\

(1) We find that large-core spheroids have $V$-band absolute magnitude
$M_{V} \la -23.50 \pm 0.10$ mag, sizes $R_{\rm e} \ga 10-300$ kpc and
stellar masses that are typically $M_{*} \ga 10^{12}M_{\sun}$, whereas
for the relatively less luminous normal-core spheroids ($ - 20.70$ mag
$\ga M_{V} \ga - $23.60 mag), $R_{\rm e} \sim 1- 50$ kpc and
$ M_{*} \sim 8\times 10^{10} - 10^{12}M_{\sun}$. Of the 13 large-core
galaxies, seven have $R_{\rm b} \ga 1.3$ kpc.  The depleted cores and
stellar mass deficits in large-core spheroids are likely due to the
cumulative effect of multiple dry mergers and the ensuing core
scouring by coalescing, overmassive SMBH binaries with a final mass
$M_{\rm BH} \ga 10^{10} M_{\sun}$. For such galaxies, an additional
mechanism that can contribute to the large cores/stellar
mass deficits is oscillatory core passages by a (gravitational
radiation)-kicked SMBH.\\

(2) The detailed multi-component decompositions of the  large-core
galaxies reveal the bulk ($\sim 77\%$) of them exhibit low surface
brightness outer stellar halos with exponential-like stellar
distribution and physical scales of $R_{\rm e} \sim 10-300$ kpc.\\

(3) We present updated structural parameter relations for our 41
massive galaxies with a large range in galaxy luminosity
(Section~\ref{Sec4}). We find that large-core spheroids follow the
same $\mu_{V}-R_{\rm b}$, $R_{\rm b}-L_{V}$, $L_{V}-R_{\rm e}$ and
$L_{V}-n$ relations defined by the relatively less massive,
normal-core spheroids. The strong correlations between $R_{\rm b}$,
and the break surface brightness ($\mu_{\rm b}$) and the spheroid
luminosity ($L_{V}$) for core-S\'ersic galaxies are such that
$R_{\rm b} \propto \mu_{\rm b}^{0.38 \pm 0.02}$ and
$R_{\rm b} \propto L_{V}^{1.38 \pm 0.13}$. We also find a tight,
linear relation between the spheroid's luminosity and size for massive
(core-S\'ersic) ellipticals and BCGs such
that $R_{\rm e} \propto L^{1.08 \pm 0.09}_{V}$.\\

(4) We find a strong log-linear $R_{\rm b}-M_{\rm BH}$ relation for 11
sample galaxies with directly determined SMBH masses
($R_{\rm b} \propto M_{\rm BH}^{0.83 \pm 0.10}$): 3 of these 11
galaxies are large-core galaxies.\\

(5) For normal-core galaxies, the break radius $R_{\rm b}$ correlates
equally well with the directly determined SMBH mass
($M_{\rm BH,direct}$) and the SMBH masses predicted using the
$M_{\rm BH}-\sigma$ and the core-S\'ersic $M_{\rm BH}-L$ relations
($M_{\rm BH,\sigma-based}$ and $M_{{\rm BH},L{\rm-based}}$,
\citealt{2013ApJ...764..151G}).  In contrast, our large host galaxy
luminosity range has revealed significant offsets in the
$R_{\rm b}-M_{\rm BH,\sigma-based}$ and
$R_{\rm b}-M_{{\rm BH},L{\rm-based}}$ diagrams at the highest galaxy
masses (i.e., $M_{*} \ga 10^{12}M_{\sun}$). The offset is more
pronounced in the former relation. The SMBH masses of large-core
galaxies estimated from the $M_{\rm BH}-L$ relation are roughly an
order of magnitude larger than those from the $M_{\rm BH}-\sigma$
relation.  We determined that these offsets arise because the SMBH
masses in the most massive galaxies (i.e.,
$M_{*} \ga 10^{12}M_{\sun}$) are high relative to what is expected
from their velocity dispersions or bulge luminosities.  For such
galaxies, we recommend the $R_{\rm b}-M_{\rm BH,direct}$ relation
should be used for determining
the SMBH masses.\\

(6) We have measured central stellar mass deficits in  large-core
galaxies ($M_{\rm def} \ga 10^{10} M_{\sun}$), which when compared to
$M_{\rm BH,direct}$, $M_{\rm BH,\sigma-based}$ and
$M_{{\rm BH},L{\rm-based}}$ yield $M_{\rm def}/M_{\rm BH}$ ratios of
$ \sim 0.5-5$, $ \sim 10-160$ and $\sim 2-70$, respectively. While the
former ratio translates to a reasonably large merger rate (i.e.,
$\mathcal{N} \sim 1-10$), the latter two correspond to unrealistically
large number of major dry mergers ($\mathcal{N} \sim 5-320$) for the
bulk ($\sim 70$\%) of the large-core spheroids. These findings strengthen
the conclusions above: the central SMBH mass in large-core galaxies is
considerably larger than the expectations from the spheroid $\sigma$
and $L$. On the other hand, the predicted SMBH masses for the large-core
spheroids based on $R_{\rm b}$ are of order
$M_{\rm BH} \ga 10^{10} M_{\sun}$ (Table~\ref{Table5}), and typically a
factor of 1.7$-$4.5 (and 10$-$43), i.e.,  $\sim 0.6-1.7 \sigma$
  (and $\sim 3.7-15.6 \sigma$), larger than the SMBH masses
estimated using the spheroids' $L$ (and $\sigma$). Using these $R_{\rm b}$-based
SMBH masses for the large-core galaxies brings down  the merger rate to
$\mathcal{N} \sim 2-7$, in good agreement
with observations and theoretical expectations.\\

(7) We find significant correlations between $R_{\rm b}$ and galaxy
environment estimates $\Sigma_{5}$ and $\Sigma_{10}$, i.e.,
r~$\sim 0.54-0.57$ and r~$\sim 0.32$, respectively.
Large-core galaxies in high-density regions tend to
exhibit larger $R_{\rm b}$ than those in relatively low-density
environments. Our findings also reveal that more massive SMBHs are
hosted by large-core galaxies that reside in denser environments. In
contrast, the galaxy merger rate exhibits no significant dependence
the environment estimates, we therefore rule out the excessive amount
of major mergers that we obtained for a couple of large-core galaxies
($\mathcal{N} \sim 5-320$)
being due to higher local projected environmental densities.\\

(8) Our results hint that large-core spheroids are more likely to
experience a higher proportion of dry major mergers than the
normal-core spheroids (see e.g., \citealt{2013ApJ...768...29V}).
Although our sample only contains three large-core spheroids with
measured $M_{\rm BH}$, we find $\mathcal{N} \sim 6-10$ for these
 spheroids, compared to $\mathcal{N} \sim 2-4$ for the bulk
(6/8) of our normal-core spheroids with measured $M_{\rm BH}$.\\

(9) We discover a break in the core-S\'ersic $\sigma-L_{V}$ relation
occurring at \mbox{$M_{V} \sim -23.50 \pm 0.10$ mag}, not to be
confused with the change in the slope of the $\sigma-L_{V}$ relation
due to \mbox{core-S\'ersic} versus S\'ersic galaxies. We attribute
this to be the result of large-core spheroids undergoing more dry
major mergers than the relatively less massive, normal-core spheroids.

Our findings carry significant implications for studies which attempt
to predict SMBH masses in the most massive galaxies using their
spheroid luminosity ($L$) or $\sigma$.  Future high-resolution
dynamical SMBH mass measurements by modelling stellar or ionized gas
kinematics in galaxies with $M_{V} \la -23.50 \pm 0.1$ mag and
$M_{*} \ga 10^{12}M_{\sun}$ are imperative to further study the
processes that shaped the growth of the most massive galaxies and
their SMBHs.

\section{ACKNOWLEDGMENTS}

I thank the referee for their timely report and constructive
suggestions that improved the original manuscript. I am grateful to Berta Margalef-Bentabol, 
Cristina Cabello and  Mario Chamorro-Cazorla for their comments on this work. I acknowledge
support from a Spanish postdoctoral fellowship `Ayudas 1265 para la
atracci\'on del talento investigador. Modalidad 2: j\'ovenes
investigadores.' funded by Comunidad de Madrid under grant number
2016-T2/TIC-2039. I acknowledge financial support from the Spanish Ministry of Science, Innovation and Universities 
(MCIUN) under grant numbers AYA2016-75808-R and RTI2018-096188-B-I00. This research made use of APLpy, an open-source
plotting package for Python \citep{2012ascl.soft08017R} and the
NASA/IPAC Extragalactic Database (NED), which is operated by the Jet
Propulsion Laboratory, California Institute of Technology, under
contract with the National Aeronautics and Space Administration.
Based on observations made with the NASA/ESA Hubble Space Telescope,
and obtained from the Hubble Legacy Archive, which is a collaboration
between the Space Telescope Science Institute (STScI/NASA), the Space
Telescope European Coordinating Facility (ST-ECF/ESA) and the Canadian
Astronomy Data Centre (CADC/NRC/CSA).  This publication makes use of
data products from the Two Micron All Sky Survey, which is a joint
project of the University of Massachusetts and the Infrared Processing
and Analysis Center/California Institute of Technology, funded by the
National Aeronautics and Space Administration and the National Science
Foundation.

\bibliographystyle{apj}
\bibliography{Bil_Paps_biblo.bib}

\begin{thebibliography}{}
\expandafter\ifx\csname natexlab\endcsname\relax\def\natexlab#1{#1}\fi

\bibitem[{{Alamo-Mart{\'{\i}}nez} \& {Blakeslee}(2017)}]{2017ApJ...849....6A}
{Alamo-Mart{\'{\i}}nez}, K.~A., \& {Blakeslee}, J.~P. 2017, \apj, 849, 6

\bibitem[{{Baldry} {et~al.}(2012){Baldry}, {Driver}, {Loveday}, {Taylor},
  {Kelvin}, {Liske}, {Norberg}, {Robotham}, {Brough}, {Hopkins}, {Bamford},
  {Peacock}, {Bland-Hawthorn}, {Conselice}, {Croom}, {Jones}, {Parkinson},
  {Popescu}, {Prescott}, {Sharp}, \& {Tuffs}}]{2012MNRAS.421..621B}
{Baldry}, I.~K., {Driver}, S.~P., {Loveday}, J., {et~al.} 2012, \mnras, 421,
  621

\bibitem[{{Barnes}(1988)}]{1988ApJ...331..699B}
{Barnes}, J.~E. 1988, \apj, 331, 699

\bibitem[{{Barrows} {et~al.}(2012){Barrows}, {Stern}, {Madsen}, {Harrison},
  {Assef}, {Comerford}, {Cushing}, {Fassnacht}, {Gonzalez}, {Griffith},
  {Hickox}, {Kirkpatrick}, \& {Lagattuta}}]{2012ApJ...744....7B}
{Barrows}, R.~S., {Stern}, D., {Madsen}, K., {et~al.} 2012, \apj, 744, 7

\bibitem[{{Begelman} {et~al.}(1980){Begelman}, {Blandford}, \&
  {Rees}}]{1980Natur.287..307B}
{Begelman}, M.~C., {Blandford}, R.~D., \& {Rees}, M.~J. 1980, \nat, 287, 307

\bibitem[{{Bell} {et~al.}(2006){Bell}, {Phleps}, {Somerville}, {Wolf}, {Borch},
  \& {Meisenheimer}}]{2006ApJ...652..270B}
{Bell}, E.~F., {Phleps}, S., {Somerville}, R.~S., {et~al.} 2006, \apj, 652, 270

\bibitem[{{Bender} {et~al.}(2015){Bender}, {Kormendy}, {Cornell}, \&
  {Fisher}}]{2015ApJ...807...56B}
{Bender}, R., {Kormendy}, J., {Cornell}, M.~E., \& {Fisher}, D.~B. 2015, \apj,
  807, 56

\bibitem[{{Benson} {et~al.}(2003){Benson}, {Bower}, {Frenk}, {Lacey}, {Baugh},
  \& {Cole}}]{2003ApJ...599...38B}
{Benson}, A.~J., {Bower}, R.~G., {Frenk}, C.~S., {et~al.} 2003, \apj, 599, 38

\bibitem[{{Bernardi} {et~al.}(2007){Bernardi}, {Hyde}, {Sheth}, {Miller}, \&
  {Nichol}}]{2007AJ....133.1741B}
{Bernardi}, M., {Hyde}, J.~B., {Sheth}, R.~K., {Miller}, C.~J., \& {Nichol},
  R.~C. 2007, \aj, 133, 1741

\bibitem[{{Bernardi} {et~al.}(2011){Bernardi}, {Roche}, {Shankar}, \&
  {Sheth}}]{2011MNRAS.412L...6B}
{Bernardi}, M., {Roche}, N., {Shankar}, F., \& {Sheth}, R.~K. 2011, \mnras,
  412, L6

\bibitem[{{Bertin} \& {Arnouts}(1996)}]{1996A&AS..117..393B}
{Bertin}, E., \& {Arnouts}, S. 1996, \aaps, 117, 393

\bibitem[{{Bianchi} {et~al.}(2008){Bianchi}, {Chiaberge}, {Piconcelli},
  {Guainazzi}, \& {Matt}}]{2008MNRAS.386..105B}
{Bianchi}, S., {Chiaberge}, M., {Piconcelli}, E., {Guainazzi}, M., \& {Matt},
  G. 2008, \mnras, 386, 105

\bibitem[{{Binney} \& {Mamon}(1982)}]{1982MNRAS.200..361B}
{Binney}, J., \& {Mamon}, G.~A. 1982, \mnras, 200, 361

\bibitem[{{Bluck} {et~al.}(2012){Bluck}, {Conselice}, {Buitrago},
  {Gr{\"u}tzbauch}, {Hoyos}, {Mortlock}, \& {Bauer}}]{2012ApJ...747...34B}
{Bluck}, A.~F.~L., {Conselice}, C.~J., {Buitrago}, F., {et~al.} 2012, \apj,
  747, 34

\bibitem[{{Bonfini} {et~al.}(2015){Bonfini}, {Dullo}, \&
  {Graham}}]{2015ApJ...807..136B}
{Bonfini}, P., {Dullo}, B.~T., \& {Graham}, A.~W. 2015, \apj, 807, 136

\bibitem[{{Bonfini} \& {Graham}(2016)}]{2016ApJ...829...81B}
{Bonfini}, P., \& {Graham}, A.~W. 2016, \apj, 829, 81

\bibitem[{{Boylan-Kolchin} {et~al.}(2004){Boylan-Kolchin}, {Ma}, \&
  {Quataert}}]{2004ApJ...613L..37B}
{Boylan-Kolchin}, M., {Ma}, C.-P., \& {Quataert}, E. 2004, \apjl, 613, L37

\bibitem[{{Boylan-Kolchin} {et~al.}(2006){Boylan-Kolchin}, {Ma}, \&
  {Quataert}}]{2006MNRAS.369.1081B}
---. 2006, \mnras, 369, 1081

\bibitem[{{Brown} {et~al.}(2012){Brown}, {Cohen}, {Geller}, \&
  {Kenyon}}]{2012ApJ...754L...2B}
{Brown}, W.~R., {Cohen}, J.~G., {Geller}, M.~J., \& {Kenyon}, S.~J. 2012,
  \apjl, 754, L2

\bibitem[{{Burke} {et~al.}(2015){Burke}, {Hilton}, \&
  {Collins}}]{2015MNRAS.449.2353B}
{Burke}, C., {Hilton}, M., \& {Collins}, C. 2015, \mnras, 449, 2353

\bibitem[{{Burke-Spolaor}(2011)}]{2011MNRAS.410.2113B}
{Burke-Spolaor}, S. 2011, \mnras, 410, 2113

\bibitem[{{Byun} {et~al.}(1996){Byun}, {Grillmair}, {Faber}, {Ajhar},
  {Dressler}, {Kormendy}, {Lauer}, {Richstone}, \&
  {Tremaine}}]{1996AJ....111.1889B}
{Byun}, Y.-I., {Grillmair}, C.~J., {Faber}, S.~M., {et~al.} 1996, \aj, 111,
  1889

\bibitem[{{Caon} {et~al.}(1993){Caon}, {Capaccioli}, \&
  {D'Onofrio}}]{1993MNRAS.265.1013C}
{Caon}, N., {Capaccioli}, M., \& {D'Onofrio}, M. 1993, \mnras, 265, 1013

\bibitem[{{Cappellari} {et~al.}(2011){Cappellari}, {Emsellem}, {Krajnovi{\'c}},
  {McDermid}, {Serra}, {Alatalo}, {Blitz}, {Bois}, {Bournaud}, {Bureau},
  {Davies}, {Davis}, {de Zeeuw}, {Khochfar}, {Kuntschner}, {Lablanche},
  {Morganti}, {Naab}, {Oosterloo}, {Sarzi}, {Scott}, {Weijmans}, \&
  {Young}}]{2011MNRAS.416.1680C}
{Cappellari}, M., {Emsellem}, E., {Krajnovi{\'c}}, D., {et~al.} 2011, \mnras,
  416, 1680

\bibitem[{{Carollo} {et~al.}(1997){Carollo}, {Franx}, {Illingworth}, \&
  {Forbes}}]{1997ApJ...481..710C}
{Carollo}, C.~M., {Franx}, M., {Illingworth}, G.~D., \& {Forbes}, D.~A. 1997,
  \apj, 481, 710

\bibitem[{{Carter}(1977)}]{1977MNRAS.178..137C}
{Carter}, D. 1977, \mnras, 178, 137

\bibitem[{{Ciotti} {et~al.}(2007){Ciotti}, {Lanzoni}, \&
  {Volonteri}}]{2007ApJ...658...65C}
{Ciotti}, L., {Lanzoni}, B., \& {Volonteri}, M. 2007, \apj, 658, 65

\bibitem[{{Comerford} {et~al.}(2015){Comerford}, {Pooley}, {Barrows}, {Greene},
  {Zakamska}, {Madejski}, \& {Cooper}}]{2015ApJ...806..219C}
{Comerford}, J.~M., {Pooley}, D., {Barrows}, R.~S., {et~al.} 2015, \apj, 806,
  219

\bibitem[{{Conselice}(2007)}]{2007IAUS..235..381C}
{Conselice}, C.~J. 2007, in IAU Symposium, Vol. 235, Galaxy Evolution across
  the Hubble Time, ed. F.~{Combes} \& J.~{Palou{\v s}}, 381--384

\bibitem[{{Cooper} {et~al.}(2015){Cooper}, {Gao}, {Guo}, {Frenk}, {Jenkins},
  {Springel}, \& {White}}]{2015MNRAS.451.2703C}
{Cooper}, A.~P., {Gao}, L., {Guo}, Q., {et~al.} 2015, \mnras, 451, 2703

\bibitem[{{Crane} {et~al.}(1993){Crane}, {Stiavelli}, {King}, {Deharveng},
  {Albrecht}, {Barbieri}, {Blades}, {Boksenberg}, {Disney}, {Jakobsen},
  {Kamperman}, {Machetto}, {Mackay}, {Paresce}, {Weigelt}, {Baxter},
  {Greenfield}, {Jedrzejewski}, {Nota}, \& {Sparks}}]{1993AJ....106.1371C}
{Crane}, P., {Stiavelli}, M., {King}, I.~R., {et~al.} 1993, \aj, 106, 1371

\bibitem[{{Davies} {et~al.}(1983){Davies}, {Efstathiou}, {Fall}, {Illingworth},
  \& {Schechter}}]{1983ApJ...266...41D}
{Davies}, R.~L., {Efstathiou}, G., {Fall}, S.~M., {Illingworth}, G., \&
  {Schechter}, P.~L. 1983, \apj, 266, 41

\bibitem[{{Davis} {et~al.}(2019){Davis}, {Graham}, \&
  {Cameron}}]{2019ApJ...873...85D}
{Davis}, B.~L., {Graham}, A.~W., \& {Cameron}, E. 2019, \apj, 873, 85

\bibitem[{{de la Rosa} {et~al.}(2016){de la Rosa}, {La Barbera}, {Ferreras},
  {S{\'a}nchez Almeida}, {Dalla Vecchia}, {Mart{\'{\i}}nez-Valpuesta}, \&
  {Stringer}}]{2016MNRAS.457.1916D}
{de la Rosa}, I.~G., {La Barbera}, F., {Ferreras}, I., {et~al.} 2016, \mnras,
  457, 1916

\bibitem[{{De Lucia} \& {Blaizot}(2007)}]{2007MNRAS.375....2D}
{De Lucia}, G., \& {Blaizot}, J. 2007, \mnras, 375, 2

\bibitem[{{de Ruiter} {et~al.}(2005){de Ruiter}, {Parma}, {Capetti}, {Fanti},
  {Morganti}, \& {Santantonio}}]{2005A&A...439..487D}
{de Ruiter}, H.~R., {Parma}, P., {Capetti}, A., {et~al.} 2005, \aap, 439, 487

\bibitem[{{de Vaucouleurs}(1948)}]{1948AnAp...11..247D}
{de Vaucouleurs}, G. 1948, Annales d'Astrophysique, 11, 247

\bibitem[{{de Vaucouleurs} {et~al.}(1991){de Vaucouleurs}, {de Vaucouleurs},
  {Corwin}, {Buta}, {Paturel}, \& {Fouqu{\'e}}}]{1991rc3..book.....D}
{de Vaucouleurs}, G., {de Vaucouleurs}, A., {Corwin}, Jr., H.~G., {et~al.}
  1991, {Third Reference Catalogue of Bright Galaxies. Volume I: Explanations
  and references. Volume II: Data for galaxies between 0$^{h}$ and 12$^{h}$.
  Volume III: Data for galaxies between 12$^{h}$ and 24$^{h}$.}

\bibitem[{{di Tullio}(1979)}]{1979A&AS...37..591D}
{di Tullio}, G.~A. 1979, \aaps, 37, 591

\bibitem[{{Donzelli} {et~al.}(2011){Donzelli}, {Muriel}, \&
  {Madrid}}]{2011ApJS..195...15D}
{Donzelli}, C.~J., {Muriel}, H., \& {Madrid}, J.~P. 2011, \apjs, 195, 15

\bibitem[{{Dressler}(1981)}]{1981ApJ...243...26D}
{Dressler}, A. 1981, \apj, 243, 26

\bibitem[{{Dubinski}(1998)}]{1998ApJ...502..141D}
{Dubinski}, J. 1998, \apj, 502, 141

\bibitem[{{Dullo} \& {Graham}(2012)}]{2012ApJ...755..163D}
{Dullo}, B.~T., \& {Graham}, A.~W. 2012, \apj, 755, 163

\bibitem[{{Dullo} \& {Graham}(2013)}]{2013ApJ...768...36D}
---. 2013, \apj, 768, 36

\bibitem[{{Dullo} \& {Graham}(2014)}]{2014MNRAS.444.2700D}
---. 2014, \mnras, 444, 2700

\bibitem[{{Dullo} \& {Graham}(2015)}]{2015ApJ...798...55D}
---. 2015, \apj, 798, 55

\bibitem[{{Dullo} {et~al.}(2017){Dullo}, {Graham}, \&
  {Knapen}}]{2017MNRAS.471.2321D}
{Dullo}, B.~T., {Graham}, A.~W., \& {Knapen}, J.~H. 2017, \mnras, 471, 2321

\bibitem[{{Dullo} {et~al.}(2016){Dullo}, {Mart{\'{\i}}nez-Lombilla}, \&
  {Knapen}}]{2016MNRAS.462.3800D}
{Dullo}, B.~T., {Mart{\'{\i}}nez-Lombilla}, C., \& {Knapen}, J.~H. 2016,
  \mnras, 462, 3800

\bibitem[{{Dullo} {et~al.}(2018){Dullo}, {Knapen}, {Williams}, {Beswick},
  {Bendo}, {Baldi}, {Argo}, {McHardy}, {Muxlow}, \&
  {Westcott}}]{2018MNRAS.475.4670D}
{Dullo}, B.~T., {Knapen}, J.~H., {Williams}, D.~R.~A., {et~al.} 2018, \mnras,
  475, 4670

\bibitem[{{Dullo} {et~al.}(2019){Dullo}, {Chamorro-Cazorla}, {Gil de Paz},
  {Castillo-Morales}, {Gallego}, {Carrasco}, {Iglesias-P{\'a}ramo}, {Cedazo},
  {Garc{\'{\i}}a-Vargas}, {Pascual}, {Cardiel}, {P{\'e}rez-Calpena},
  {G{\'o}mez-Alvarez}, {Mart{\'{\i}}nez-Delgado}, \&
  {Catal{\'a}n-Torrecilla}}]{2019ApJ...871....9D}
{Dullo}, B.~T., {Chamorro-Cazorla}, M., {Gil de Paz}, A., {et~al.} 2019, \apj,
  871, 9

\bibitem[{{Duncan} {et~al.}(2019){Duncan}, {Conselice}, {Mundy}, {Bell},
  {Donley}, {Galametz}, {Guo}, {Grogin}, {Hathi}, {Kartaltepe}, {Kocevski},
  {Koekemoer}, {P{\'e}rez-Gonz{\'a}lez}, {Mantha}, {Snyder}, \&
  {Stefanon}}]{2019arXiv190312188D}
{Duncan}, K., {Conselice}, C.~J., {Mundy}, C., {et~al.} 2019, arXiv e-prints,
  arXiv:1903.12188

\bibitem[{{Ebisuzaki} {et~al.}(1991){Ebisuzaki}, {Makino}, \&
  {Okumura}}]{1991Natur.354..212E}
{Ebisuzaki}, T., {Makino}, J., \& {Okumura}, S.~K. 1991, \nat, 354, 212

\bibitem[{{Edwards} \& {Patton}(2012)}]{2012MNRAS.425..287E}
{Edwards}, L.~O.~V., \& {Patton}, D.~R. 2012, \mnras, 425, 287

\bibitem[{{Efstathiou} {et~al.}(1988){Efstathiou}, {Ellis}, \&
  {Peterson}}]{1988MNRAS.232..431E}
{Efstathiou}, G., {Ellis}, R.~S., \& {Peterson}, B.~A. 1988, \mnras, 232, 431

\bibitem[{{Faber} \& {Jackson}(1976)}]{1976ApJ...204..668F}
{Faber}, S.~M., \& {Jackson}, R.~E. 1976, \apj, 204, 668

\bibitem[{{Faber} {et~al.}(1997){Faber}, {Tremaine}, {Ajhar}, {Byun},
  {Dressler}, {Gebhardt}, {Grillmair}, {Kormendy}, {Lauer}, \&
  {Richstone}}]{1997AJ....114.1771F}
{Faber}, S.~M., {Tremaine}, S., {Ajhar}, E.~A., {et~al.} 1997, \aj, 114, 1771

\bibitem[{{Feigelson} \& {Babu}(1992)}]{1992ApJ...397...55F}
{Feigelson}, E.~D., \& {Babu}, G.~J. 1992, \apj, 397, 55

\bibitem[{{Ferrarese} \& {Ford}(2005)}]{2005SSRv..116..523F}
{Ferrarese}, L., \& {Ford}, H. 2005, \ssr, 116, 523

\bibitem[{{Ferrarese} \& {Merritt}(2000)}]{2000ApJ...539L...9F}
{Ferrarese}, L., \& {Merritt}, D. 2000, \apjl, 539, L9

\bibitem[{{Ferrarese} {et~al.}(1994){Ferrarese}, {van den Bosch}, {Ford},
  {Jaffe}, \& {O'Connell}}]{1994AJ....108.1598F}
{Ferrarese}, L., {van den Bosch}, F.~C., {Ford}, H.~C., {Jaffe}, W., \&
  {O'Connell}, R.~W. 1994, \aj, 108, 1598

\bibitem[{{Ferrarese} {et~al.}(2006){Ferrarese}, {C{\^o}t{\'e}}, {Jord{\'a}n},
  {Peng}, {Blakeslee}, {Piatek}, {Mei}, {Merritt}, {Milosavljevi{\'c}},
  {Tonry}, \& {West}}]{2006ApJS..164..334F}
{Ferrarese}, L., {C{\^o}t{\'e}}, P., {Jord{\'a}n}, A., {et~al.} 2006, \apjs,
  164, 334

\bibitem[{{Fukugita} {et~al.}(1995){Fukugita}, {Shimasaku}, \&
  {Ichikawa}}]{1995PASP..107..945F}
{Fukugita}, M., {Shimasaku}, K., \& {Ichikawa}, T. 1995, \pasp, 107, 945

\bibitem[{{Gebhardt} {et~al.}(2011){Gebhardt}, {Adams}, {Richstone}, {Lauer},
  {Faber}, {G{\"u}ltekin}, {Murphy}, \& {Tremaine}}]{2011ApJ...729..119G}
{Gebhardt}, K., {Adams}, J., {Richstone}, D., {et~al.} 2011, \apj, 729, 119

\bibitem[{{Gebhardt} {et~al.}(1996){Gebhardt}, {Richstone}, {Ajhar}, {Lauer},
  {Byun}, {Kormendy}, {Dressler}, {Faber}, {Grillmair}, \&
  {Tremaine}}]{1996AJ....112..105G}
{Gebhardt}, K., {Richstone}, D., {Ajhar}, E.~A., {et~al.} 1996, \aj, 112, 105

\bibitem[{{Gebhardt} {et~al.}(2000){Gebhardt}, {Bender}, {Bower}, {Dressler},
  {Faber}, {Filippenko}, {Green}, {Grillmair}, {Ho}, {Kormendy}, {Lauer},
  {Magorrian}, {Pinkney}, {Richstone}, \& {Tremaine}}]{2000ApJ...539L..13G}
{Gebhardt}, K., {Bender}, R., {Bower}, G., {et~al.} 2000, \apjl, 539, L13

\bibitem[{{Gebhardt} {et~al.}(2003){Gebhardt}, {Richstone}, {Tremaine},
  {Lauer}, {Bender}, {Bower}, {Dressler}, {Faber}, {Filippenko}, {Green},
  {Grillmair}, {Ho}, {Kormendy}, {Magorrian}, \&
  {Pinkney}}]{2003ApJ...583...92G}
{Gebhardt}, K., {Richstone}, D., {Tremaine}, S., {et~al.} 2003, \apj, 583, 92

\bibitem[{{Gonzalez} {et~al.}(2003){Gonzalez}, {Zabludoff}, \&
  {Zaritsky}}]{2003Ap&SS.285...67G}
{Gonzalez}, A.~H., {Zabludoff}, A.~I., \& {Zaritsky}, D. 2003, \apss, 285, 67

\bibitem[{{Gonzalez} {et~al.}(2005){Gonzalez}, {Zabludoff}, \&
  {Zaritsky}}]{2005ApJ...618..195G}
---. 2005, \apj, 618, 195

\bibitem[{{Goulding} {et~al.}(2019){Goulding}, {Pardo}, {Greene}, {Mingarelli},
  {Nyland}, \& {Strauss}}]{2019arXiv190703757G}
{Goulding}, A.~D., {Pardo}, K., {Greene}, J.~E., {et~al.} 2019, arXiv e-prints,
  arXiv:1907.03757

\bibitem[{{Graham}(2004)}]{2004ApJ...613L..33G}
{Graham}, A.~W. 2004, \apjl, 613, L33

\bibitem[{{Graham}(2013)}]{2013pss6.book...91G}
---. 2013, {Elliptical and Disk Galaxy Structure and Modern Scaling Laws}, ed.
  T.~D. {Oswalt} \& W.~C. {Keel}, 91

\bibitem[{{Graham}(2016)}]{2016ASSL..418..263G}
{Graham}, A.~W. 2016, in Astrophysics and Space Science Library, Vol. 418,
  Galactic Bulges, ed. E.~{Laurikainen}, R.~{Peletier}, \& D.~{Gadotti}, 263

\bibitem[{{Graham} \& {Driver}(2007)}]{2007ApJ...655...77G}
{Graham}, A.~W., \& {Driver}, S.~P. 2007, \apj, 655, 77

\bibitem[{{Graham} {et~al.}(2015){Graham}, {Dullo}, \&
  {Savorgnan}}]{2015ApJ...804...32G}
{Graham}, A.~W., {Dullo}, B.~T., \& {Savorgnan}, G.~A.~D. 2015, \apj, 804, 32

\bibitem[{{Graham} {et~al.}(2003){Graham}, {Erwin}, {Trujillo}, \& {Asensio
  Ramos}}]{2003AJ....125.2951G}
{Graham}, A.~W., {Erwin}, P., {Trujillo}, I., \& {Asensio Ramos}, A. 2003, \aj,
  125, 2951

\bibitem[{{Graham} \& {Scott}(2013)}]{2013ApJ...764..151G}
{Graham}, A.~W., \& {Scott}, N. 2013, \apj, 764, 151

\bibitem[{{Gualandris} \& {Merritt}(2008)}]{2008ApJ...678..780G}
{Gualandris}, A., \& {Merritt}, D. 2008, \apj, 678, 780

\bibitem[{{Held} {et~al.}(1992){Held}, {de Zeeuw}, {Mould}, \&
  {Picard}}]{1992AJ....103..851H}
{Held}, E.~V., {de Zeeuw}, T., {Mould}, J., \& {Picard}, A. 1992, \aj, 103, 851

\bibitem[{{Hills}(1988)}]{1988Natur.331..687H}
{Hills}, J.~G. 1988, \nat, 331, 687

\bibitem[{{Hilz} {et~al.}(2013){Hilz}, {Naab}, \&
  {Ostriker}}]{2013MNRAS.429.2924H}
{Hilz}, M., {Naab}, T., \& {Ostriker}, J.~P. 2013, \mnras, 429, 2924

\bibitem[{{Hilz} {et~al.}(2012){Hilz}, {Naab}, {Ostriker}, {Thomas}, {Burkert},
  \& {Jesseit}}]{2012MNRAS.425.3119H}
{Hilz}, M., {Naab}, T., {Ostriker}, J.~P., {et~al.} 2012, \mnras, 425, 3119

\bibitem[{{Hlavacek-Larrondo} {et~al.}(2012){Hlavacek-Larrondo}, {Fabian},
  {Edge}, \& {Hogan}}]{2012MNRAS.424..224H}
{Hlavacek-Larrondo}, J., {Fabian}, A.~C., {Edge}, A.~C., \& {Hogan}, M.~T.
  2012, \mnras, 424, 224

\bibitem[{{Hopkins} \& {Hernquist}(2010)}]{2010MNRAS.407..447H}
{Hopkins}, P.~F., \& {Hernquist}, L. 2010, \mnras, 407, 447

\bibitem[{{Hopkins} {et~al.}(2009){Hopkins}, {Lauer}, {Cox}, {Hernquist}, \&
  {Kormendy}}]{2009ApJS..181..486H}
{Hopkins}, P.~F., {Lauer}, T.~R., {Cox}, T.~J., {Hernquist}, L., \& {Kormendy},
  J. 2009, \apjs, 181, 486

\bibitem[{{Hyde} {et~al.}(2008){Hyde}, {Bernardi}, {Sheth}, \&
  {Nichol}}]{2008MNRAS.391.1559H}
{Hyde}, J.~B., {Bernardi}, M., {Sheth}, R.~K., \& {Nichol}, R.~C. 2008, \mnras,
  391, 1559

\bibitem[{{Jaffe} {et~al.}(1994){Jaffe}, {Ford}, {O'Connell}, {van den Bosch},
  \& {Ferrarese}}]{1994AJ....108.1567J}
{Jaffe}, W., {Ford}, H.~C., {O'Connell}, R.~W., {van den Bosch}, F.~C., \&
  {Ferrarese}, L. 1994, \aj, 108, 1567

\bibitem[{{Jedrzejewski}(1987)}]{1987MNRAS.226..747J}
{Jedrzejewski}, R.~I. 1987, \mnras, 226, 747

\bibitem[{{Jim{\'e}nez-Teja} \& {Dupke}(2016)}]{2016ApJ...820...49J}
{Jim{\'e}nez-Teja}, Y., \& {Dupke}, R. 2016, \apj, 820, 49

\bibitem[{{Johansson} {et~al.}(2012){Johansson}, {Naab}, \&
  {Ostriker}}]{2012ApJ...754..115J}
{Johansson}, P.~H., {Naab}, T., \& {Ostriker}, J.~P. 2012, \apj, 754, 115

\bibitem[{{Kauffmann} {et~al.}(1993){Kauffmann}, {White}, \&
  {Guiderdoni}}]{1993MNRAS.264..201K}
{Kauffmann}, G., {White}, S.~D.~M., \& {Guiderdoni}, B. 1993, \mnras, 264, 201

\bibitem[{{Khosroshahi} {et~al.}(2006){Khosroshahi}, {Ponman}, \&
  {Jones}}]{2006MNRAS.372L..68K}
{Khosroshahi}, H.~G., {Ponman}, T.~J., \& {Jones}, L.~R. 2006, \mnras, 372, L68

\bibitem[{{King} \& {Nealon}(2019)}]{2019MNRAS.487.4827K}
{King}, A., \& {Nealon}, R. 2019, \mnras, 487, 4827

\bibitem[{{King}(1978)}]{1978ApJ...222....1K}
{King}, I.~R. 1978, \apj, 222, 1

\bibitem[{{King} \& {Minkowski}(1966)}]{1966ApJ...143.1002K}
{King}, I.~R., \& {Minkowski}, R. 1966, \apj, 143, 1002

\bibitem[{{Kochanek} {et~al.}(2001){Kochanek}, {Pahre}, {Falco}, {Huchra},
  {Mader}, {Jarrett}, {Chester}, {Cutri}, \& {Schneider}}]{2001ApJ...560..566K}
{Kochanek}, C.~S., {Pahre}, M.~A., {Falco}, E.~E., {et~al.} 2001, \apj, 560,
  566

\bibitem[{{Komossa} {et~al.}(2003){Komossa}, {Burwitz}, {Hasinger}, {Predehl},
  {Kaastra}, \& {Ikebe}}]{2003ApJ...582L..15K}
{Komossa}, S., {Burwitz}, V., {Hasinger}, G., {et~al.} 2003, \apjl, 582, L15

\bibitem[{{Kormendy}(1999)}]{1999ASPC..182..124K}
{Kormendy}, J. 1999, in Astronomical Society of the Pacific Conference Series,
  Vol. 182, Galaxy Dynamics - A Rutgers Symposium, ed. D.~R. {Merritt},
  M.~{Valluri}, \& J.~A. {Sellwood}

\bibitem[{{Kormendy} \& {Bender}(2013)}]{2013ApJ...769L...5K}
{Kormendy}, J., \& {Bender}, R. 2013, \apjl, 769, L5

\bibitem[{{Kormendy} {et~al.}(1994){Kormendy}, {Dressler}, {Byun}, {Faber},
  {Grillmair}, {Lauer}, {Richstone}, \& {Tremaine}}]{1994ESOC...49..147K}
{Kormendy}, J., {Dressler}, A., {Byun}, Y.~I., {et~al.} 1994, in European
  Southern Observatory Conference and Workshop Proceedings, Vol.~49, European
  Southern Observatory Conference and Workshop Proceedings, ed. G.~{Meylan} \&
  P.~{Prugniel}, 147

\bibitem[{{Kormendy} {et~al.}(2009){Kormendy}, {Fisher}, {Cornell}, \&
  {Bender}}]{2009ApJS..182..216K}
{Kormendy}, J., {Fisher}, D.~B., {Cornell}, M.~E., \& {Bender}, R. 2009, \apjs,
  182, 216

\bibitem[{{Kormendy} \& {Ho}(2013)}]{2013ARA&A..51..511K}
{Kormendy}, J., \& {Ho}, L.~C. 2013, \araa, 51, 511

\bibitem[{{Kormendy} \& {Richstone}(1995)}]{1995ARA&A..33..581K}
{Kormendy}, J., \& {Richstone}, D. 1995, \araa, 33, 581

\bibitem[{{Krajnovi{\'c}} {et~al.}(2018){Krajnovi{\'c}}, {Emsellem}, {den
  Brok}, {Marino}, {Schmidt}, {Steinmetz}, \&
  {Weilbacher}}]{2018MNRAS.477.5327K}
{Krajnovi{\'c}}, D., {Emsellem}, E., {den Brok}, M., {et~al.} 2018, \mnras,
  477, 5327

\bibitem[{{Kulkarni} \& {Loeb}(2012)}]{2012MNRAS.422.1306K}
{Kulkarni}, G., \& {Loeb}, A. 2012, \mnras, 422, 1306

\bibitem[{{Laine} {et~al.}(2003){Laine}, {van der Marel}, {Lauer}, {Postman},
  {O'Dea}, \& {Owen}}]{2003AJ....125..478L}
{Laine}, S., {van der Marel}, R.~P., {Lauer}, T.~R., {et~al.} 2003, \aj, 125,
  478

\bibitem[{{Laporte} {et~al.}(2013){Laporte}, {White}, {Naab}, \&
  {Gao}}]{2013MNRAS.435..901L}
{Laporte}, C.~F.~P., {White}, S.~D.~M., {Naab}, T., \& {Gao}, L. 2013, \mnras,
  435, 901

\bibitem[{{Lauer}(1985)}]{1985ApJS...57..473L}
{Lauer}, T.~R. 1985, \apjs, 57, 473

\bibitem[{{Lauer} {et~al.}(1995){Lauer}, {Ajhar}, {Byun}, {Dressler}, {Faber},
  {Grillmair}, {Kormendy}, {Richstone}, \& {Tremaine}}]{1995AJ....110.2622L}
{Lauer}, T.~R., {Ajhar}, E.~A., {Byun}, Y.-I., {et~al.} 1995, \aj, 110, 2622

\bibitem[{{Lauer} {et~al.}(2007{\natexlab{a}}){Lauer}, {Gebhardt}, {Faber},
  {Richstone}, {Tremaine}, {Kormendy}, {Aller}, {Bender}, {Dressler},
  {Filippenko}, {Green}, \& {Ho}}]{2007ApJ...664..226L}
{Lauer}, T.~R., {Gebhardt}, K., {Faber}, S.~M., {et~al.} 2007{\natexlab{a}},
  \apj, 664, 226

\bibitem[{{Lauer} {et~al.}(2007{\natexlab{b}}){Lauer}, {Faber}, {Richstone},
  {Gebhardt}, {Tremaine}, {Postman}, {Dressler}, {Aller}, {Filippenko},
  {Green}, {Ho}, {Kormendy}, {Magorrian}, \& {Pinkney}}]{2007ApJ...662..808L}
{Lauer}, T.~R., {Faber}, S.~M., {Richstone}, D., {et~al.} 2007{\natexlab{b}},
  \apj, 662, 808

\bibitem[{{Li} {et~al.}(2011){Li}, {Ho}, {Barth}, \&
  {Peng}}]{2011ApJS..197...22L}
{Li}, Z.-Y., {Ho}, L.~C., {Barth}, A.~J., \& {Peng}, C.~Y. 2011, \apjs, 197, 22

\bibitem[{{Lidman} {et~al.}(2013){Lidman}, {Iacobuta}, {Bauer}, {Barrientos},
  {Cerulo}, {Couch}, {Delaye}, {Demarco}, {Ellingson}, {Faloon}, {Gilbank},
  {Huertas-Company}, {Mei}, {Meyers}, {Muzzin}, {Noble}, {Nantais}, {Rettura},
  {Rosati}, {S{\'a}nchez-Janssen}, {Strazzullo}, {Webb}, {Wilson}, {Yan}, \&
  {Yee}}]{2013MNRAS.433..825L}
{Lidman}, C., {Iacobuta}, G., {Bauer}, A.~E., {et~al.} 2013, \mnras, 433, 825

\bibitem[{{Liu} {et~al.}(2009){Liu}, {Mao}, {Deng}, {Xia}, \&
  {Wen}}]{2009MNRAS.396.2003L}
{Liu}, F.~S., {Mao}, S., {Deng}, Z.~G., {Xia}, X.~Y., \& {Wen}, Z.~L. 2009,
  \mnras, 396, 2003

\bibitem[{{Liu} {et~al.}(2019){Liu}, {Hou}, {Li}, {Guo}, {Kong}, {Shen},
  {Wrobel}, {Peng}, \& {Nyland}}]{2019arXiv190710639L}
{Liu}, X., {Hou}, M., {Li}, Z., {et~al.} 2019, arXiv e-prints, arXiv:1907.10639

\bibitem[{{L{\'o}pez-Cruz} {et~al.}(2014){L{\'o}pez-Cruz}, {A{\~n}orve},
  {Birkinshaw}, {Worrall}, {Ibarra-Medel}, {Barkhouse}, {Torres-Papaqui}, \&
  {Motta}}]{2014ApJ...795L..31L}
{L{\'o}pez-Cruz}, O., {A{\~n}orve}, C., {Birkinshaw}, M., {et~al.} 2014, \apjl,
  795, L31

\bibitem[{{Lugger}(1984)}]{1984ApJ...286..106L}
{Lugger}, P.~M. 1984, \apj, 286, 106

\bibitem[{{Madrid} \& {Donzelli}(2016)}]{2016ApJ...819...50M}
{Madrid}, J.~P., \& {Donzelli}, C.~J. 2016, \apj, 819, 50

\bibitem[{{Magorrian} {et~al.}(1998){Magorrian}, {Tremaine}, {Richstone},
  {Bender}, {Bower}, {Dressler}, {Faber}, {Gebhardt}, {Green}, {Grillmair},
  {Kormendy}, \& {Lauer}}]{1998AJ....115.2285M}
{Magorrian}, J., {Tremaine}, S., {Richstone}, D., {et~al.} 1998, \aj, 115, 2285

\bibitem[{{Malumuth} \& {Kirshner}(1981)}]{1981ApJ...251..508M}
{Malumuth}, E.~M., \& {Kirshner}, R.~P. 1981, \apj, 251, 508

\bibitem[{{Man} {et~al.}(2016){Man}, {Zirm}, \& {Toft}}]{2016ApJ...830...89M}
{Man}, A.~W.~S., {Zirm}, A.~W., \& {Toft}, S. 2016, \apj, 830, 89

\bibitem[{{Marconi} \& {Hunt}(2003)}]{2003ApJ...589L..21M}
{Marconi}, A., \& {Hunt}, L.~K. 2003, \apjl, 589, L21

\bibitem[{{Matkovi{\'c}} \& {Guzm{\'a}n}(2005)}]{2005MNRAS.362..289M}
{Matkovi{\'c}}, A., \& {Guzm{\'a}n}, R. 2005, \mnras, 362, 289

\bibitem[{{McConnell} \& {Ma}(2013)}]{2013ApJ...764..184M}
{McConnell}, N.~J., \& {Ma}, C.-P. 2013, \apj, 764, 184

\bibitem[{{McConnell} {et~al.}(2011){McConnell}, {Ma}, {Gebhardt}, {Wright},
  {Murphy}, {Lauer}, {Graham}, \& {Richstone}}]{2011Natur.480..215M}
{McConnell}, N.~J., {Ma}, C.-P., {Gebhardt}, K., {et~al.} 2011, \nat, 480, 215

\bibitem[{{McConnell} {et~al.}(2012){McConnell}, {Ma}, {Murphy}, {Gebhardt},
  {Lauer}, {Graham}, {Wright}, \& {Richstone}}]{2012ApJ...756..179M}
{McConnell}, N.~J., {Ma}, C.-P., {Murphy}, J.~D., {et~al.} 2012, \apj, 756, 179

\bibitem[{{McNamara} {et~al.}(2009){McNamara}, {Kazemzadeh}, {Rafferty},
  {B{\^i}rzan}, {Nulsen}, {Kirkpatrick}, \& {Wise}}]{2009ApJ...698..594M}
{McNamara}, B.~R., {Kazemzadeh}, F., {Rafferty}, D.~A., {et~al.} 2009, \apj,
  698, 594

\bibitem[{{Mehrgan} {et~al.}(2019){Mehrgan}, {Thomas}, {Saglia}, {Mazzalay},
  {Erwin}, {Bender}, {Kluge}, \& {Fabricius}}]{2019arXiv190710608M}
{Mehrgan}, K., {Thomas}, J., {Saglia}, R., {et~al.} 2019, arXiv e-prints,
  arXiv:1907.10608

\bibitem[{{Merritt}(2006)}]{2006ApJ...648..976M}
{Merritt}, D. 2006, \apj, 648, 976

\bibitem[{{Merritt} {et~al.}(2004){Merritt}, {Milosavljevi{\'c}}, {Favata},
  {Hughes}, \& {Holz}}]{2004ApJ...607L...9M}
{Merritt}, D., {Milosavljevi{\'c}}, M., {Favata}, M., {Hughes}, S.~A., \&
  {Holz}, D.~E. 2004, \apjl, 607, L9

\bibitem[{{Mezcua} {et~al.}(2018){Mezcua}, {Hlavacek-Larrondo}, {Lucey},
  {Hogan}, {Edge}, \& {McNamara}}]{2018MNRAS.474.1342M}
{Mezcua}, M., {Hlavacek-Larrondo}, J., {Lucey}, J.~R., {et~al.} 2018, \mnras,
  474, 1342

\bibitem[{{Milosavljevi{\'c}} \& {Merritt}(2001)}]{2001ApJ...563...34M}
{Milosavljevi{\'c}}, M., \& {Merritt}, D. 2001, \apj, 563, 34

\bibitem[{{Milosavljevi{\'c}} {et~al.}(2002){Milosavljevi{\'c}}, {Merritt},
  {Rest}, \& {van den Bosch}}]{2002MNRAS.331L..51M}
{Milosavljevi{\'c}}, M., {Merritt}, D., {Rest}, A., \& {van den Bosch}, F.~C.
  2002, \mnras, 331, L51

\bibitem[{{Montes} \& {Trujillo}(2018)}]{2018MNRAS.474..917M}
{Montes}, M., \& {Trujillo}, I. 2018, \mnras, 474, 917

\bibitem[{{Morgan} \& {Lesh}(1965)}]{1965ApJ...142.1364M}
{Morgan}, W.~W., \& {Lesh}, J.~R. 1965, \apj, 142, 1364

\bibitem[{{Mundy} {et~al.}(2017){Mundy}, {Conselice}, {Duncan}, {Almaini},
  {H{\"a}u{\ss}ler}, \& {Hartley}}]{2017MNRAS.470.3507M}
{Mundy}, C.~J., {Conselice}, C.~J., {Duncan}, K.~J., {et~al.} 2017, \mnras,
  470, 3507

\bibitem[{{Naab} {et~al.}(2009){Naab}, {Johansson}, \&
  {Ostriker}}]{2009ApJ...699L.178N}
{Naab}, T., {Johansson}, P.~H., \& {Ostriker}, J.~P. 2009, \apjl, 699, L178

\bibitem[{{Nipoti} {et~al.}(2003){Nipoti}, {Londrillo}, \&
  {Ciotti}}]{2003MNRAS.342..501N}
{Nipoti}, C., {Londrillo}, P., \& {Ciotti}, L. 2003, \mnras, 342, 501

\bibitem[{{Oegerle} \& {Hoessel}(1991)}]{1991ApJ...375...15O}
{Oegerle}, W.~R., \& {Hoessel}, J.~G. 1991, \apj, 375, 15

\bibitem[{{Oemler}(1974)}]{1974ApJ...194....1O}
{Oemler}, Jr., A. 1974, \apj, 194, 1

\bibitem[{{O'Leary} \& {Loeb}(2008)}]{2008MNRAS.383...86O}
{O'Leary}, R.~M., \& {Loeb}, A. 2008, \mnras, 383, 86

\bibitem[{{Oser} {et~al.}(2012){Oser}, {Naab}, {Ostriker}, \&
  {Johansson}}]{2012ApJ...744...63O}
{Oser}, L., {Naab}, T., {Ostriker}, J.~P., \& {Johansson}, P.~H. 2012, \apj,
  744, 63

\bibitem[{{Oser} {et~al.}(2010){Oser}, {Ostriker}, {Naab}, {Johansson}, \&
  {Burkert}}]{2010ApJ...725.2312O}
{Oser}, L., {Ostriker}, J.~P., {Naab}, T., {Johansson}, P.~H., \& {Burkert}, A.
  2010, \apj, 725, 2312

\bibitem[{{Paturel} {et~al.}(2003){Paturel}, {Petit}, {Prugniel}, {Theureau},
  {Rousseau}, {Brouty}, {Dubois}, \& {Cambr{\'e}sy}}]{2003A&A...412...45P}
{Paturel}, G., {Petit}, C., {Prugniel}, P., {et~al.} 2003, \aap, 412, 45

\bibitem[{{Phipps} {et~al.}(2019){Phipps}, {Bogdan}, {Lovisari}, {Kovacs},
  {Volonteri}, \& {Dubois}}]{2019arXiv190309965P}
{Phipps}, F., {Bogdan}, A., {Lovisari}, L., {et~al.} 2019, arXiv e-prints,
  arXiv:1903.09965

\bibitem[{{Pierini} {et~al.}(2008){Pierini}, {Zibetti}, {Braglia},
  {B{\"o}hringer}, {Finoguenov}, {Lynam}, \& {Zhang}}]{2008A&A...483..727P}
{Pierini}, D., {Zibetti}, S., {Braglia}, F., {et~al.} 2008, \aap, 483, 727

\bibitem[{{Pillepich} {et~al.}(2018){Pillepich}, {Nelson}, {Hernquist},
  {Springel}, {Pakmor}, {Torrey}, {Weinberger}, {Genel}, {Naiman}, {Marinacci},
  \& {Vogelsberger}}]{2018MNRAS.475..648P}
{Pillepich}, A., {Nelson}, D., {Hernquist}, L., {et~al.} 2018, \mnras, 475, 648

\bibitem[{{Porter} {et~al.}(1991){Porter}, {Schneider}, \&
  {Hoessel}}]{1991AJ....101.1561P}
{Porter}, A.~C., {Schneider}, D.~P., \& {Hoessel}, J.~G. 1991, \aj, 101, 1561

\bibitem[{{Postman} \& {Lauer}(1995)}]{1995ApJ...440...28P}
{Postman}, M., \& {Lauer}, T.~R. 1995, \apj, 440, 28

\bibitem[{{Postman} {et~al.}(2012){Postman}, {Lauer}, {Donahue}, {Graves},
  {Coe}, {Moustakas}, {Koekemoer}, {Bradley}, {Ford}, {Grillo}, {Zitrin},
  {Lemze}, {Broadhurst}, {Moustakas}, {Ascaso}, {Medezinski}, \&
  {Kelson}}]{2012ApJ...756..159P}
{Postman}, M., {Lauer}, T.~R., {Donahue}, M., {et~al.} 2012, \apj, 756, 159

\bibitem[{{Quinlan} \& {Hernquist}(1997)}]{1997NewA....2..533Q}
{Quinlan}, G.~D., \& {Hernquist}, L. 1997, \mnras, 2, 533

\bibitem[{{Rantala} {et~al.}(2018){Rantala}, {Johansson}, {Naab}, {Thomas}, \&
  {Frigo}}]{2018ApJ...864..113R}
{Rantala}, A., {Johansson}, P.~H., {Naab}, T., {Thomas}, J., \& {Frigo}, M.
  2018, \apj, 864, 113

\bibitem[{{Ravindranath} {et~al.}(2001){Ravindranath}, {Ho}, {Peng},
  {Filippenko}, \& {Sargent}}]{2001AJ....122..653R}
{Ravindranath}, S., {Ho}, L.~C., {Peng}, C.~Y., {Filippenko}, A.~V., \&
  {Sargent}, W.~L.~W. 2001, \aj, 122, 653

\bibitem[{{Redmount} \& {Rees}(1989)}]{1989ComAp..14..165R}
{Redmount}, I.~H., \& {Rees}, M.~J. 1989, Comments on Astrophysics, 14, 165

\bibitem[{{Rest} {et~al.}(2001){Rest}, {van den Bosch}, {Jaffe}, {Tran},
  {Tsvetanov}, {Ford}, {Davies}, \& {Schafer}}]{2001AJ....121.2431R}
{Rest}, A., {van den Bosch}, F.~C., {Jaffe}, W., {et~al.} 2001, \aj, 121, 2431

\bibitem[{{Richings} {et~al.}(2011){Richings}, {Uttley}, \&
  {K{\"o}rding}}]{2011MNRAS.415.2158R}
{Richings}, A.~J., {Uttley}, P., \& {K{\"o}rding}, E. 2011, \mnras, 415, 2158

\bibitem[{{Richstone} {et~al.}(1998){Richstone}, {Ajhar}, {Bender}, {Bower},
  {Dressler}, {Faber}, {Filippenko}, {Gebhardt}, {Green}, {Ho}, {Kormendy},
  {Lauer}, {Magorrian}, \& {Tremaine}}]{1998Natur.395A..14R}
{Richstone}, D., {Ajhar}, E.~A., {Bender}, R., {et~al.} 1998, \nat, 395, A14

\bibitem[{{Robitaille} \& {Bressert}(2012)}]{2012ascl.soft08017R}
{Robitaille}, T., \& {Bressert}, E. 2012, {APLpy: Astronomical Plotting Library
  in Python}, Astrophysics Source Code Library, ascl:1208.017

\bibitem[{{Rodriguez} {et~al.}(2006){Rodriguez}, {Taylor}, {Zavala}, {Peck},
  {Pollack}, \& {Romani}}]{2006ApJ...646...49R}
{Rodriguez}, C., {Taylor}, G.~B., {Zavala}, R.~T., {et~al.} 2006, \apj, 646, 49

\bibitem[{{Rodriguez-Gomez} {et~al.}(2016){Rodriguez-Gomez}, {Pillepich},
  {Sales}, {Genel}, {Vogelsberger}, {Zhu}, {Wellons}, {Nelson}, {Torrey},
  {Springel}, {Ma}, \& {Hernquist}}]{2016MNRAS.458.2371R}
{Rodriguez-Gomez}, V., {Pillepich}, A., {Sales}, L.~V., {et~al.} 2016, \mnras,
  458, 2371

\bibitem[{{Rusli} {et~al.}(2013){Rusli}, {Erwin}, {Saglia}, {Thomas},
  {Fabricius}, {Bender}, \& {Nowak}}]{2013AJ....146..160R}
{Rusli}, S.~P., {Erwin}, P., {Saglia}, R.~P., {et~al.} 2013, \aj, 146, 160

\bibitem[{{Sahu} {et~al.}(2019){Sahu}, {Graham}, \&
  {Davis}}]{2019arXiv190806838S}
{Sahu}, N., {Graham}, A.~W., \& {Davis}, B.~L. 2019, arXiv e-prints,
  arXiv:1908.06838

\bibitem[{{Savorgnan} \& {Graham}(2015)}]{2015MNRAS.446.2330S}
{Savorgnan}, G.~A.~D., \& {Graham}, A.~W. 2015, \mnras, 446, 2330

\bibitem[{{Schlafly} \& {Finkbeiner}(2011)}]{2011ApJ...737..103S}
{Schlafly}, E.~F., \& {Finkbeiner}, D.~P. 2011, \apj, 737, 103

\bibitem[{{Schombert}(1986)}]{1986ApJS...60..603S}
{Schombert}, J.~M. 1986, \apjs, 60, 603

\bibitem[{{Schweizer}(1982)}]{1982ApJ...252..455S}
{Schweizer}, F. 1982, \apj, 252, 455

\bibitem[{{Seigar} {et~al.}(2007){Seigar}, {Graham}, \&
  {Jerjen}}]{2007MNRAS.378.1575S}
{Seigar}, M.~S., {Graham}, A.~W., \& {Jerjen}, H. 2007, \mnras, 378, 1575

\bibitem[{{S\'ersic}(1963)}]{1963BAAA....6...41S}
{S\'ersic}, J.~L. 1963, Boletin de la Asociacion Argentina de Astronomia La
  Plata Argentina, 6, 41

\bibitem[{{S\'ersic}(1968)}]{1968adga.book.....S}
---. 1968, {Atlas de Galaxias Australes}

\bibitem[{{Sheth} {et~al.}(2003){Sheth}, {Bernardi}, {Schechter}, {Burles},
  {Eisenstein}, {Finkbeiner}, {Frieman}, {Lupton}, {Schlegel}, {Subbarao},
  {Shimasaku}, {Bahcall}, {Brinkmann}, \& {Ivezi{\'c}}}]{2003ApJ...594..225S}
{Sheth}, R.~K., {Bernardi}, M., {Schechter}, P.~L., {et~al.} 2003, \apj, 594,
  225

\bibitem[{{Thomas} {et~al.}(2016){Thomas}, {Ma}, {McConnell}, {Greene},
  {Blakeslee}, \& {Janish}}]{2016Natur.532..340T}
{Thomas}, J., {Ma}, C.-P., {McConnell}, N.~J., {et~al.} 2016, \nat, 532, 340

\bibitem[{{Thomas} {et~al.}(2014){Thomas}, {Saglia}, {Bender}, {Erwin}, \&
  {Fabricius}}]{2014ApJ...782...39T}
{Thomas}, J., {Saglia}, R.~P., {Bender}, R., {Erwin}, P., \& {Fabricius}, M.
  2014, \apj, 782, 39

\bibitem[{{Toomre} \& {Toomre}(1972)}]{1972ApJ...178..623T}
{Toomre}, A., \& {Toomre}, J. 1972, \apj, 178, 623

\bibitem[{{Trujillo} {et~al.}(2001){Trujillo}, {Aguerri}, {Cepa}, \&
  {Guti{\'e}rrez}}]{2001MNRAS.328..977T}
{Trujillo}, I., {Aguerri}, J.~A.~L., {Cepa}, J., \& {Guti{\'e}rrez}, C.~M.
  2001, \mnras, 328, 977

\bibitem[{{Trujillo} {et~al.}(2004){Trujillo}, {Erwin}, {Asensio Ramos}, \&
  {Graham}}]{2004AJ....127.1917T}
{Trujillo}, I., {Erwin}, P., {Asensio Ramos}, A., \& {Graham}, A.~W. 2004, \aj,
  127, 1917

\bibitem[{{van den Bosch} {et~al.}(1994){van den Bosch}, {Ferrarese}, {Jaffe},
  {Ford}, \& {O'Connell}}]{1994AJ....108.1579V}
{van den Bosch}, F.~C., {Ferrarese}, L., {Jaffe}, W., {Ford}, H.~C., \&
  {O'Connell}, R.~W. 1994, \aj, 108, 1579

\bibitem[{{Veale} {et~al.}(2017){Veale}, {Ma}, {Thomas}, {Greene}, {McConnell},
  {Walsh}, {Ito}, {Blakeslee}, \& {Janish}}]{2017MNRAS.464..356V}
{Veale}, M., {Ma}, C.-P., {Thomas}, J., {et~al.} 2017, \mnras, 464, 356

\bibitem[{{Volonteri} \& {Ciotti}(2013)}]{2013ApJ...768...29V}
{Volonteri}, M., \& {Ciotti}, L. 2013, \apj, 768, 29

\bibitem[{{West} {et~al.}(2017){West}, {de Propris}, {Bremer}, \&
  {Phillipps}}]{2017NatAs...1E.157W}
{West}, M.~J., {de Propris}, R., {Bremer}, M.~N., \& {Phillipps}, S. 2017,
  Nature Astronomy, 1, 0157

\bibitem[{{White} \& {Rees}(1978)}]{1978MNRAS.183..341W}
{White}, S.~D.~M., \& {Rees}, M.~J. 1978, \mnras, 183, 341

\bibitem[{{Worthey}(1994)}]{1994ApJS...95..107W}
{Worthey}, G. 1994, \apjs, 95, 107

\bibitem[{{Wu} {et~al.}(2015){Wu}, {Wang}, {Fan}, {Yi}, {Zuo}, {Bian}, {Jiang},
  {McGreer}, {Wang}, {Yang}, {Yang}, {Thompson}, \&
  {Beletsky}}]{2015Natur.518..512W}
{Wu}, X.-B., {Wang}, F., {Fan}, X., {et~al.} 2015, \nat, 518, 512

\bibitem[{{Young} {et~al.}(1978){Young}, {Westphal}, {Kristian}, {Wilson}, \&
  {Landauer}}]{1978ApJ...221..721Y}
{Young}, P.~J., {Westphal}, J.~A., {Kristian}, J., {Wilson}, C.~P., \&
  {Landauer}, F.~P. 1978, \apj, 221, 721

\bibitem[{{Zibetti} {et~al.}(2005){Zibetti}, {White}, {Schneider}, \&
  {Brinkmann}}]{2005MNRAS.358..949Z}
{Zibetti}, S., {White}, S.~D.~M., {Schneider}, D.~P., \& {Brinkmann}, J. 2005,
  \mnras, 358, 949

\bibitem[{{Zolotov} {et~al.}(2009){Zolotov}, {Willman}, {Brooks}, {Governato},
  {Brook}, {Hogg}, {Quinn}, \& {Stinson}}]{2009ApJ...702.1058Z}
{Zolotov}, A., {Willman}, B., {Brooks}, A.~M., {et~al.} 2009, \apj, 702, 1058

\end{thebibliography}
\section{Appendix }\label{A}
\begin{appendices}
\section{Appendix A}\label{AppA}

In Fig.~\ref{FigA1}, we show the multi-component decomposition of the major-axis surface
brightness profiles of the large-core galaxies (Table~\ref{Table1}).

\renewcommand\thefigure{\thesection\arabic{figure}} 
\setcounter{figure}{0}
\begin{figure*}
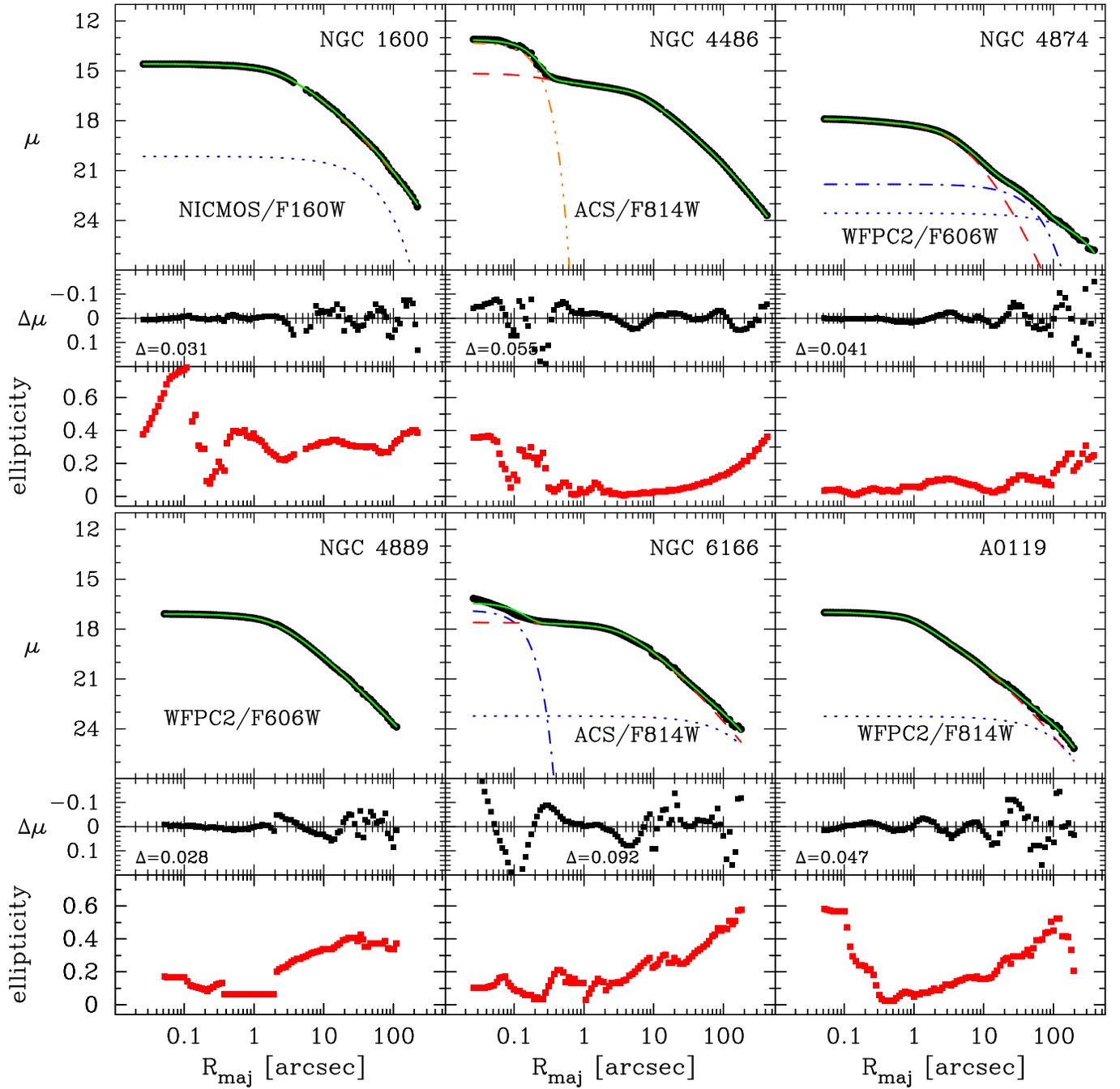

\includegraphics[angle=270,scale=0.725]{GRPI.ps}
\includegraphics[angle=270,scale=0.725]{GRPII.ps}
\caption{Fits to the major-axis surface brightness profiles of our sample
  of 12 core-S\'ersic galaxies (see Table~\ref{Table1}).  The red dashed
  curves indicate the core-S\'ersic model, while the blue dotted
  curves show the outer stellar halo.  Additional nuclear light
  components such as AGN and star clusters were modelled using either
  a Gaussian (brown, triple dot-dashed curve) or a S\'ersic function (blue dot-dashed curve) . The
  solid green curves represent the complete fit to the profiles. The
  fit rms residuals and ellipticity ($\epsilon$ = $1-b/a$) are given
  in the lower panels. }
\label{FigA1} 
 \end{figure*}

\setcounter{figure}{0}
\begin{figure*}
\includegraphics[angle=270,scale=0.725]{GRPIII.ps}
\includegraphics[angle=270,scale=0.725]{GRPIV.ps}
\caption{\it{continued}}

 \end{figure*}

\begin{figure*}
\hspace*{-.0949cm}   
\includegraphics[angle=0,scale=0.18030299354]{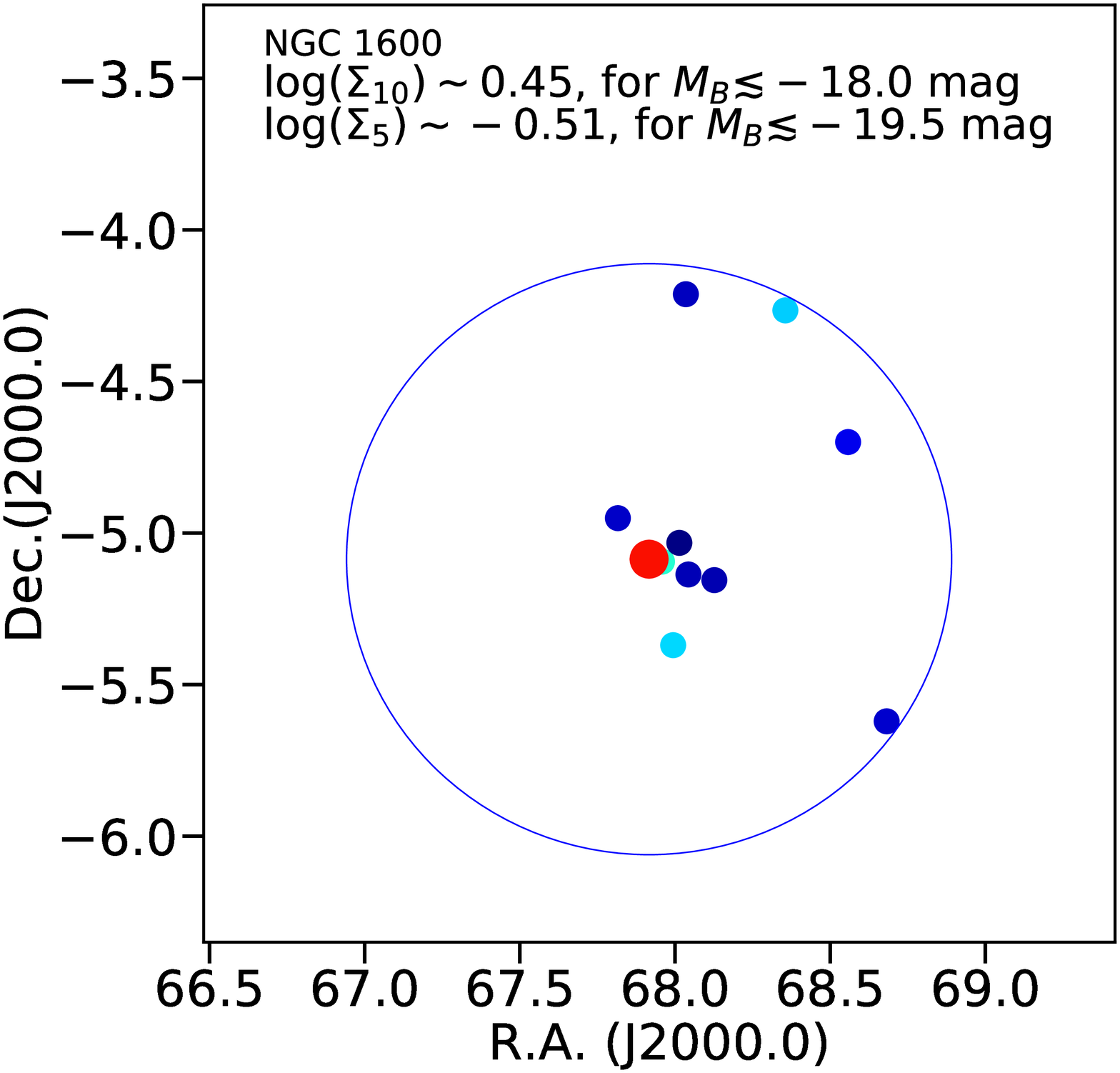}
\put(-146.5,113.85){\color{black}{{\scriptsize $R_{\rm b} \sim $ 0.65 kpc}}}
\hspace*{-.2853072599cm}   
\includegraphics[angle=0,scale=0.178039080354]{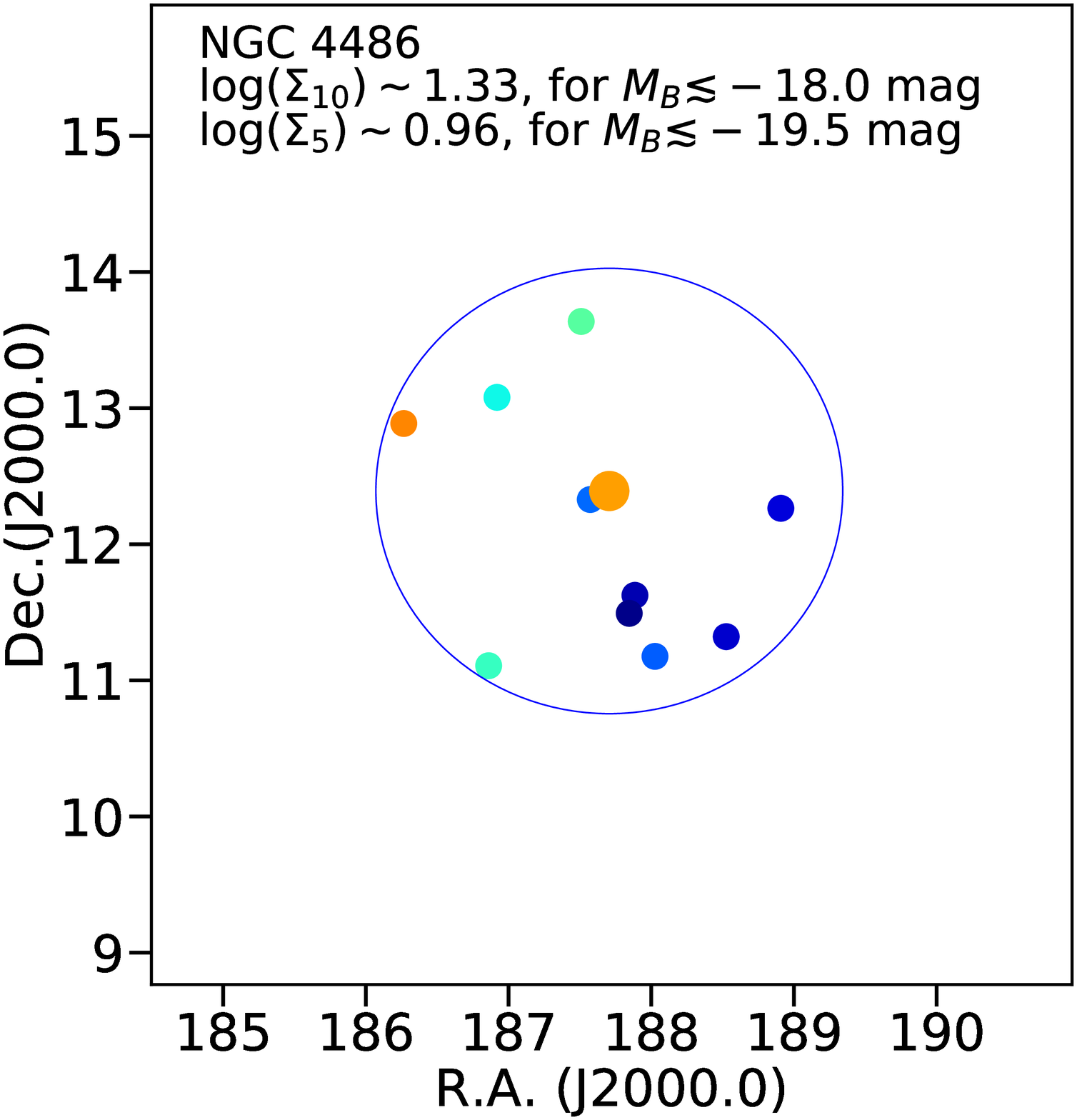}
\put(-143.3,113.85){\color{black}{{\scriptsize $R_{\rm b} \sim $ 0.64 kpc}}}
\hspace*{-.685399cm}  
\includegraphics[angle=0,scale=0.1756430020354]{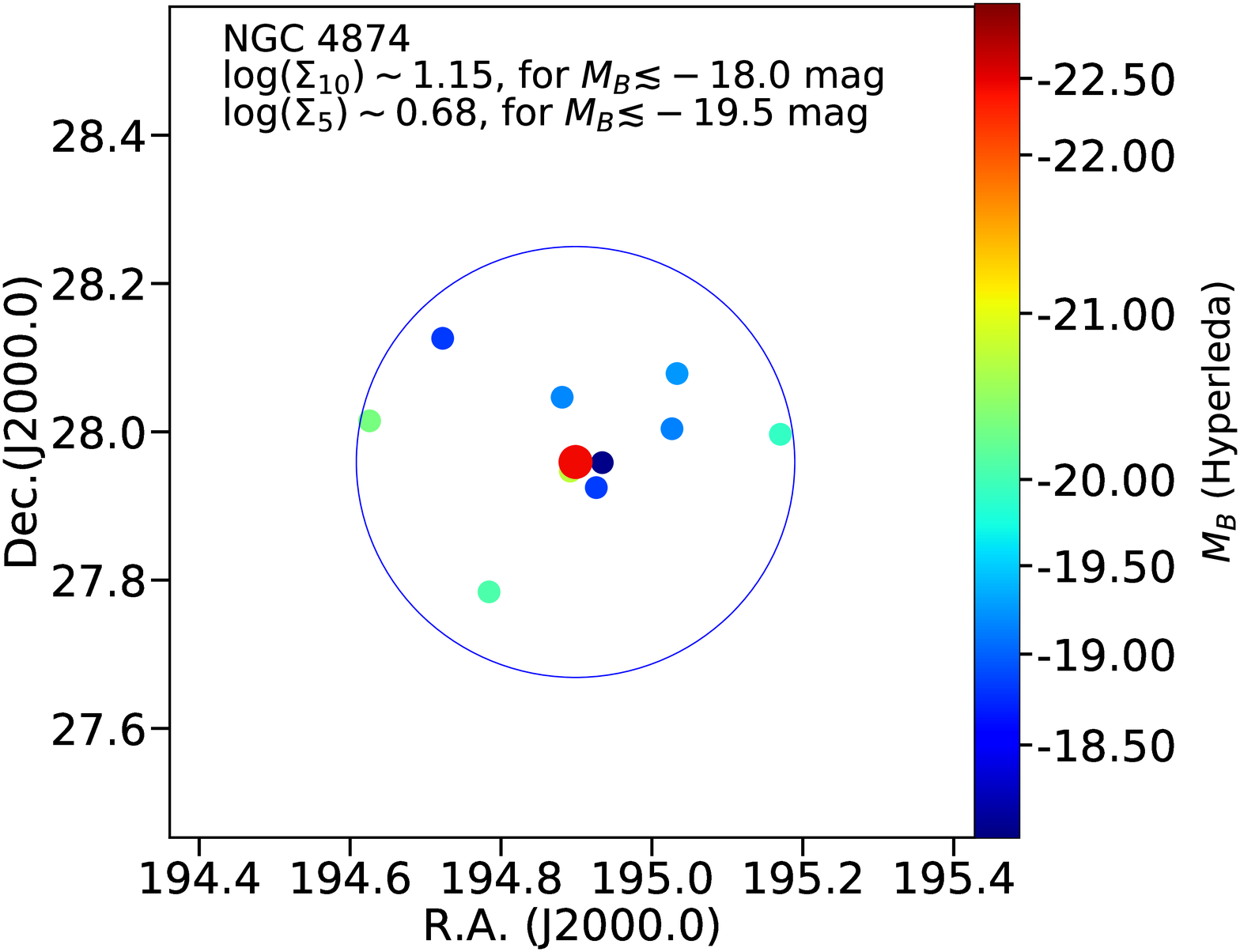}
\put(-143.88,113.05){\color{black}{{\scriptsize $R_{\rm b} \sim $ 1.63 kpc}}}\\
\hspace*{-.0949cm}   
\includegraphics[angle=0,scale=0.18030299354]{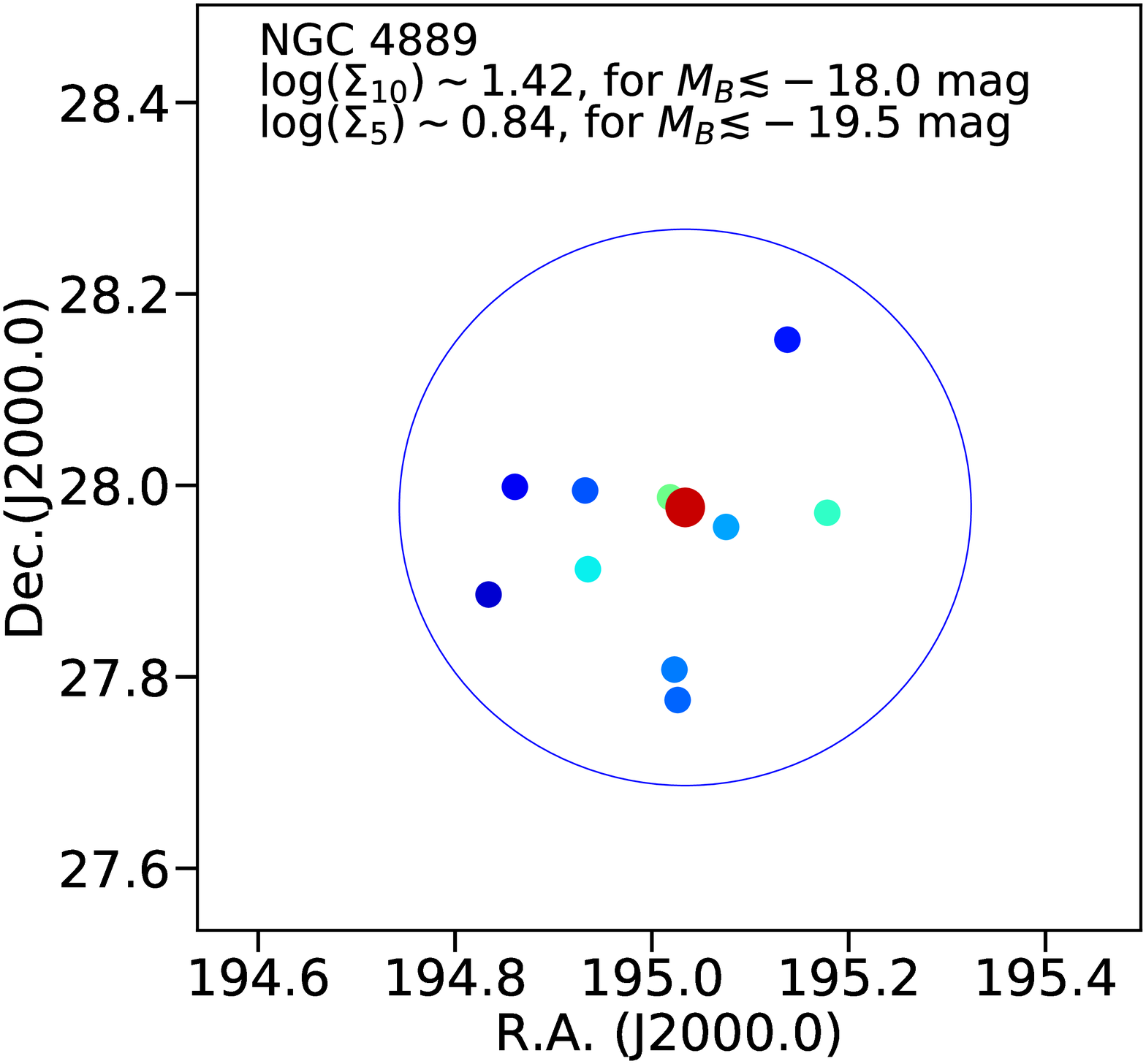}
\put(-147.9,110.05){\color{black}{{\scriptsize $R_{\rm b} \sim $ 0.86 kpc}}}
\hspace*{-.2853072599cm}  
\includegraphics[angle=0,scale=0.17478030299354]{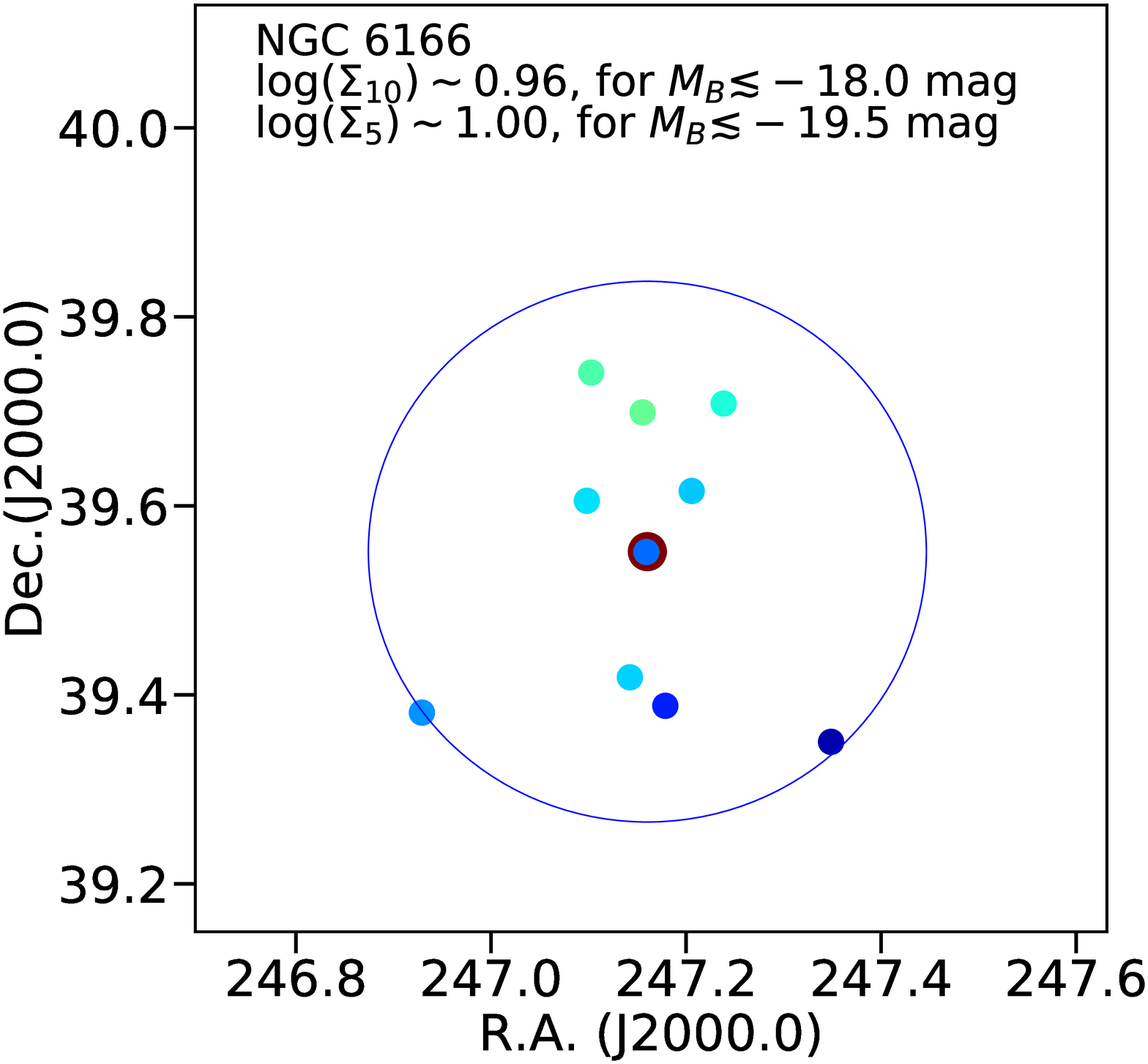}
\put(-142.273,109.05){\color{black}{{\scriptsize $R_{\rm b} \sim $ 2.11 kpc}}}
\hspace*{-.685399cm}   
\includegraphics[angle=0,scale=0.17428030299354]{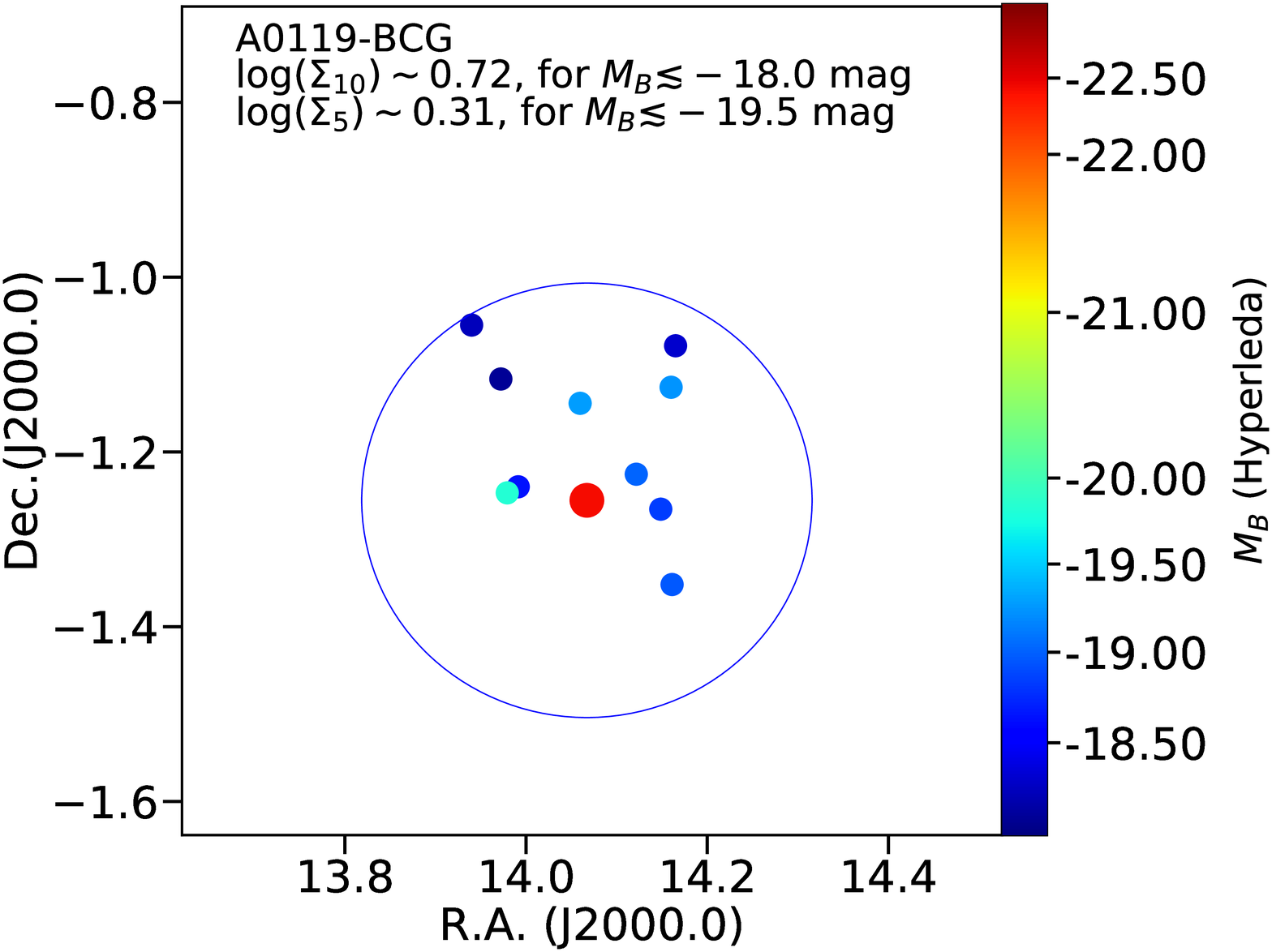}
\put(-143.4,109.05){\color{black}{{\scriptsize $R_{\rm b} \sim $ 0.67 kpc}}}\\
\hspace*{-.0949cm}   
\includegraphics[angle=0,scale=0.1788030299354]{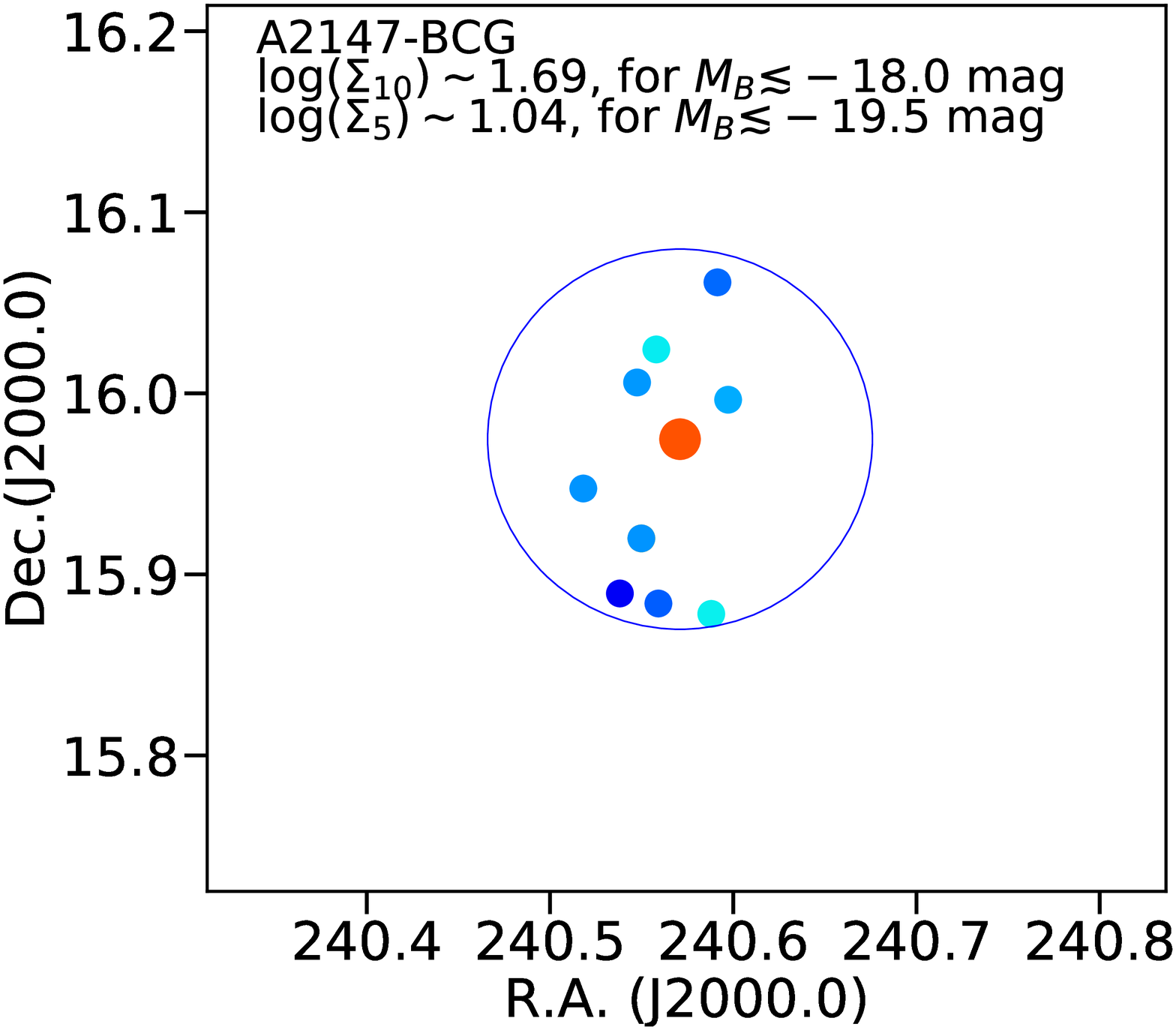}
\put(-142.91,99.85){\color{black}{{\scriptsize $R_{\rm b} \sim $ 1.28 kpc}}}
\hspace*{-.2853072599cm}  
\includegraphics[angle=0,scale=0.1760078030299354]{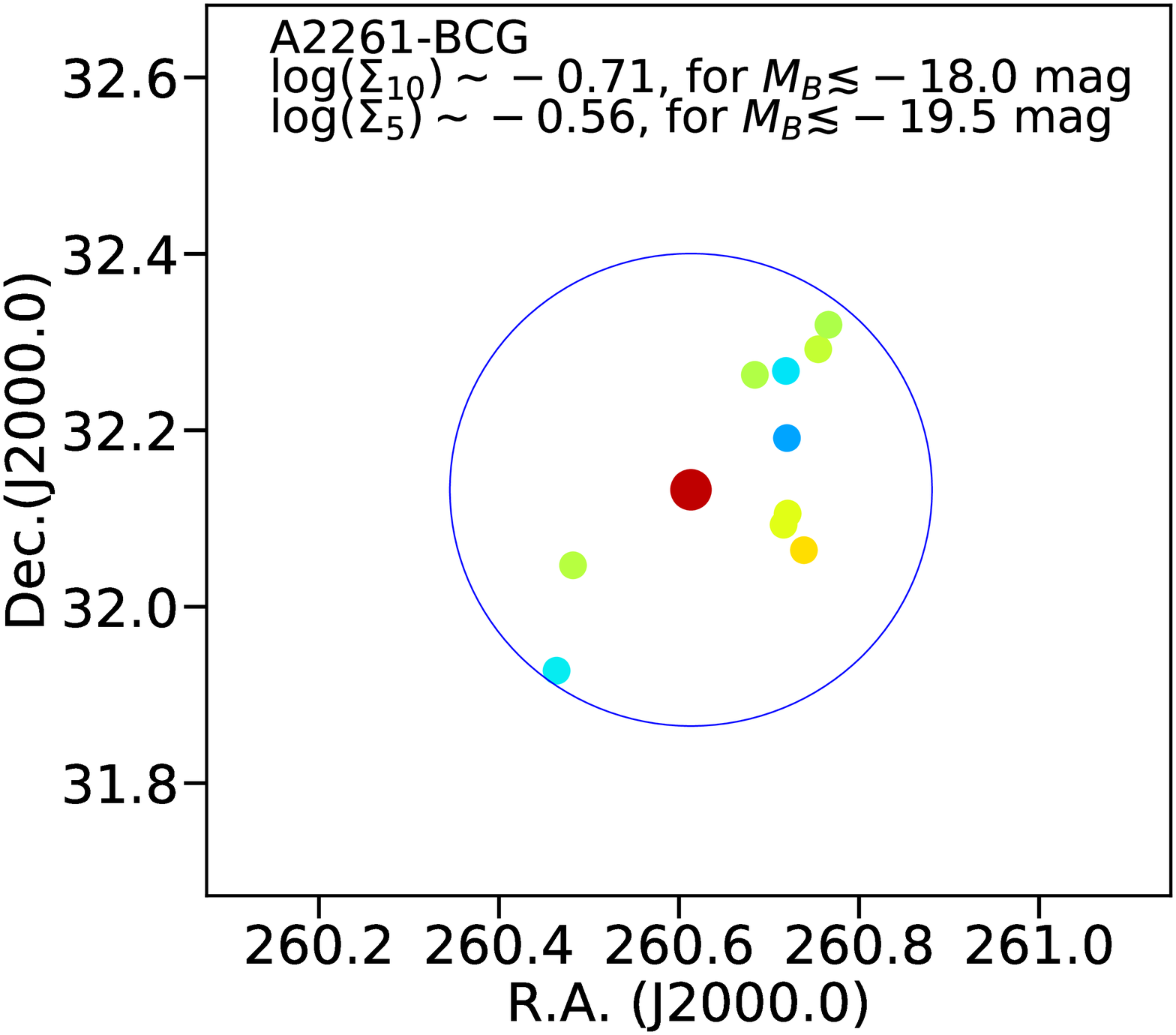}
\put(-141.75,99.85){\color{black}{{\scriptsize $R_{\rm b} \sim $ 2.71 kpc}}}
\hspace*{-.685399cm}  
\includegraphics[angle=0,scale=0.178030299354]{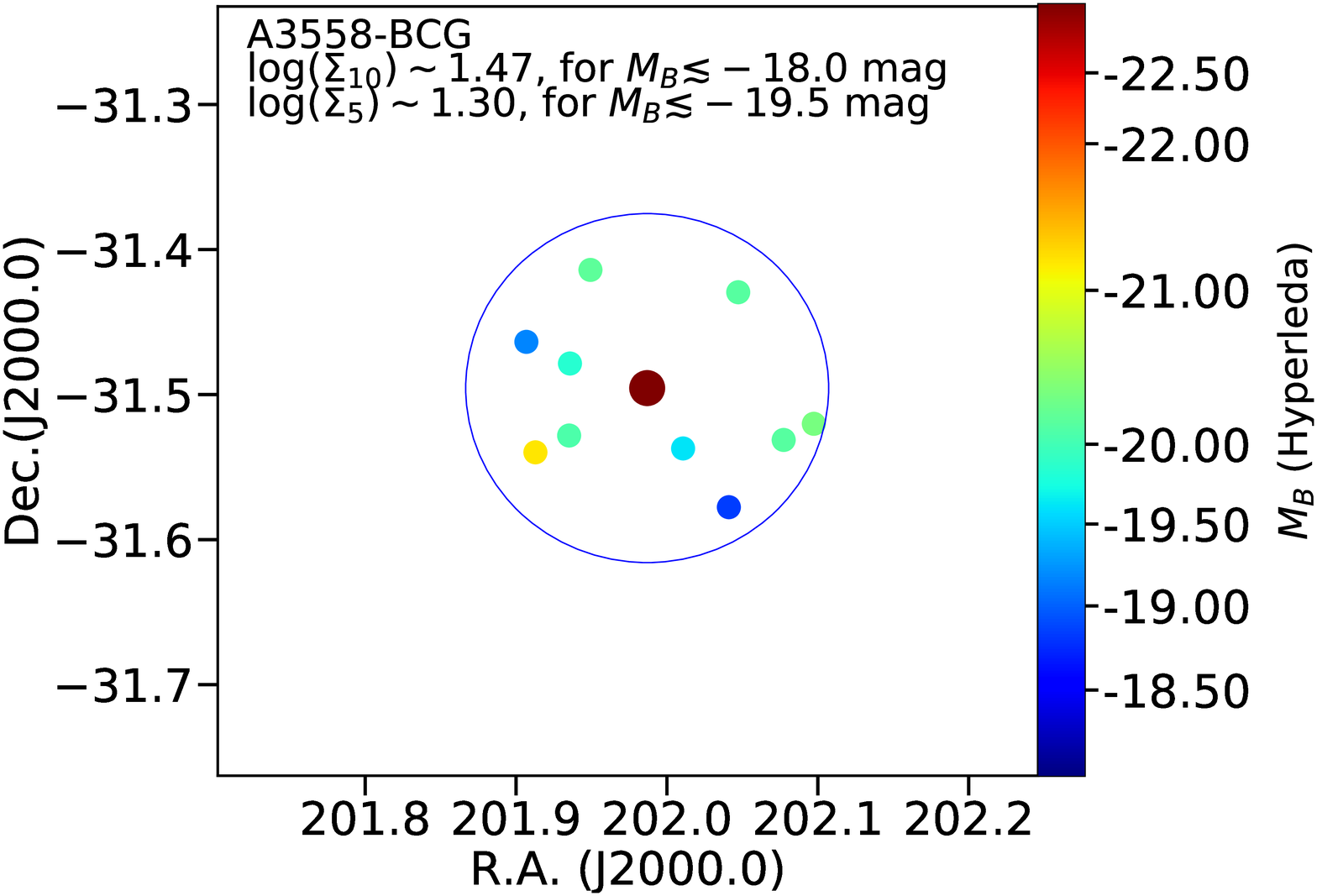}
\put(-141.9925,99.85){\color{black}{{\scriptsize $R_{\rm b} \sim $ 1.30 kpc}}}\\
\hspace*{-.0949cm}   
\includegraphics[angle=0,scale=0.176030299354]{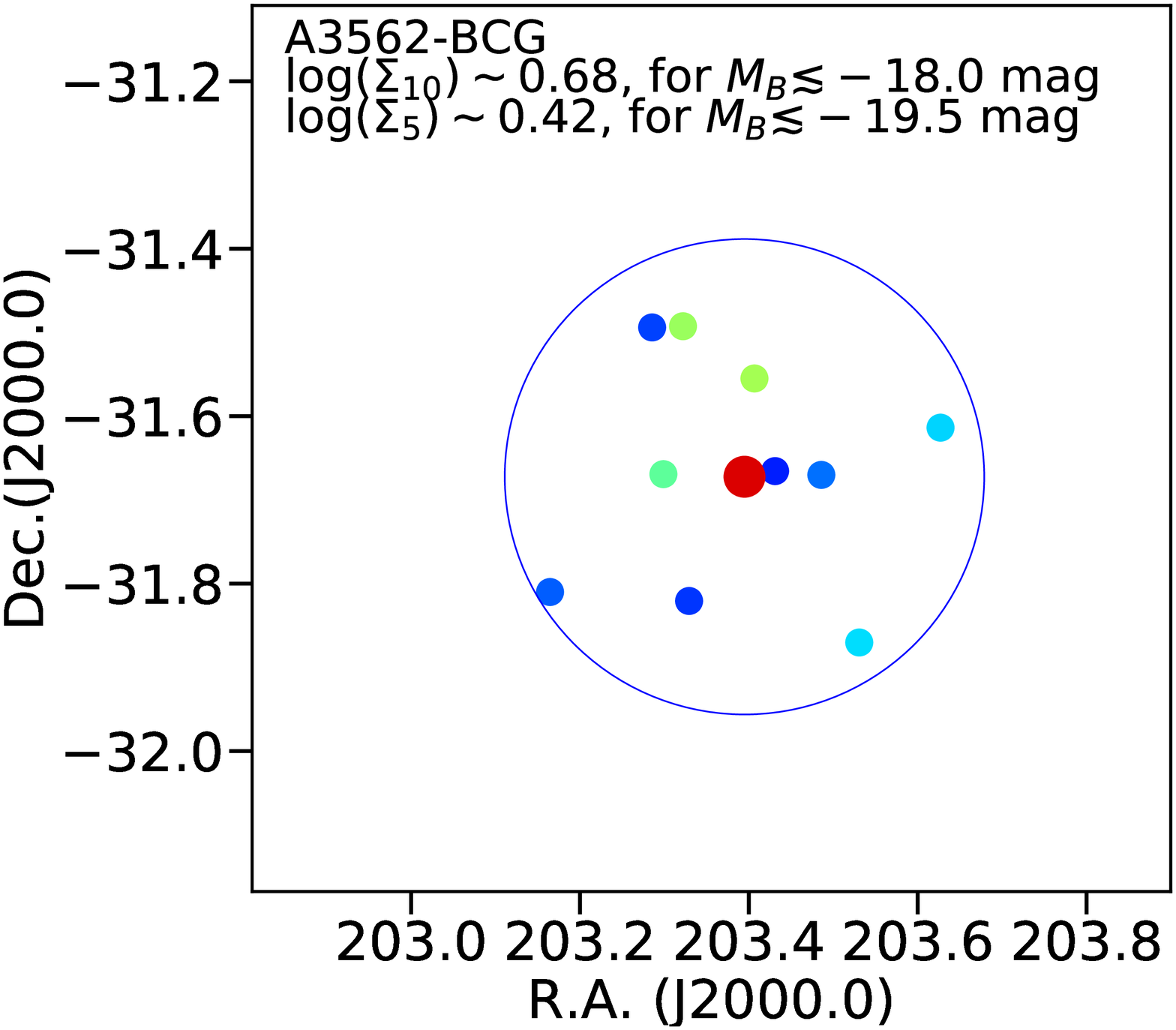}
\put(-138.2,97.11){\color{black}{{\scriptsize $R_{\rm b} \sim $ 0.64 kpc}}}
\hspace*{-.2353072599cm}  
\includegraphics[angle=0,scale=0.165980078030299354]{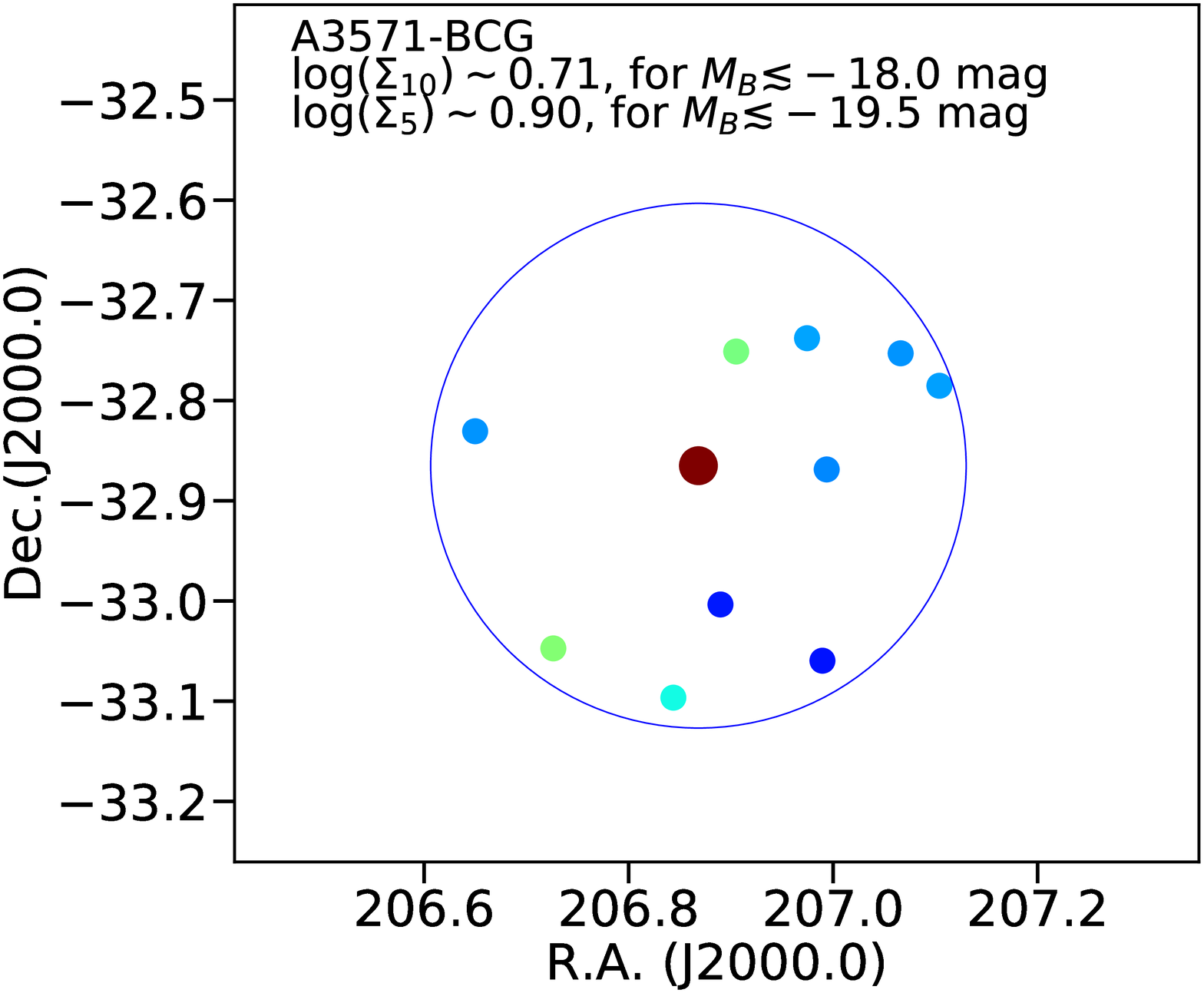}
\put(-142.79,96.81){\color{black}{{\scriptsize $R_{\rm b} \sim $ 1.70 kpc}}}
\hspace*{-.31539753072599cm}  
\includegraphics[angle=0,scale=0.17580078030299354]{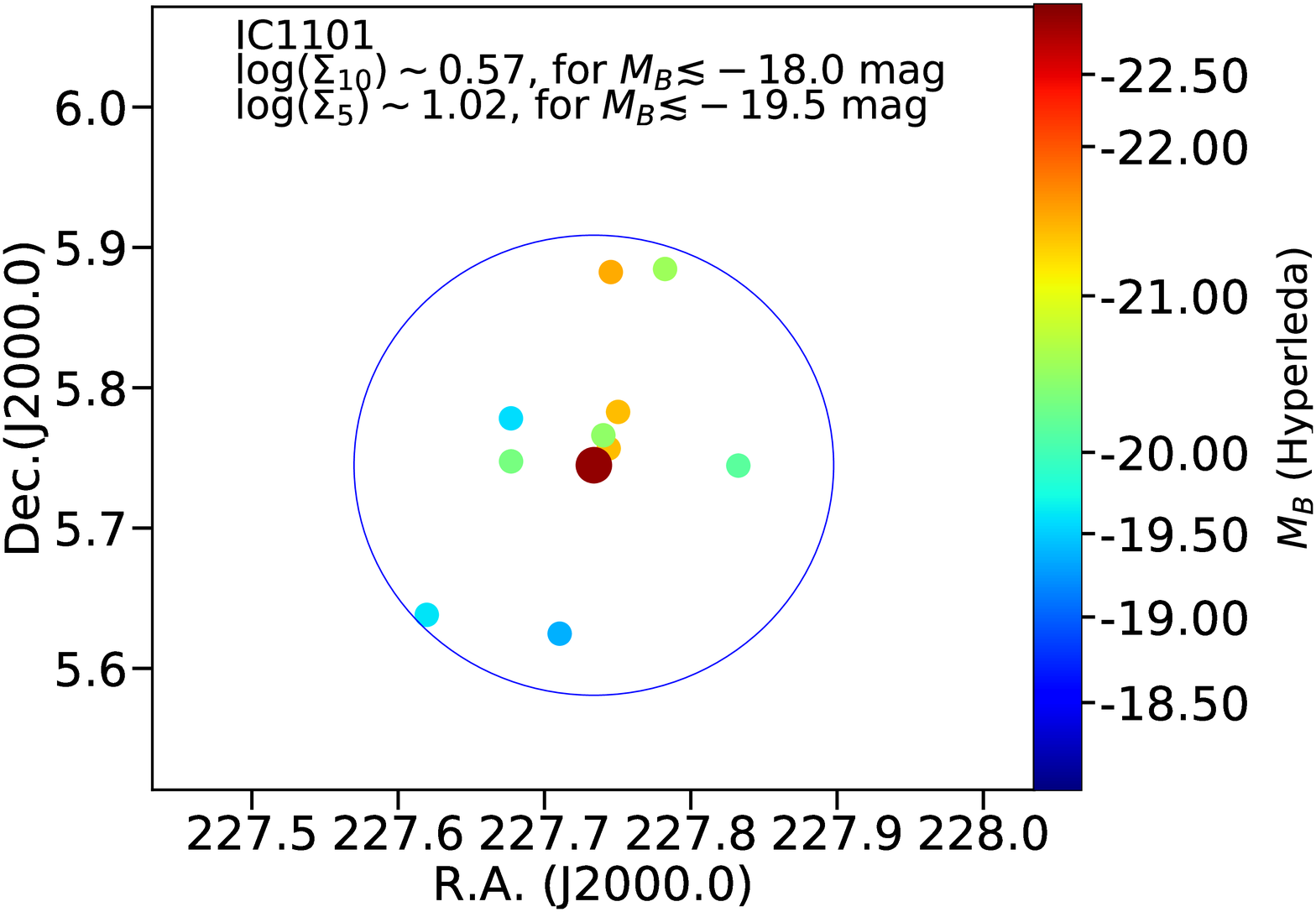}
\put(-141.9,98.611){\color{black}{{\scriptsize $R_{\rm b} \sim $ 4.20 kpc}}}
\caption{Spatial distribution of our large-core galaxies (big filled
  circles) and their 10 nearest neighbours with $M_{B} \la -18.0$ mag
  and $| V_{\rm hel,large-core} -V_{\rm hel,neighbour}|<$ 300 km
  s$^{-1}$ (small filled circles). The spatial distribution, nearest
  neighbour identification and recession velocities ($V_{\rm hel}$) are based on NED,
  while the $B$-band absolute galaxy magnitudes are from Hyperleda. We
  excluded the sample galaxy 4C +74.13 with no robust data for its
  nearest neighbours in NED.  The blue solid circles, centred on the
  large-core galaxies, enclose the 10 nearest neighbours.
  $\Sigma_{10}$=${\rm N_{gal}}$/($\pi R_{10}^{2}$) and
  $\Sigma_{5}$=${\rm N_{gal}}$/($\pi R_{5}^{2}$) are the surface
  density measurements in Mpc$^{-2}$, where the radius $R_{5}$
  ($R_{10}$), centred on the large-core galaxies, encloses the 5 (10)
  nearest neighbours with $M_{B} \la -19.5$ mag ( $M_{B} \la -18.0$
  mag), see \citet[their Section~3.1]{2011MNRAS.416.1680C}.  The
  caveat here is that galaxies with $M_{B} \ga -19.5$
  mag may be too faint to be detected at  the distances of    4C
  +74.13 (D $\sim$ 925 Mpc), A2261-BCG (D $\sim$ 959
  Mpc)  and  IC~1101 (D $\sim$ 363 Mpc). Therefore, we caution
  that for A2261-BCG and IC~1101,  the $\Sigma_{10}$ values are likely biased
  toward low values compared to other
  large-core galaxies shown here with D$\la$ 213 Mpc. }
\label{FigAII} 
 \end{figure*}

\section{Appendix B}\label{AppB}

Spatial distribution of our large-core galaxies and their nearest
neighbours (Fig.~\ref{FigAII}).

\end{appendices}

\label{lastpage}
\end{document}